%% file: NatureMethods.tex
\documentclass[pdflatex,sn-nature, twocolumn]{sn-jnl}

\usepackage{graphicx}%
\usepackage{multirow}%
\usepackage{amsmath,amssymb,amsfonts}%
\usepackage{amsthm}%
\usepackage{mathrsfs}%
\usepackage[title]{appendix}%
\usepackage{xcolor}%
\usepackage{textcomp}%
\usepackage{manyfoot}%
\usepackage{booktabs}%
\usepackage{algorithm}%
\usepackage{algorithmicx}%
\usepackage{algpseudocode}%
\usepackage{listings}%
\setcitestyle{super,open={},close={}}

\theoremstyle{thmstyleone}%
\theoremstyle{thmstyletwo}%

\theoremstyle{thmstylethree}%

\input{preambule.tex}
\input{commands.tex}

\def\ps{0.5}
\def\ms{1pt}
\def\lw{1pt}

\def\sz{5.5cm}

\begin{document}

\title{CryoLithe: Rapid Cryo-ET Reconstruction via Transform-Localized Deep Learning}

\author[1]{\fnm{Vinith} \sur{Kishore}}
\author[2]{\fnm{Valentin} \sur{Debarnot}}
\author[1]{\fnm{Amir} \sur{Khorashadizadeh}}
\author[3]{\fnm{Ricardo D.} \sur{Righetto}}
\author[3]{\fnm{Benjamin D.} \sur{Engel}}
\author[1]{\fnm{Ivan} \sur{Dokmani\'c}}

\affil[1]{\orgdiv{Department of Mathematics and Computer Science}, \orgname{University of Basel}, \orgaddress{\city{Basel}, \postcode{4051}, \country{Switzerland}}}
\affil[2]{\orgname{INSA‐Lyon, Universite Claude Bernard Lyon 1, CNRS, Inserm, CREATIS UMR 5220, U1294}, \orgaddress{\city{Lyon}, \country{France}}}
\affil[3]{\orgdiv{Biozentrum}, \orgname{University of Basel}, \orgaddress{\city{Basel}, \postcode{4056}, \country{Switzerland}}}


\abstract{
Cryo-electron tomography (cryo-ET) enables 3D visualization of cellular structures. Accurate reconstruction of high-resolution volumes is complicated by the very low signal-to-noise ratio and a restricted range of sample tilts. Recent self-supervised deep learning approaches, which post-process initial reconstructions by filtered backprojection (FBP), have significantly improved reconstruction quality with respect to signal processing iterative algorithms, but they are slow, taking dozens of hours for an expert to reconstruct a tomogram and demand large memory. 
We present \method{}, an end-to-end network that directly estimates the volume from an aligned tilt series. 
\method{} achieves denoising and missing wedge correction comparable or better than state-of-the-art self-supervised deep learning approaches such as \icecream{}, Cryo-CARE, IsoNet or DeepDeWedge, while being two orders of magnitude faster. 
To achieve this, we implement a local, memory-efficient reconstruction network.
We demonstrate that leveraging transform-domain locality makes our network robust to distribution shifts, enabling effective supervised training and giving excellent results on real data---without retraining or fine-tuning.
\method{} reconstructions facilitate downstream cryo-ET analysis, including segmentation and subtomogram averaging and is openly available: \git.

}

\keywords{Cryogenic Electron Tomography, Cryo-ET, Denoising, Missing Wedge, Deep Learning, Supervised Learning.}



\maketitle


\addcontentsline{toc}{section}{Main}
Cryogenic electron tomography (cryo-ET) visualizes biological molecules and intracellular structures in their native 3D environments at nanometer resolution. Unlike single-particle cryo-electron microscopy (cryo-EM), nuclear magnetic resonance (NMR), or X-ray crystallography, cryo-ET captures the full intracellular context that is crucial for understanding biological function. This  bridges the gap between molecular imaging and cellular analysis, offering high-resolution insights into the native cell organization \cite{navarro2022quantitative,mccafferty2024integrating}.

Cryo-ET has recently experienced rapid progress thanks to advancements across the entire imaging pipeline. Improvements in sample preparation—such as cryo-focused ion beam (FIB) automation and cryo-lift-out techniques—have enabled more consistent and targeted sample thinning \cite{schaffer2019cryo,klumpe2021modular, kelley2022waffle}. Automated data collection strategies like PACE-tomo have streamlined tilt-series acquisition, increasing throughput and reproducibility \cite{eisenstein2023parallel}. Powerful self-supervised denoising methods such as Cryo-CARE \cite{buchholz2019cryo}, IsoNet \cite{liu2022isotropic},  DeepDeWedge \cite{wiedemann2024deep} and \icecream{} \cite{kishore2025icecream} have dramatically improved the quality of noisy reconstructions by leveraging data redundancy without requiring clean ground truth volumes. High-resolution subtomogram averaging methods, including those implemented in M \cite{tegunov2021multi} and RELION-5 \cite{burt2024image}, have pushed the limits of in situ structure determination by enabling accurate alignment and classification of subvolumes.

The cryo-ET pipeline begins with biological samples being vitrified and then imaged with a transmission electron microscope to obtain projections at a sequence of tilt angles. The sample deteriorates with each exposure, which limits the electron dose and, in turn, results in very low signal-to-noise ratio. Early reconstruction algorithms such as the filtered backprojection (FBP) rely on analytic properties of the imaging operator \cite{kak2001principles,harauz1986exact}. Recently self-supervised deep learning methods have proven very effective in denoising and improving such reconstructions. These methods can be trained with only noisy raw data, without noiseless ground truth required by supervised methods which is generally unavailable in biological imaging. Among the most popular is Cryo-CARE \cite{buchholz2019cryo} which relies on the Noise2Noise framework \cite{lehtinen2018noise2noise}. Topaz-Denoise similarly uses Noise2Noise to train a general denoiser on a large set of volumes \cite{bepler2020topaz}. 

\begin{figure*}
	\begin{subfigure}[c]{0.48\textwidth}
		\centering
		\includegraphics[scale=0.25]{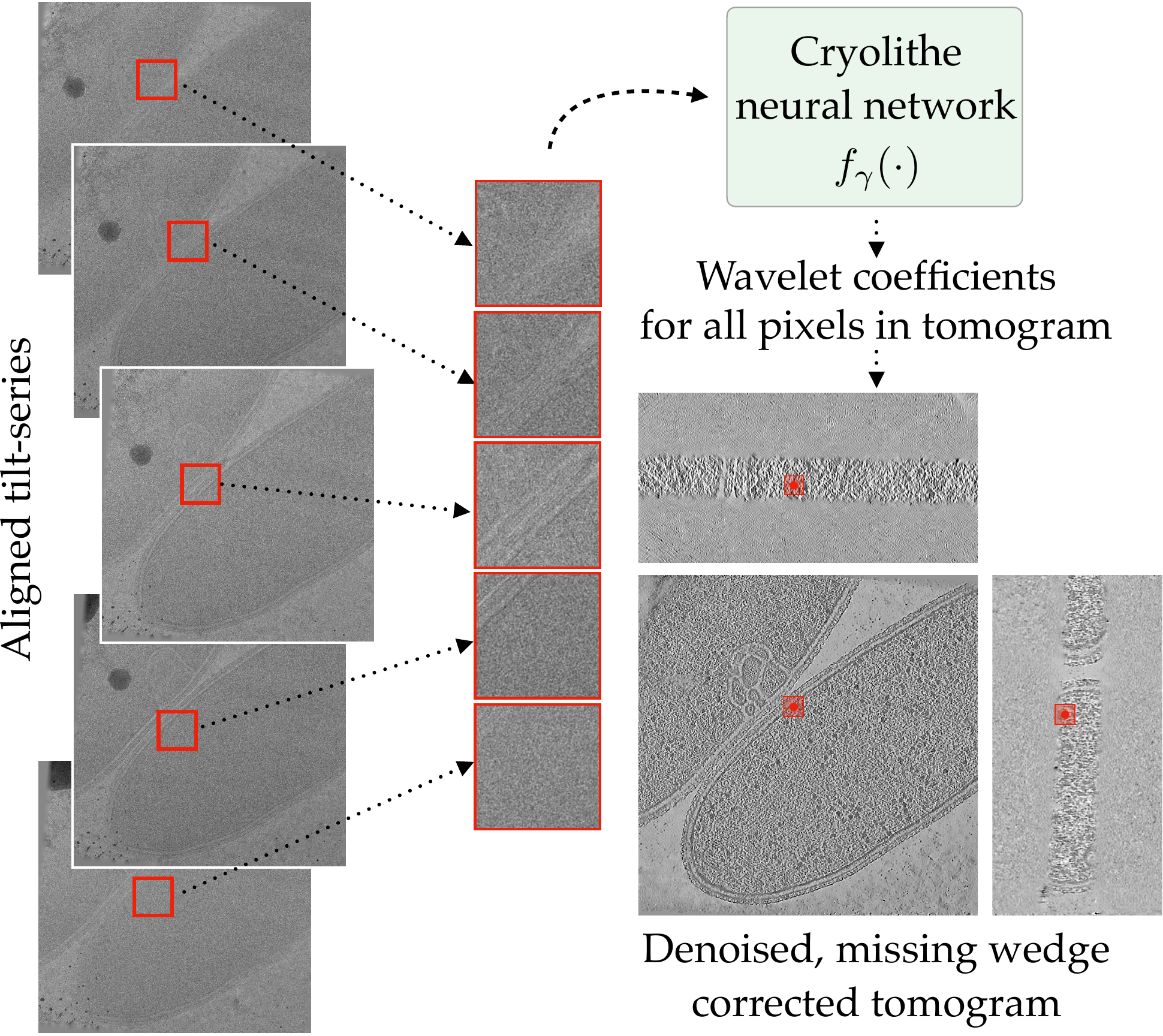}
		\caption{\method{} estimates a local portion of the unknown volume density (red cube) by processing corresponding localized information directly in the aligned tilt series (red squares). The patch information is transformed by a pre-trained neural network that estimates the corresponding portion of the tomogram. To fasten inference, \method{} predicts wavelet coefficients that are used to render the final tomogram after an inverse wavelet transform.} \label{fig:cryoET_process}
	\end{subfigure}\hfill
\begin{tabular}[c]{@{}c@{}}
\begin{subfigure}[c]{0.5\textwidth}
\centering
          \begin{tikzpicture}
            \begin{axis}[
            width=0.7\linewidth,
            height=0.5\linewidth,
            axis lines = left,
            scaled ticks=false,
            xmode=log,
            xmin = 2, xmax= 10, ymin= 0, ymax = 26000,
            xtick={2, 3, 4, 6, 8},
            xticklabels={$2048 \times 2048 \times 1024$,$1366 \times 1366 \times 682$, $1024 \times 1024 \times 512$, $684 \times 684 \times 342$,$512 \times 512 \times 256$},
            ytick={3600, 10800, 18000, 25200},
            yticklabels={1, 3, 5, 7},
            yticklabel style={
            /pgf/number format/fixed,
            /pgf/number format/precision=5},
            xticklabel style= {rotate=15,anchor=north east, font=\tiny},
			grid=major, 
			grid style={dashed,gray!30}, 
			xlabel= {\notsotiny{Tomogram size}},
			ylabel={\notsotiny{Processign time (hours)}},
            legend style={at={(1.,0.5)}, legend cell align=right, align=right, draw=none,font=\notsotiny}]
			\addplot[mark=\isoCaremark, mark size=\ms, line width=\lw,  color=\isoCareColor] coordinates{ (2,23530)(3,18704)(4,18257) (6,18036) (8,17980) };
			\addlegendentry{FBP+Cryo-CARE+IsoNet}
            \addplot[mark=\methodWmark, mark size=\ms, line width=\lw,  color=\methodWColor] coordinates{ (2,2702.5)(3,806.2)(4,341.2) (6,45) (8,45) };
            \addlegendentry{\method{}}
            \end{axis}
            \end{tikzpicture} 
            \vspace{-0.2cm}
            \caption{  Computational time against target tomogram size. \vspace{-0.3cm}}
            \end{subfigure} \\
            \noalign{\bigskip}%
\begin{subfigure}[c]{0.5\textwidth}
\centering
          \begin{tikzpicture}
            \begin{axis}[
            width=0.7\linewidth,
            height=0.5\linewidth,
            axis lines = left,
            scaled ticks=false,
            xmin = 4, xmax= 40, ymin= 0, ymax = 600,
            xtick={4, 8, 12,16, 24, 32, 3},
            ytick={ 300, 600, 1200, 1800, 3000},
            yticklabels={5, 10,  20, 30,50},
            yticklabel style={
            /pgf/number format/fixed,
            /pgf/number format/precision=5},
			grid=major, 
			grid style={dashed,gray!30}, 
			xlabel= {\notsotiny{GPU memory.}},
			ylabel={\notsotiny{Processign time (minutes)}},
            legend style={at={(1.,1.)}, legend cell align=right, align=right, draw=none,font=\notsotiny}]
  
            \addplot[mark=\methodWmark, mark size=\ms, line width=\lw,  color=\methodWColor] coordinates{ (4,366.18) (8,289.54) (12, 286.26) (16, 285.25) (24, 287.83) (32, 285.89 )  (36,290.03 )};
            \end{axis}
            \node[ rotate=90] at (0,-1.4) {};
            \end{tikzpicture} 
            \vspace{-0.5cm}
            \caption{Inference time versus memory available on the GPU for \method{}.}
            \end{subfigure}
  \end{tabular}
    \caption{\method{} is a supervised deep learning method that is able to reconstruct tomograms in a couple of minutes without any parameter tuning. Existing self-supervised deep learning methods may require from hours to day and the expertise for training one or several neural networks. 
    Our proposed localized architecture, inspired by the tomography imaging operator, scales favorably with any GPU whatever its memory capacity.  }
    \label{fig:time}
\end{figure*}

Beyond high noise, another challenge is that due to the sample geometry and tilt stage mechanics, the range of tilt angles is limited, typically between $-60^{\circ}$ and $60^{\circ}$, with a step of $2^{\circ}$ or $3^{\circ}$.  This  results in the notorious ``missing wedge problem''--- a wedge-shaped gap of missing information in the reciprocal Fourier space \cite{shkolnisky2012viewing}---and to the corresponding missing wedge artifacts in the FBP reconstruction.
A promising recent approach to the missing wedge problem is again by self-supervised learning which exploits the presence of cellular features such as proteins and membranes in random orientations within a tomogram. The first method to do this in cryo-ET was IsoNet which has been shown to perform well in practice, especially when combined with a denoising algorithm \cite{liu2022isotropic}. The more recent DeepDeWedge method jointly learns to denoise and correct the missing wedge, combining Noise2Noise and IsoNet-like attributes in a single framework \cite{wiedemann2024deep}. 
Unfortunately, using self-supervised learning requires training a deep neural network to cryo-ET data, which is rendered difficult by their critical dependence on several parameters and their long training time (between 6 and 24 hours on a GeForce RTX 4090 GPU with 24 GB memory). In this paper, we introduce a rapid reconstruction method that produces enhanced tomograms in a couple of minutes on a standard GPU, without any parameter tuning, see Fig.~\ref{fig:time}.

A first fundamental downside of all the above denoising and missing wedge correction approaches is that they are post-processing methods. They are applied to the initial reconstructions obtained by FBP. The fixed FBP, however, is a suboptimal step that introduces artifacts and spatially correlated noise due to information being delocalized \cite{quinto2009electron}. Here, we introduce a denoising and missing wedge correction method that directly processes the aligned tilt series.

The second limiting aspect of existing approaches is that they rely on self-supervised deep learning.
The development of supervised methods for cryo-ET is stymied by the absence of ground truth data. In an attempt to get around this problem, several simulators have been created to generate realistic data \cite{purnell2023rapid,hendriksen2021tomosipo,harastani2024template,gubins2020shrec}. They leverage known structures from the protein data bank \cite{burley2017protein} to create protein-dense synthetic volumes. These simulators have been used in deep learning methods for segmentation of membranes \cite{martinez2023simulating, lamm2024membrain}, for template matching \cite{harastani2024template}, and to train tomogram denoising networks \cite{purnell2023rapid}. But using simulated volumes to train reconstruction networks has been hampered by the complexity of the full cryo-ET forward model and the inadequate approximation of real cellular volumes by a simple random distribution of proteins in the cytosol. Training on simulated volumes and measurements generalizes poorly to messy and complex real data. In this paper, we show that it is possible to circumvent the need to obtain synthetic ground truth by selecting a small number of experimental tilt series for which a high-quality tomogram reconstruction is available.

We introduce \method{}, an end-to-end neural network that directly estimates the volume from an aligned tilt series. \method{} is based on a physics-informed localized neural network architecture specifically designed to exploit the property of the tomography imaging operator. Transform-domain locality greatly alleviates overfitting, reduces the memory footprint and enables end-to-end training, resulting in the first general-purpose neural network for cryo-ET reconstruction that can be used off-the-shelf to directly process any aligned tilt series.
In order to train \method{}, we carefully design a training set by combining several packages to obtain denoised, missing-wedge corrected tomograms from real data. 
All things considered, we show that once trained, \method{} performs consistently on qualitatively distinct test volume distributions.

\def\ps{0.5}

\addcontentsline{toc}{section}{Results}
\section*{Results}

\subsection*{Training with pairs of tilt series and tomograms} 
The training dataset is central to any supervised learning algorithm. \method{} is by design highly robust to previously unseen data, even when it differs from the training set. This is due to the localized neural network architecture that is based on the physics of the acquisition system, see Methods.
While \method{} can be trained with a small number of training examples, their quality is essential, as captured by the well-known adage: "garbage in, garbage out".
For this reason, alongside the package, we release a carefully curated training dataset, ranging from aligned tilt series to sharp, denoised 3D tomograms from EMPIAR-11830. This unique dataset of 114 paired tilt series and tomograms provides a valuable foundation for developing new learning-based algorithms in cryo-ET. 
Notice that only tilt series from EMPIAR-11830 have been used for training. Despite this, \method{} remains highly effective on tomograms from different microscopes, with different pixel sizes, or even with different samples.
Because the cryo-ET acquisition process is difficult to model—particularly the non-linear deformations caused by ice, scattering, noise, beam-induced motion, stage drift, and other factors—we found that training on real tilt series data is crucial for achieving optimal performance. To generate high-quality reconstructions from these tilt series, we combined FBP with either \icecream{} \cite{kishore2025icecream} (88 volumes) or Cryo-CARE and IsoNet (26 volumes), both state-of-the-art self-supervised deep learning tools. Example of training volumes are shown in Fig.~\ref{fig:mosaic-training}.

\subsection*{Reconstruction of diverse cryo-ET datasets}
\method{} is trained on a rich set of tomograms from diverse cryo-ET datasets, covering a wide range of biological structures and imaging conditions, with and without gold fiducials.
To verify that the performance of our method remains similar across various biological samples imaged with different microscopes and acquisition settings, we display in Fig.~\ref{fig:mosaic-2} a panel of various \method{} reconstructions from tilt series downloaded from EMPIAR. The FBP reconstructions are displayed in Appendix, in Fig.~\ref{fig:mosaic-2-fbp}. 
Processing information is displayed in Table \ref{table:mosaic-2}. The processing time is evaluated on a GeForce RTX 4090 GPU with 24 GB memory.

\begin{table*}
    \centering
    \begin{tabular}{@{}c@{}c@{}c@{}}
        \overlayerimage{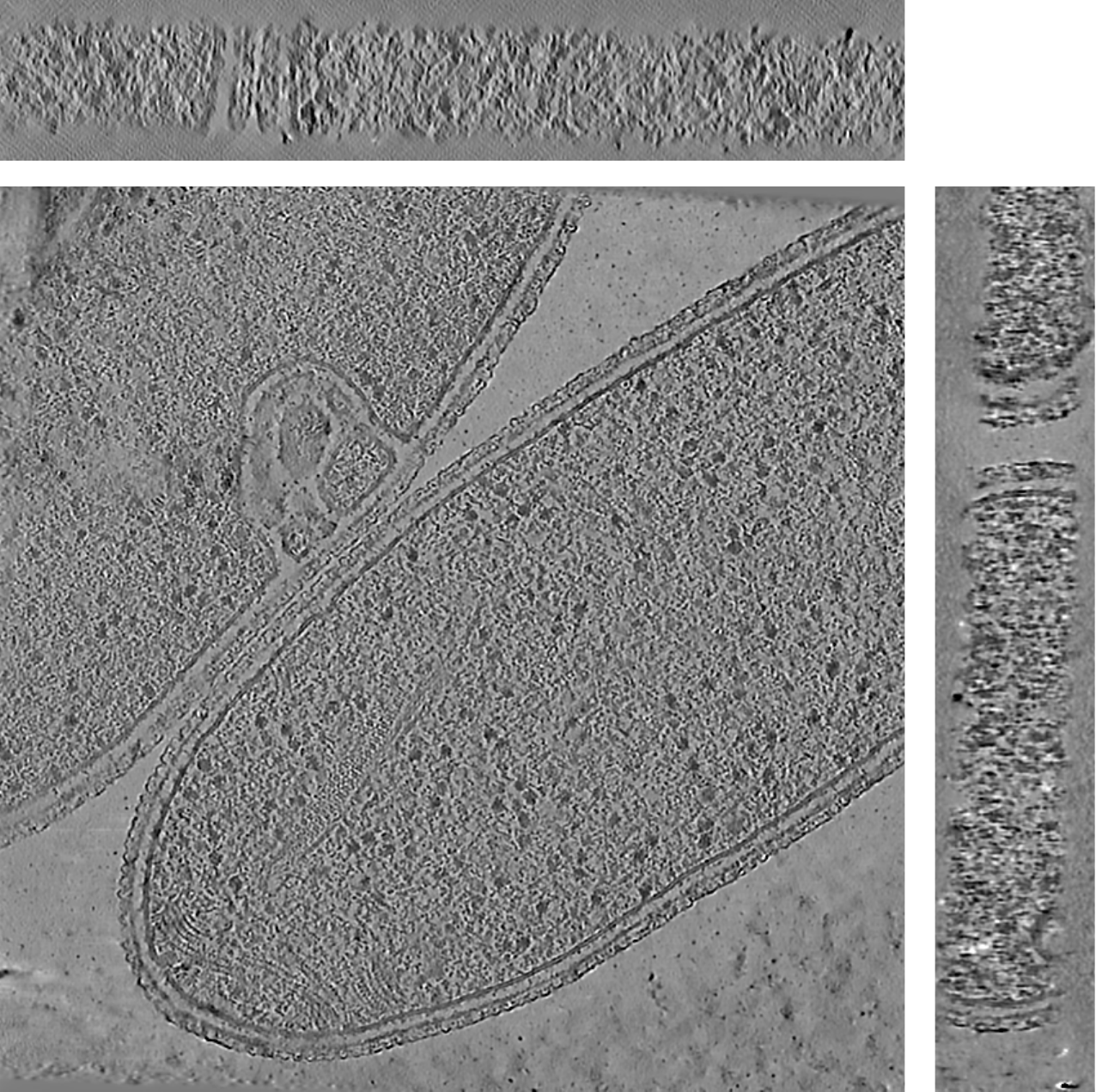}{\sz}{0.247}{100 nm}{A.} &
        \hspace{0.5mm} 
        \overlayerimage{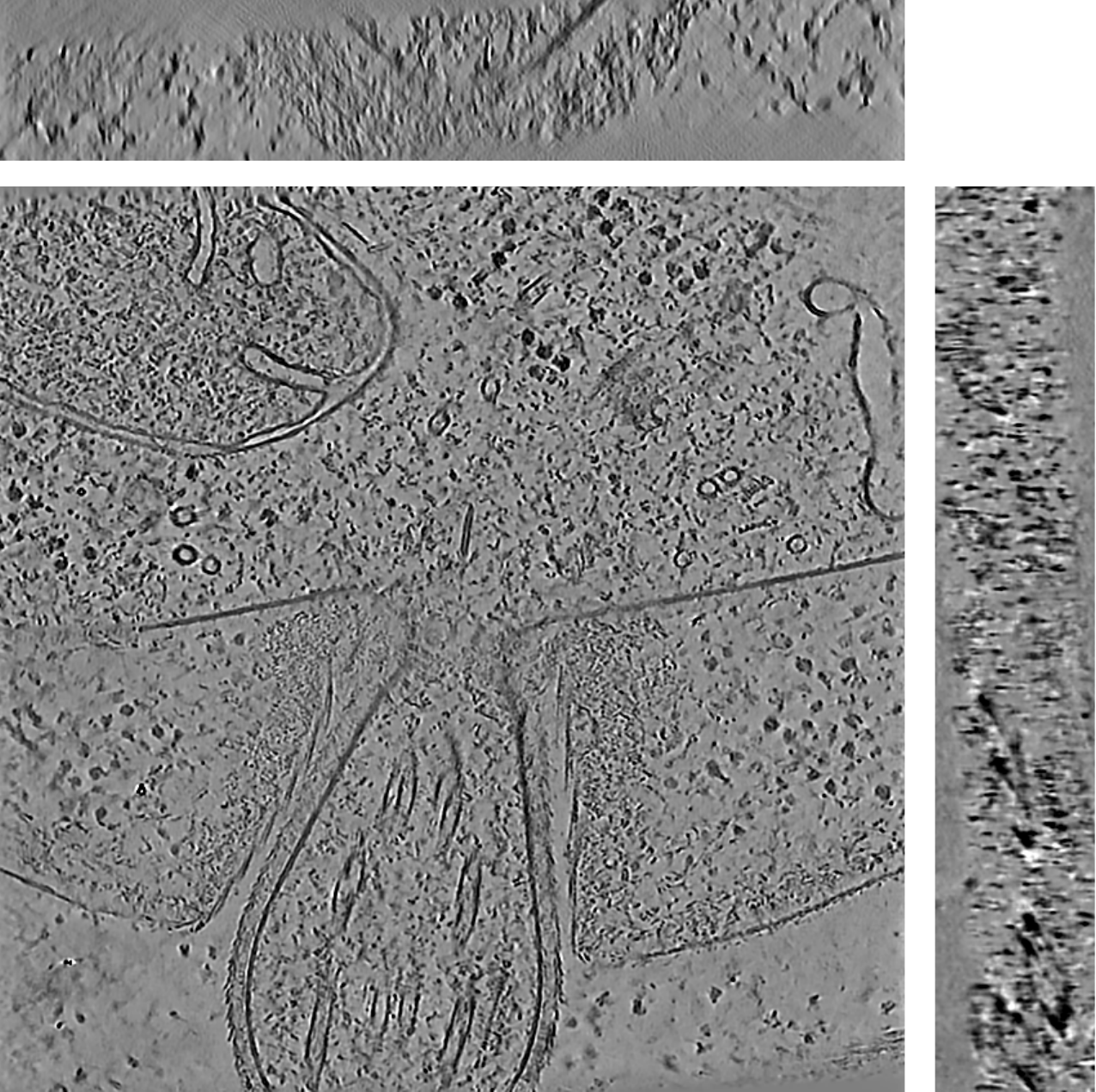}{\sz}{0.255}{100 nm}{B.} &
        \hspace{0.5mm} 
        \overlayerimage{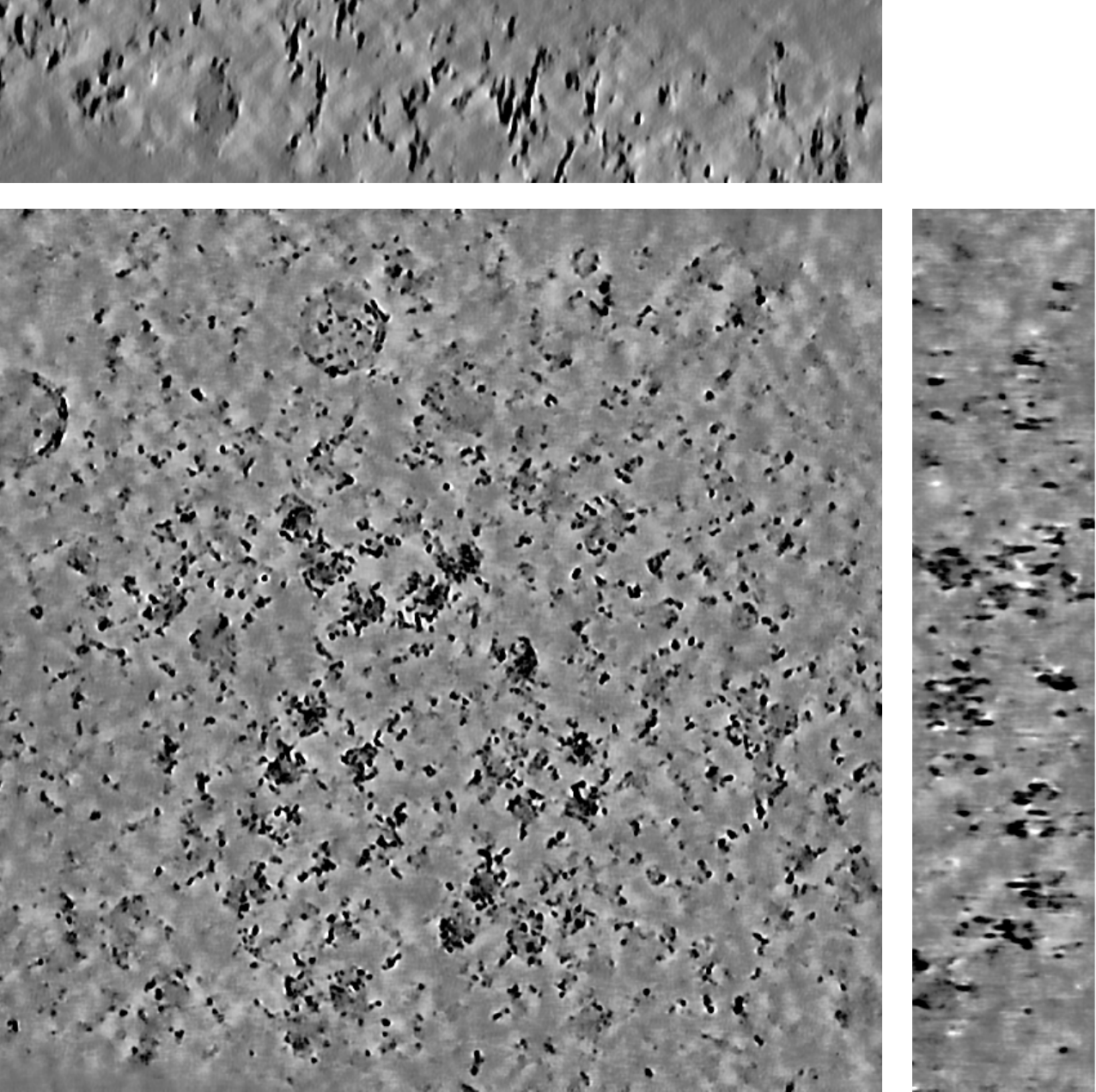}{\sz}{0.4177}{50 nm}{C.}  \\
        \overlayerimage{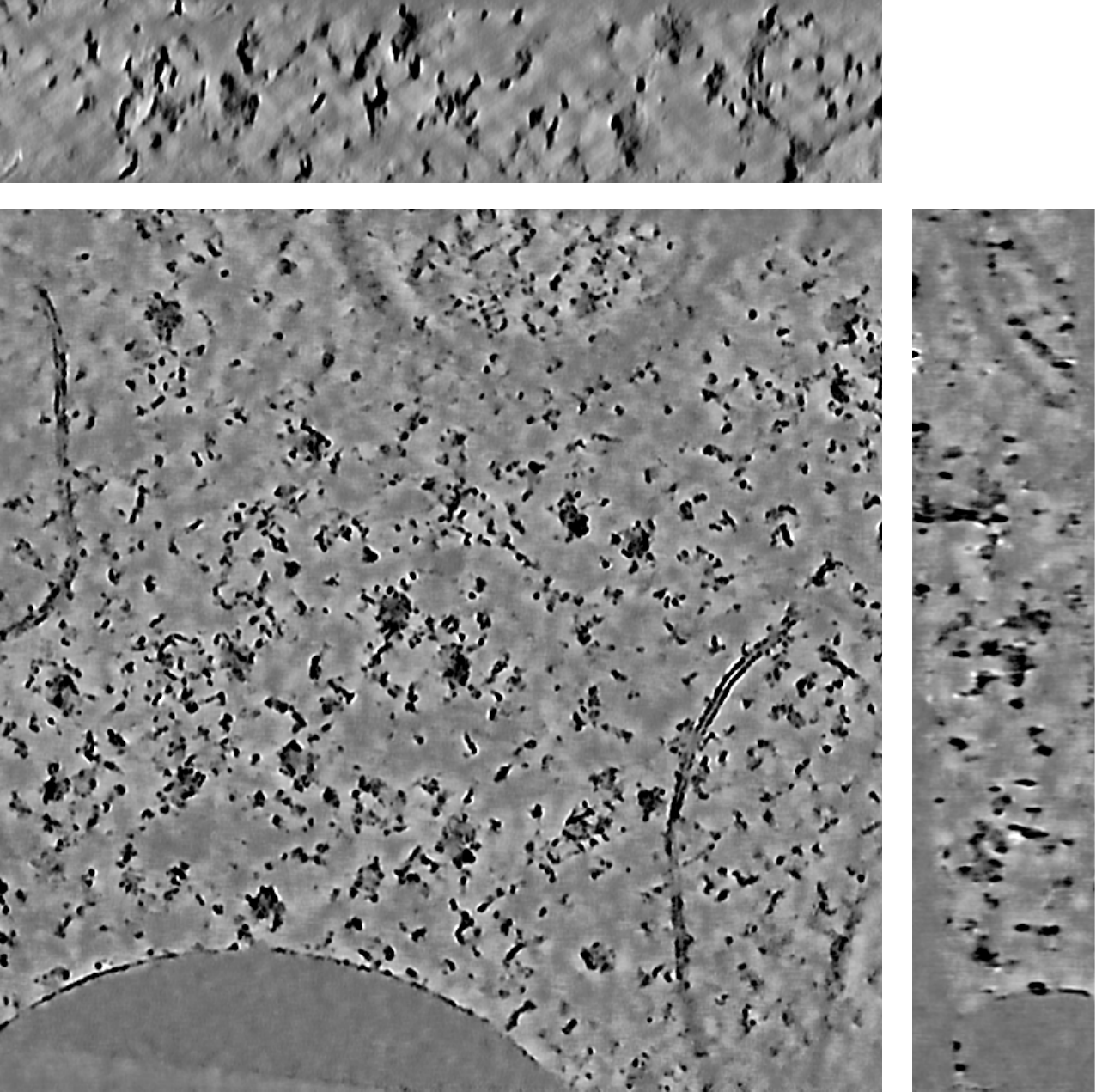}{\sz}{0.2956}{100 nm}{D.}  &
        \hspace{0.5mm} 
        \overlayerimage{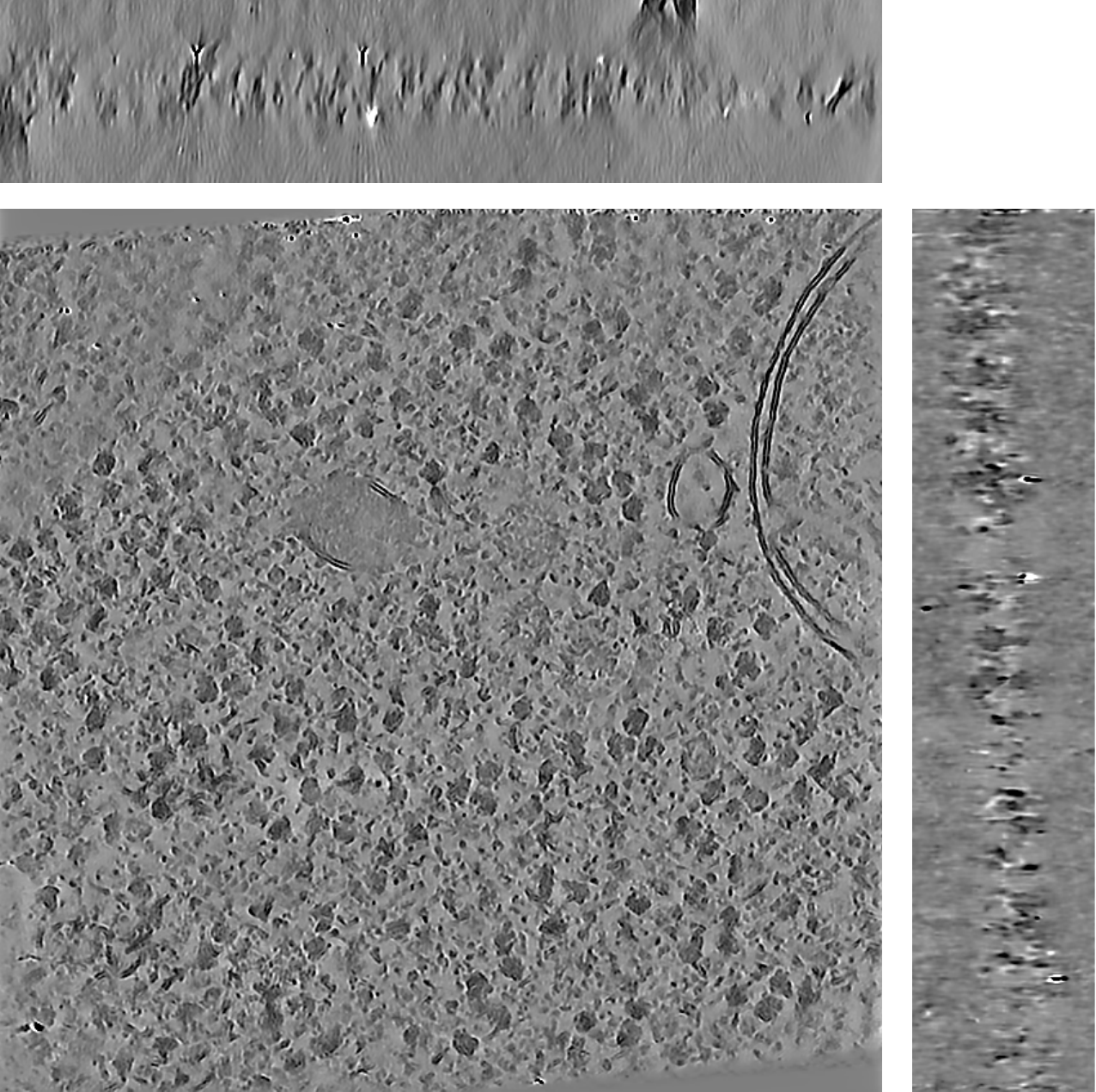}{\sz}{0.488}{50 nm}{E.}  &
        \hspace{0.5mm} 
        \overlayerimage{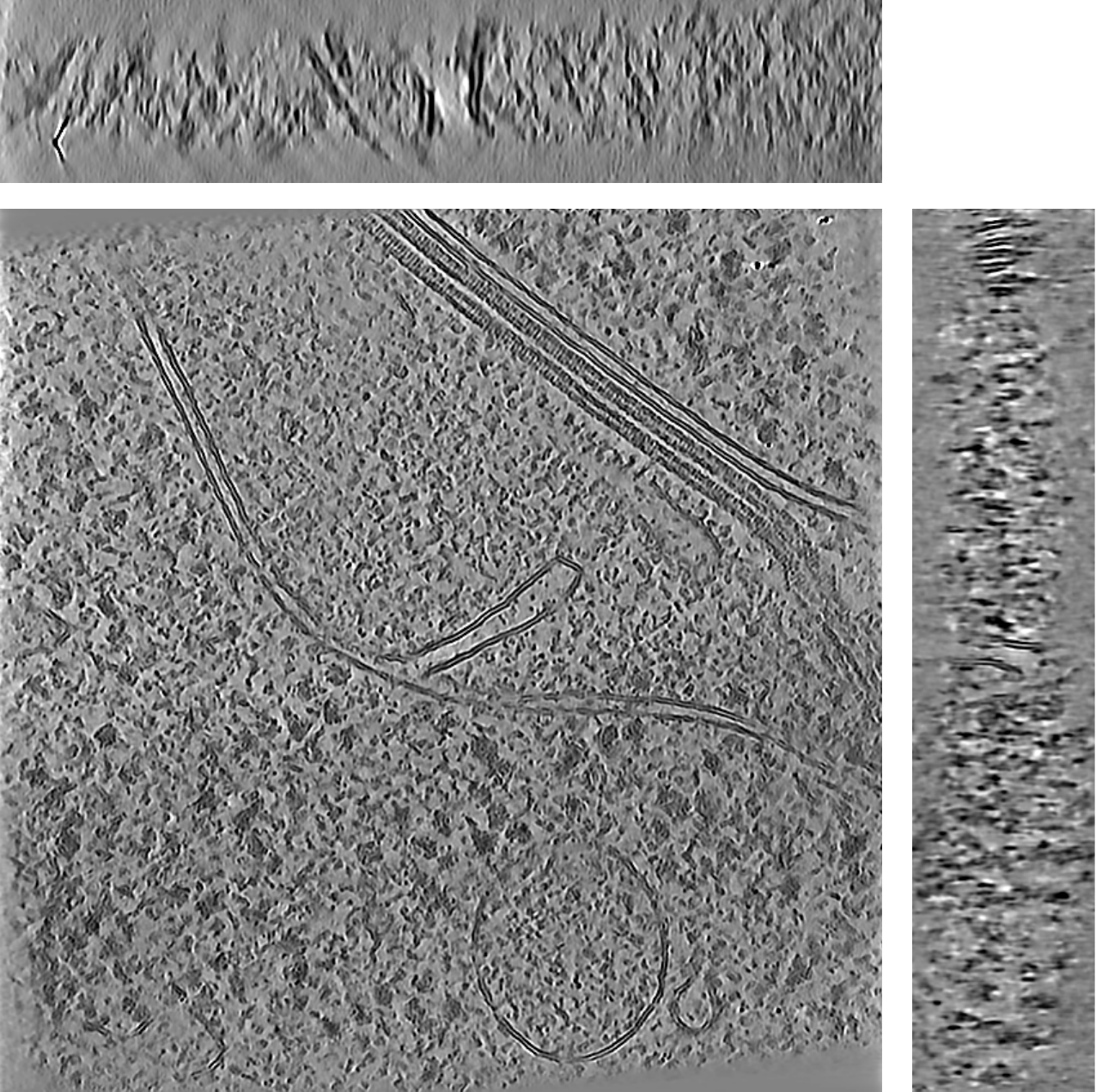}{\sz}{0.306}{100 nm}{F.}  \\
        \overlayerimage{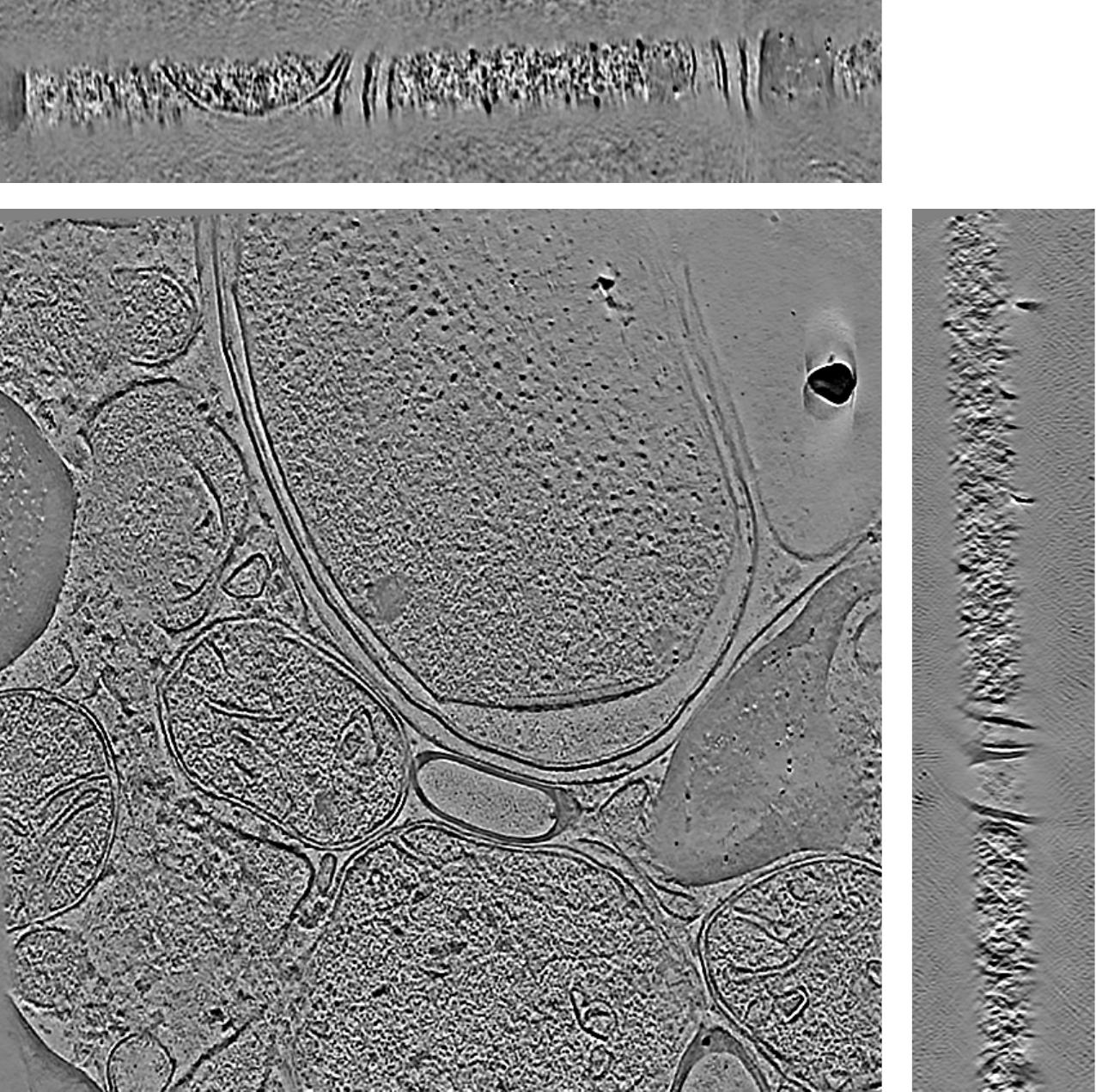}{\sz}{0.178}{100 nm}{G.} &
        \hspace{0.5mm} 
        \overlayerimage{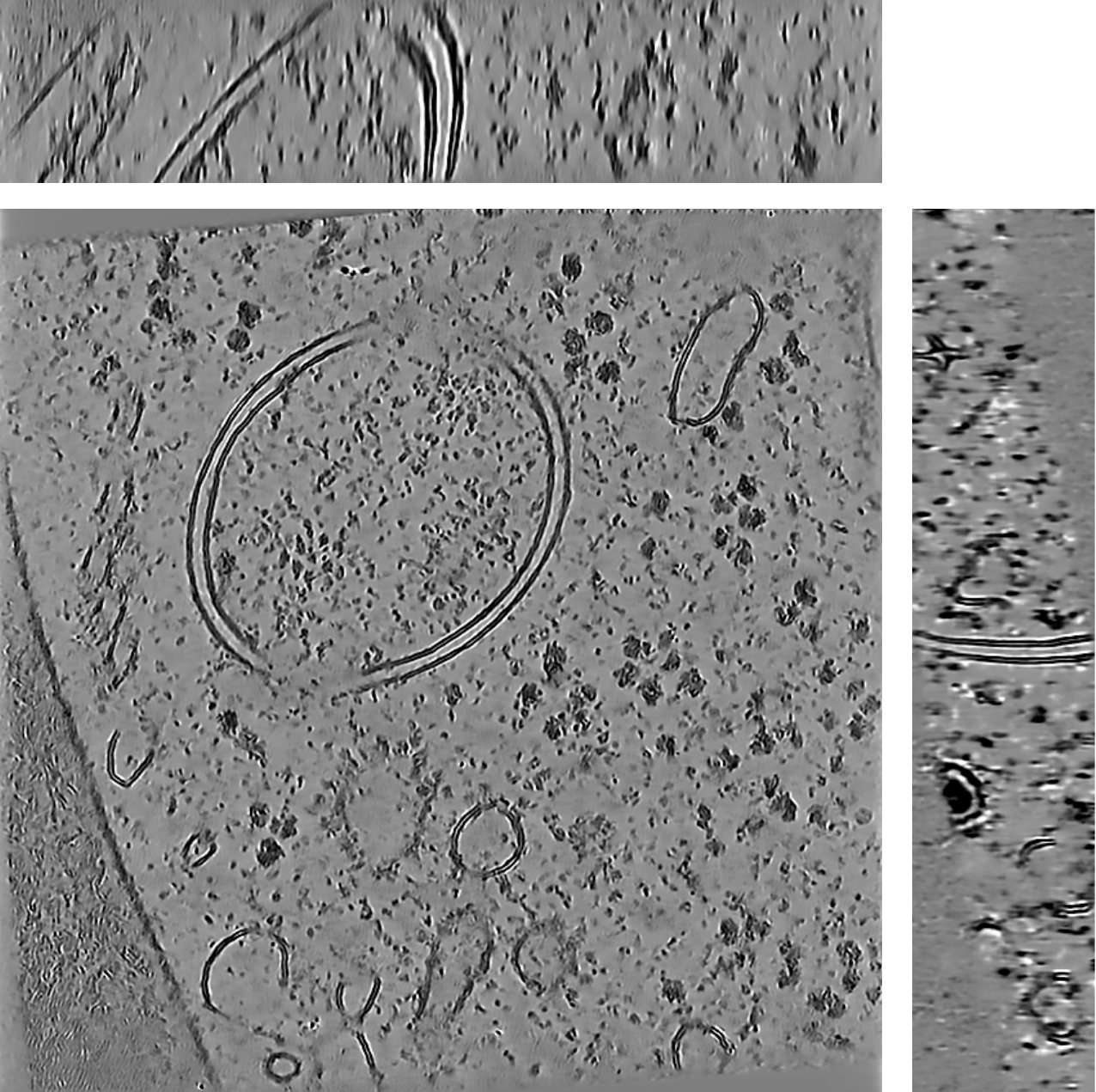}{\sz}{0.4102}{100 nm}{H.}  &
        \hspace{0.5mm} 
        \overlayerimage{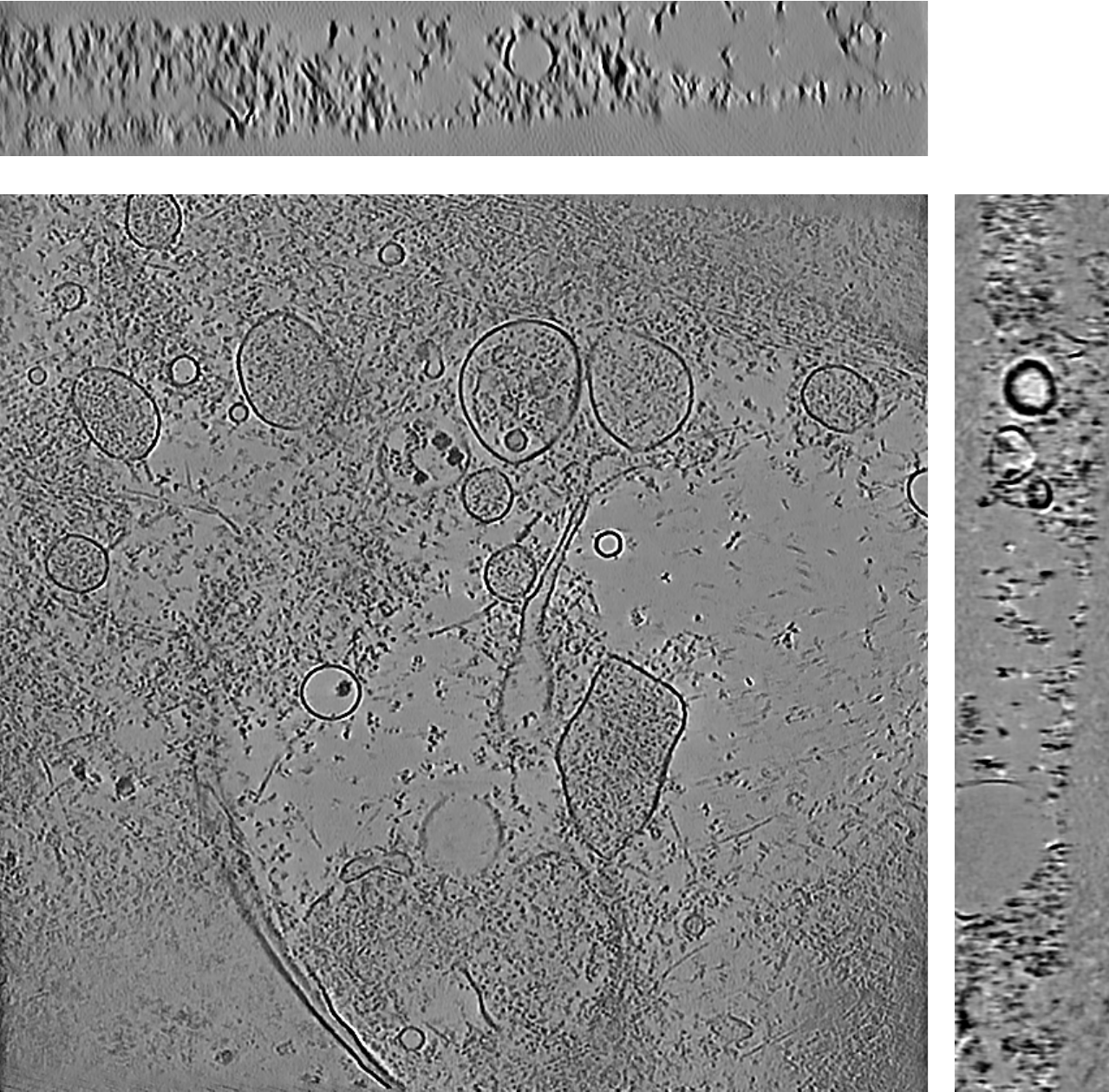}{\sz}{0.316}{50 nm}{I.}  \\
    \end{tabular}
    \captionof{figure}{
    	Orthogonal slices though a variety of tomograms reconstructed with  \method{}, representing diverse biological contexts and imaging conditions. \method{} can process arbitrary aligned tilt series and produce denoised and missing wedge corrected tomograms in 4 to 8 minutes on a GeForce RTX 4090 GPU with 24 GB memory. 
    The reconstructed tomograms are cropped to form a square image on the $x-y$ axis.  Table \ref{table:mosaic-2} provides information about the original data and their processing. The corresponding FBP reconstructions are displayed in Appendix, in Fig.~\ref{fig:mosaic-2-fbp}. \label{fig:mosaic-2}}
\end{table*}

\subsection*{Quantitative comparison}
We quantitatively validate the performance of \method{} on tilt series not seen during training, but obtained from the same dataset (EMPIAR-11830). These test tilt series were specifically selected to ensure strong performance of the baseline pipeline (FBP+\icecream{} and FBP+Cryo-CARE+IsoNet). As shown in Figure~\ref{fig:real-test}, we report the Fourier Shell Correlation (FSC) between the reference tomogram (obtained with FBP+Cryo-CARE+IsoNet) and alternative reconstructions. The data have been pre-processed with Slabify to extract only the subregion that corresponds to biological content. The FSC curve indicates that \method{} closely matches the reference reconstruction up to a resolution of 15.7 \nano\meter{} (above the 0.5-threshold)). Notably, \method{} achieves this performance with a significant speed advantage—up to $75\times$ faster—and without any parameter tuning, highlighting its practical benefits.
We provide additional quantitative evaluation of \method{} on the test dataset, see Fig.~\ref{fig:selffsc-test}.

\def\sztmp{4.8cm}
\def\ps{0.32}
\def\scc{1024}
\begin{figure*}
	\begin{subfigure}[t]{\ps\textwidth}
	    \centering
    	\begin{tikzpicture}[spy using outlines={circle,orange,magnification=4,size=2cm, connect spies}]
    		\node[ rotate=90] at (0,0) {}; outlines={circle,orange,magnification=2,size=3cm, connect spies}]
    		\node[rotate=0, line width=0.05mm, draw=white] at (0,0) { \includegraphics[height=\sz]{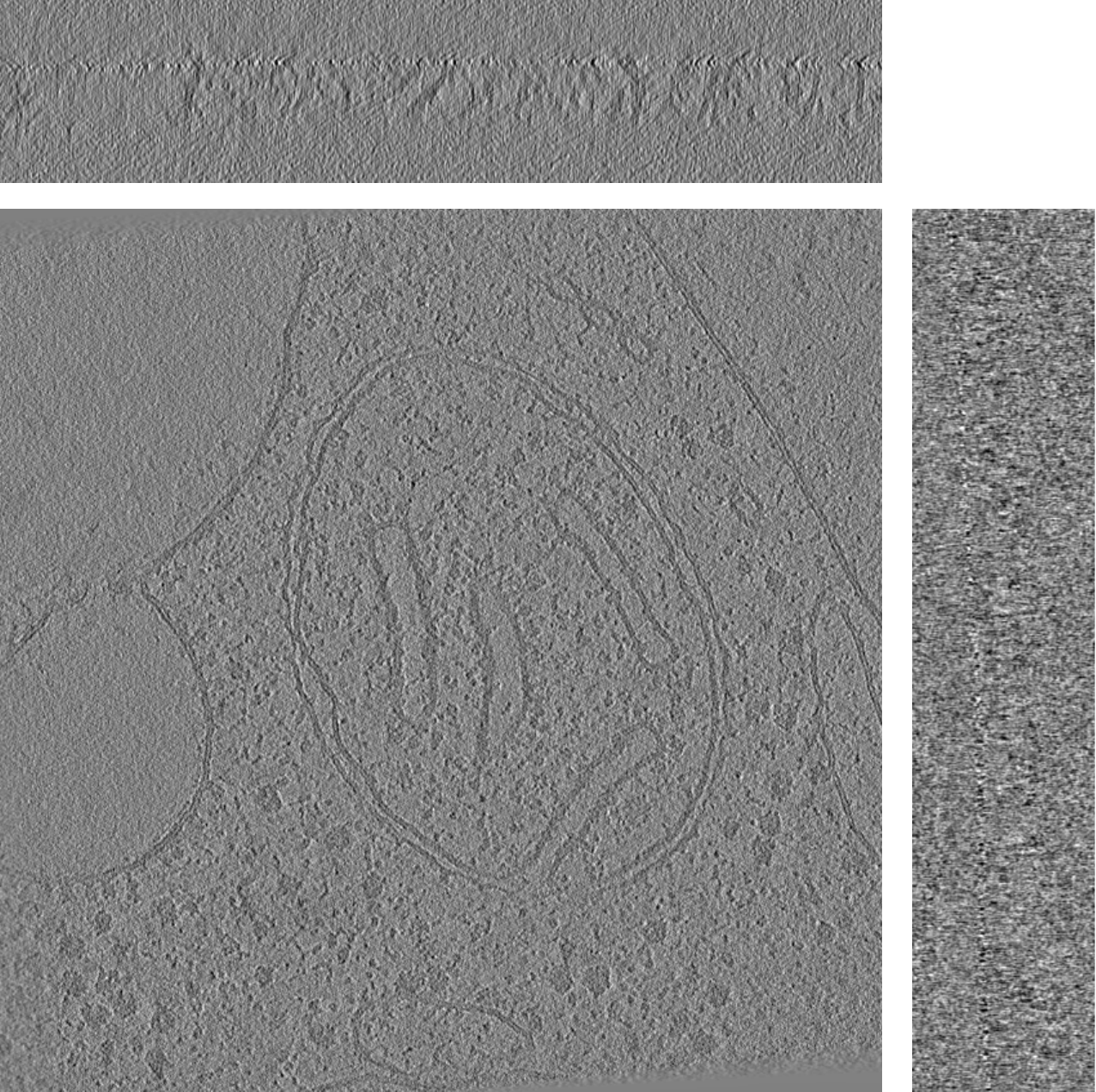}};
    		\spy on (-1.2,-0.5) in node [left] at (2.75,1.75);
		\end{tikzpicture} 
		\caption{FBP ($  4.69$sec).}
	\end{subfigure}\hfill
	\begin{subfigure}[t]{\ps\textwidth}
	    \centering
    	\begin{tikzpicture}[spy using outlines={circle,orange,magnification=4,size=2cm, connect spies}]
    		\node[ rotate=90] at (0,0) {}; outlines={circle,orange,magnification=2,size=3cm, connect spies}]
    		\node[rotate=0, line width=0.05mm, draw=white] at (0,0) { \includegraphics[height=\sz]{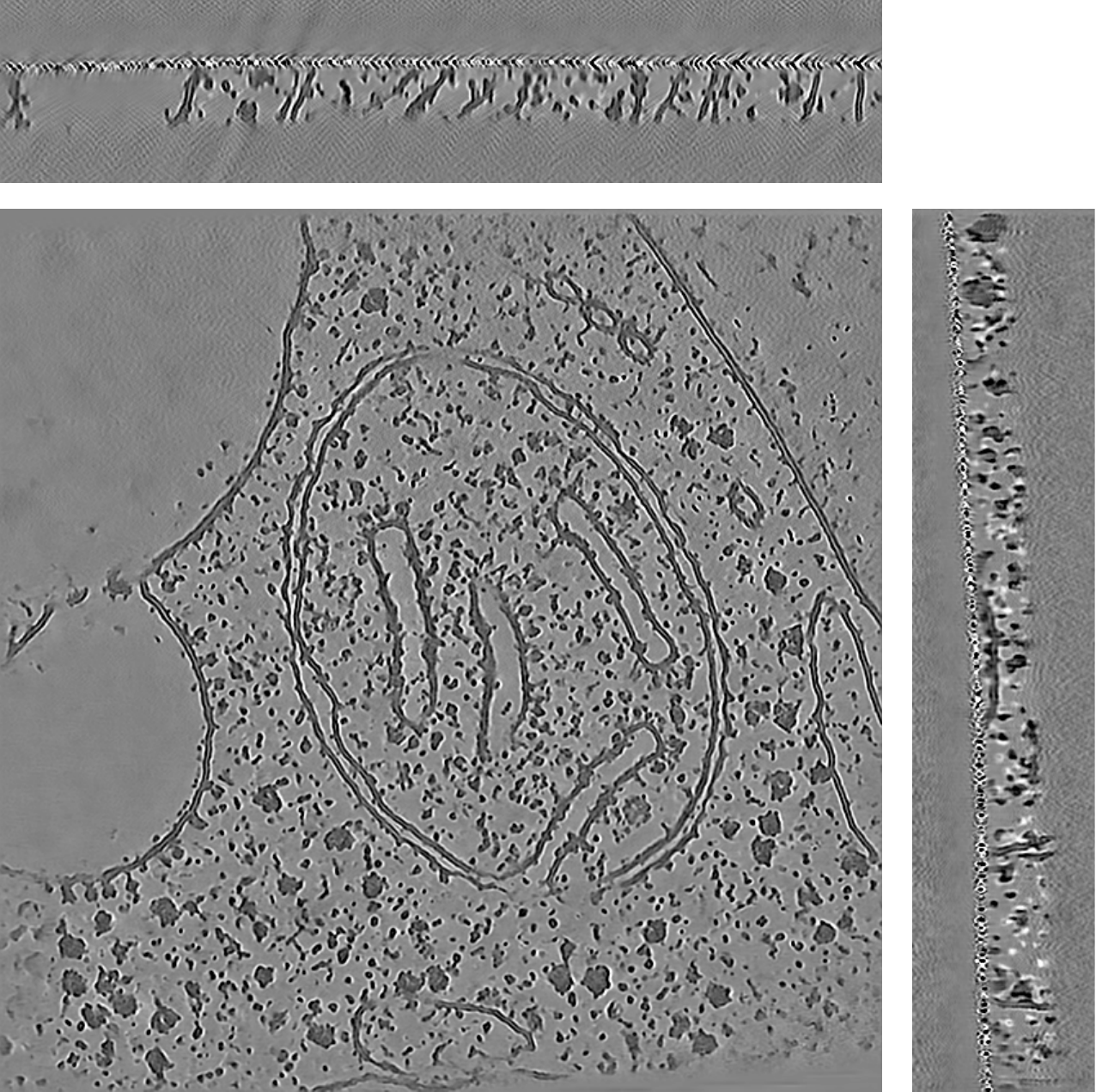}};
    		\spy on (-1.2,-0.5) in node [left] at (2.75,1.75);
		\end{tikzpicture} 
		\caption{FBP+\icecream{} ($\approx 12h$).}
	\end{subfigure}\hfill
	\begin{subfigure}[t]{\ps\textwidth}
	    \centering
    	\begin{tikzpicture}[spy using outlines={circle,orange,magnification=4,size=2cm, connect spies}]
    		\node[ rotate=90] at (0,0) {}; outlines={circle,orange,magnification=2,size=3cm, connect spies}]
    		\node[rotate=0, line width=0.05mm, draw=white] at (0,0) { \includegraphics[height=\sz]{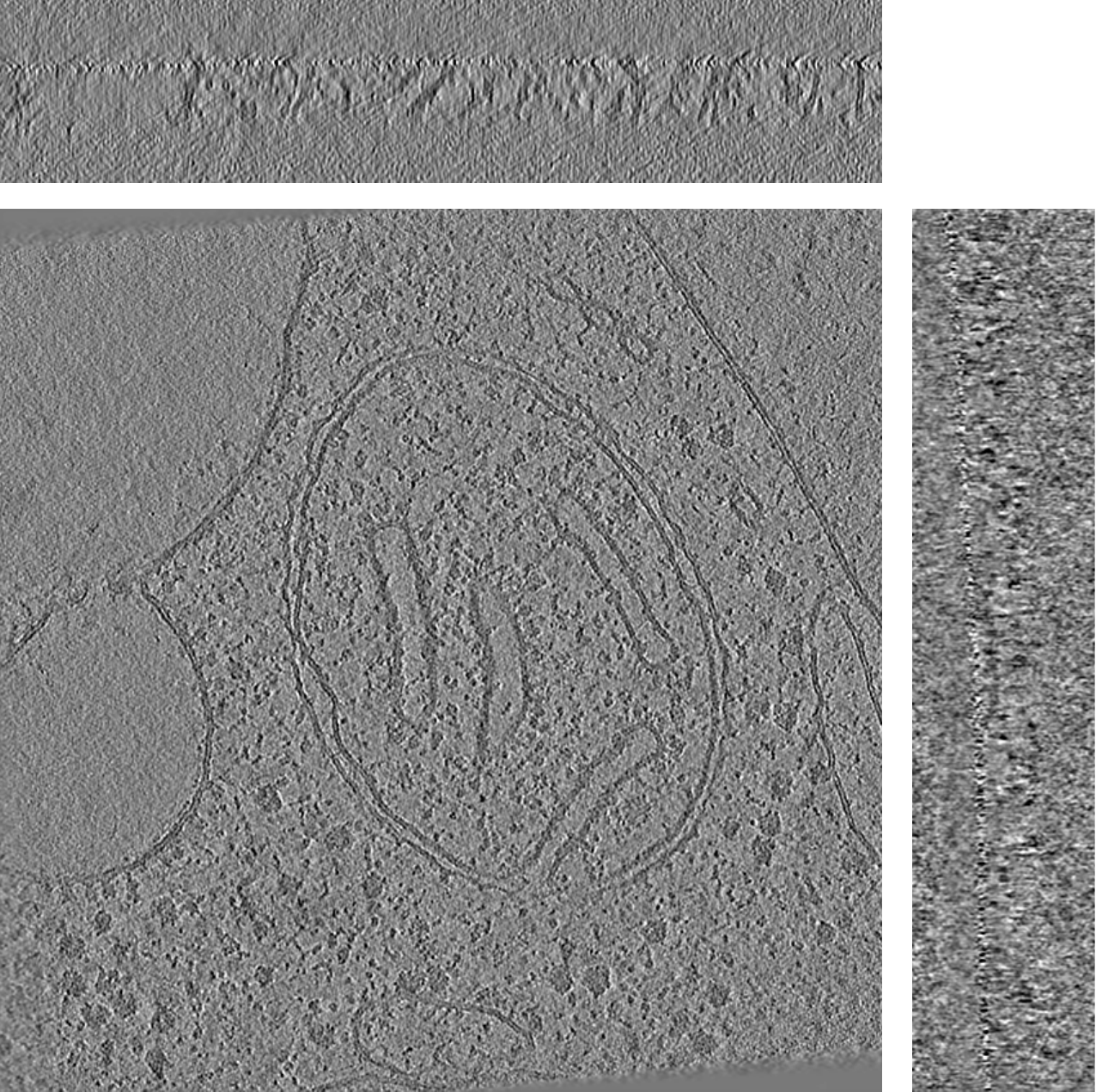}};
    		\spy on (-1.2,-0.5) in node [left] at (2.75,1.75);
		\end{tikzpicture} 
		\caption{Topaz (2min 58sec).}
	\end{subfigure}\hfill
	\begin{subfigure}[t]{\ps\textwidth}
	    \centering
    	\begin{tikzpicture}[spy using outlines={circle,orange,magnification=4,size=2.cm, connect spies}]
    		\node[rotate=0, line width=0.05mm, draw=white] at (0,0) { \includegraphics[height=\sz]{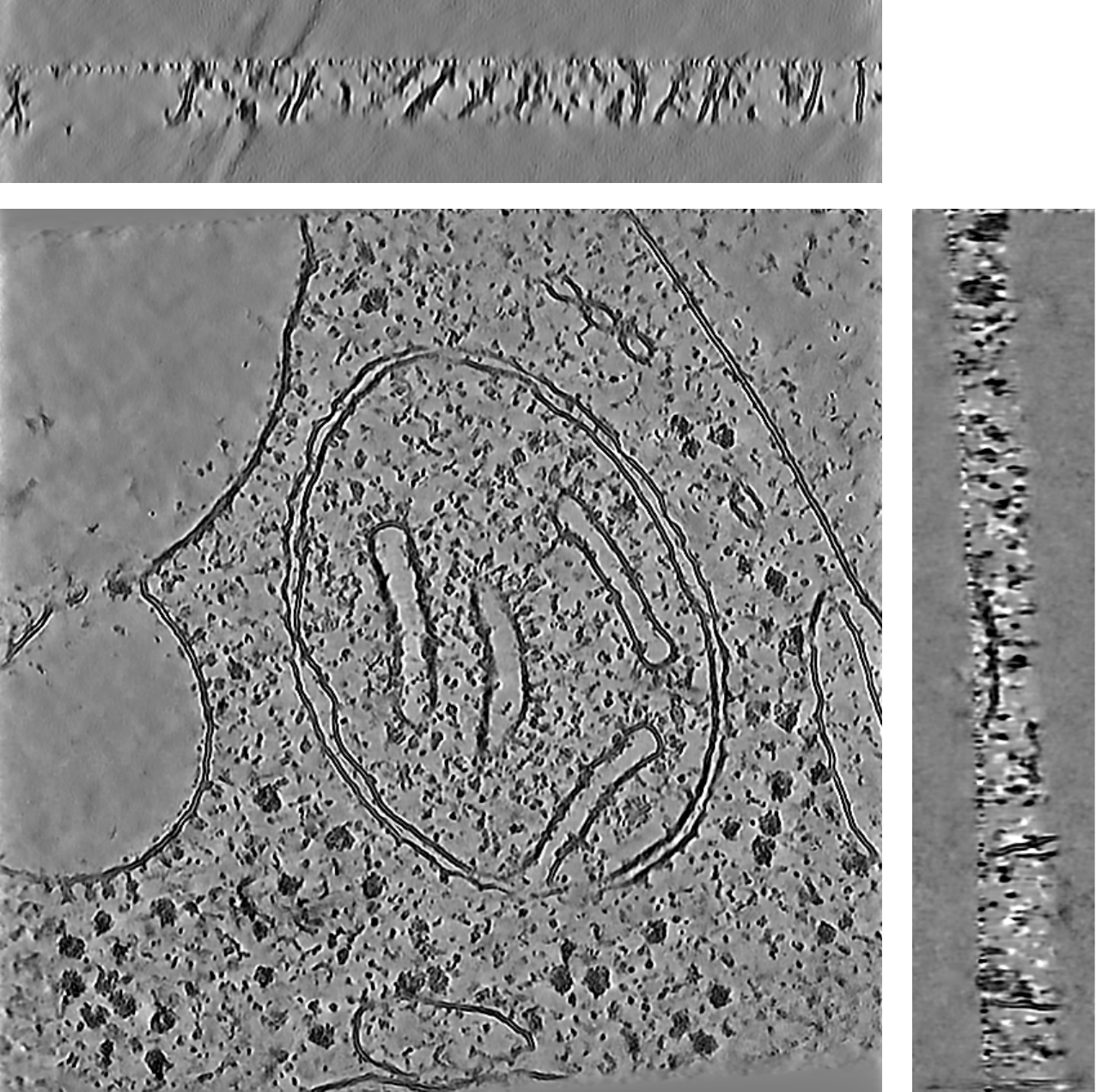}};
    		\spy on (-1.2,-0.5) in node [left] at (2.75,1.75);
		\end{tikzpicture} 
		\caption{\method (5min 16sec).}
	\end{subfigure}\hfill
	    \centering
        \begin{subfigure}[t]{0.64\textwidth}
	    \centering
          \begin{tikzpicture}
            \begin{axis}[
            width=0.8\linewidth, 
            height=0.5\linewidth, 
			grid=major, 
			xtick={0.1,0.2,0.3,0.4, 0.5},
			xticklabels={78.4, 39.2, 26.1, 19.6, 15.7},
            xmin = -0.0, xmax= 0.5,
			grid style={dashed,gray!30}, 
			xlabel= {{\tiny{Resolution (\AA)}}},
			ylabel=FSC,
            legend style={at={(0.84,0.87)}, legend cell align=right, align=right, draw=none,font=\tiny}]
           \addplot[mark=\methodWmark, mark size=\ms, line width=\lw,  mark repeat=60, color=\methodWColor] table [x expr=\coordindex/\scc, y=OURS, col sep=comma] {Figure3/fsc_val.txt};
            \addlegendentry{\method{}}
            \addplot[mark=\FBPmark, mark size=\ms, line width=\lw,  mark repeat=60, color=\FBPColor] table [x expr=\coordindex/\scc, y=FBP, col sep=comma] {Figure3/fsc_val.txt};
            \addlegendentry{FBP}
           \addplot[mark=\isoCaremark, mark size=\ms, line width=\lw,  mark repeat=60, color=\isoCareColor] table [x expr=\coordindex/\scc, y=TOPAZ, col sep=comma] {Figure3/fsc_val.txt};
            \addlegendentry{Topaz}
            \end{axis}
            \end{tikzpicture}
		\caption{FSC Curve.}
	\end{subfigure}\hfill
\caption{\method{} reaches similar FSC performance (above the 0.5-threshold) than state-of-the-art reconstruction algorithms on test data, but significantly faster (up to 75x) and without any parameter tuning. This makes \method{} an easy to use technique for high quality 3D reconstruction. FBP+\icecream{} volume is used as reference to compute the FSC curves. }\label{fig:real-test}
\end{figure*}

\subsection*{Visual comparison on new tilt series}
We expect \method{} to perform similarly to the baseline (FBP+\icecream{} or FBP+Cryo-CARE+IsoNet) on the test data (similar distribution as the training data). The \method{} algorithm is fundamentally different from the baseline which trains itself on the raw data. 
We now apply \method{} to a dataset where self-supervised methods struggle to improve over FBP reconstruction.  
Specifically, we use tilt series from EMPIAR-12262 \cite{ishemgulova2024endosome}, which contains non-infected cos-7 cells sampled at 5.525 \AA{}/pixel.
Crowded cellular environments are challenging to reconstruct accurately, as a lot of overlapping features are projected onto the tilt series, and it becomes difficult to disentangle them.
We emphasize that, compared to the training set, this dataset represents a completely different biological sample and has been acquired from a different microscope and detector. 

\method{} achieves performance comparable to or better than the self-supervised baseline (FBP+\icecream{}), as shown in Figure~\ref{fig:challenge_data}. This highlights the ability of our localized physically inspired architecture in generalizing beyond the training distribution, particularly where the self-supervised baseline struggles to produce high-quality reconstructions. Of importance, \method{} achieves this performance without any parameter tuning and with a significant speed advantage, because it does not require retraining. Visually, it exhibits cleaner denoising in empty regions of the cellular volume.
We provide additional reconstructions of the proposed approach on a dataset containing purified \textit{S. cerevisiae} 80S Ribosomes (EMPIAR-10045) in Fig.~\ref{fig:real-recons-ribo}.

\def\xx{0}
\def\yy{0}
\def\zx{-0.1}
\def\zy{1.2}
\def\ps{0.32}
\begin{figure*}
	\centering
        \begin{subfigure}[t]{\ps\textwidth}
        	    \centering
        	    \begin{tikzpicture}[spy using outlines={circle,orange,magnification=5,size=2.5cm, connect spies}]
            		\node[ rotate=90] at (0*\xx,-0*\yy) {}; outlines={circle,orange,magnification=2,size=3cm, connect spies}]
            		\node[rotate=0, line width=0.05mm, draw=white] at (0*\xx,-0.*\yy) { \includegraphics[height=\sz]{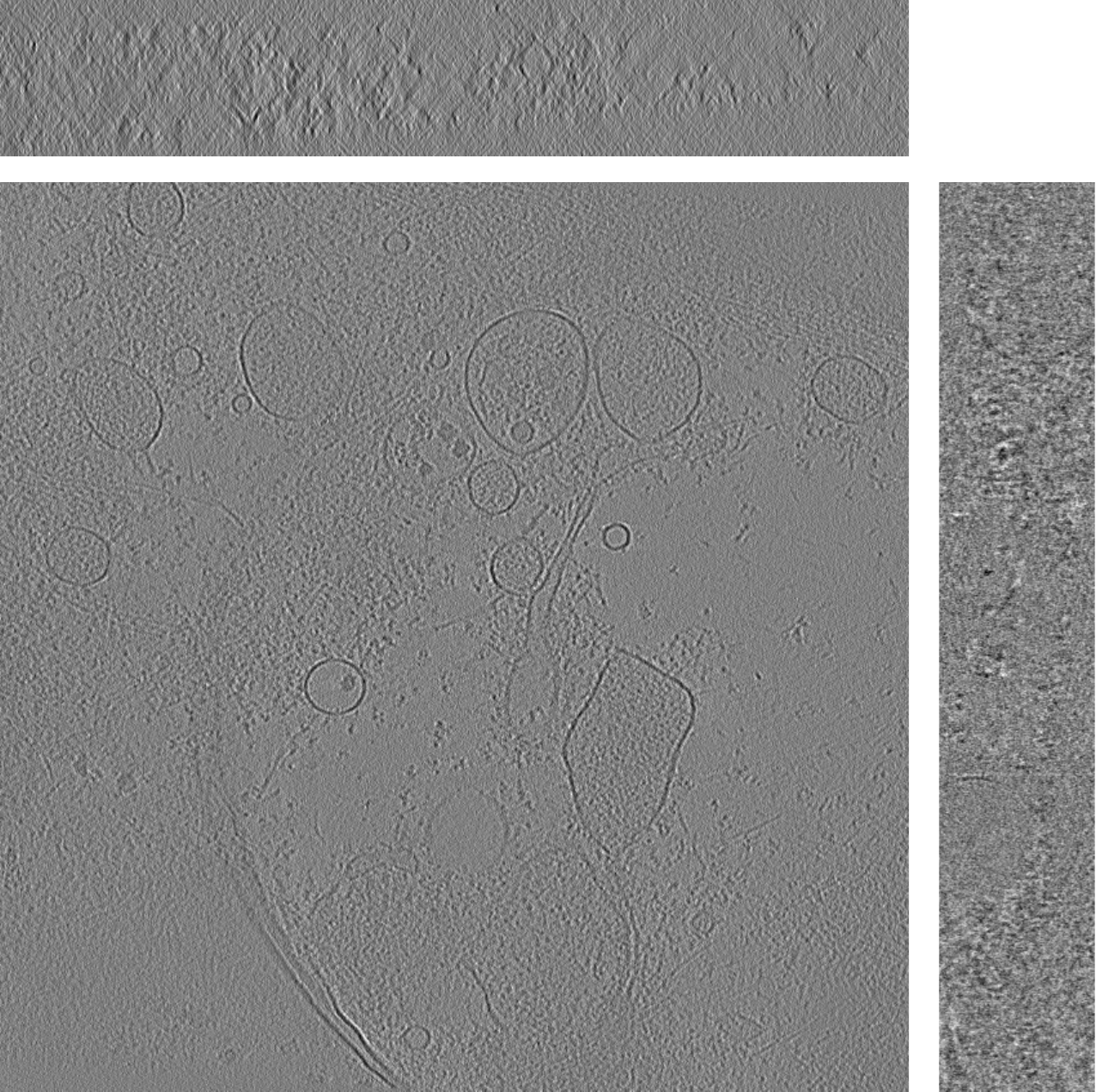}};
            		\spy on (-0.9,-0.2) in node [left] at (2.5,1.4);
        		\end{tikzpicture} 
        		\caption{FBP ($\approx 5min$).}
        	\end{subfigure}\hfill
        \begin{subfigure}[t]{\ps\textwidth}
        	    \centering
        	    \begin{tikzpicture}[spy using outlines={circle,orange,magnification=5,size=2.5cm, connect spies}]
            		\node[ rotate=90] at (0*\xx,-0*\yy) {}; outlines={circle,orange,magnification=2,size=3cm, connect spies}]
            		\node[rotate=0, line width=0.05mm, draw=white] at (0*\xx,-0.*\yy) { \includegraphics[height=\sz]{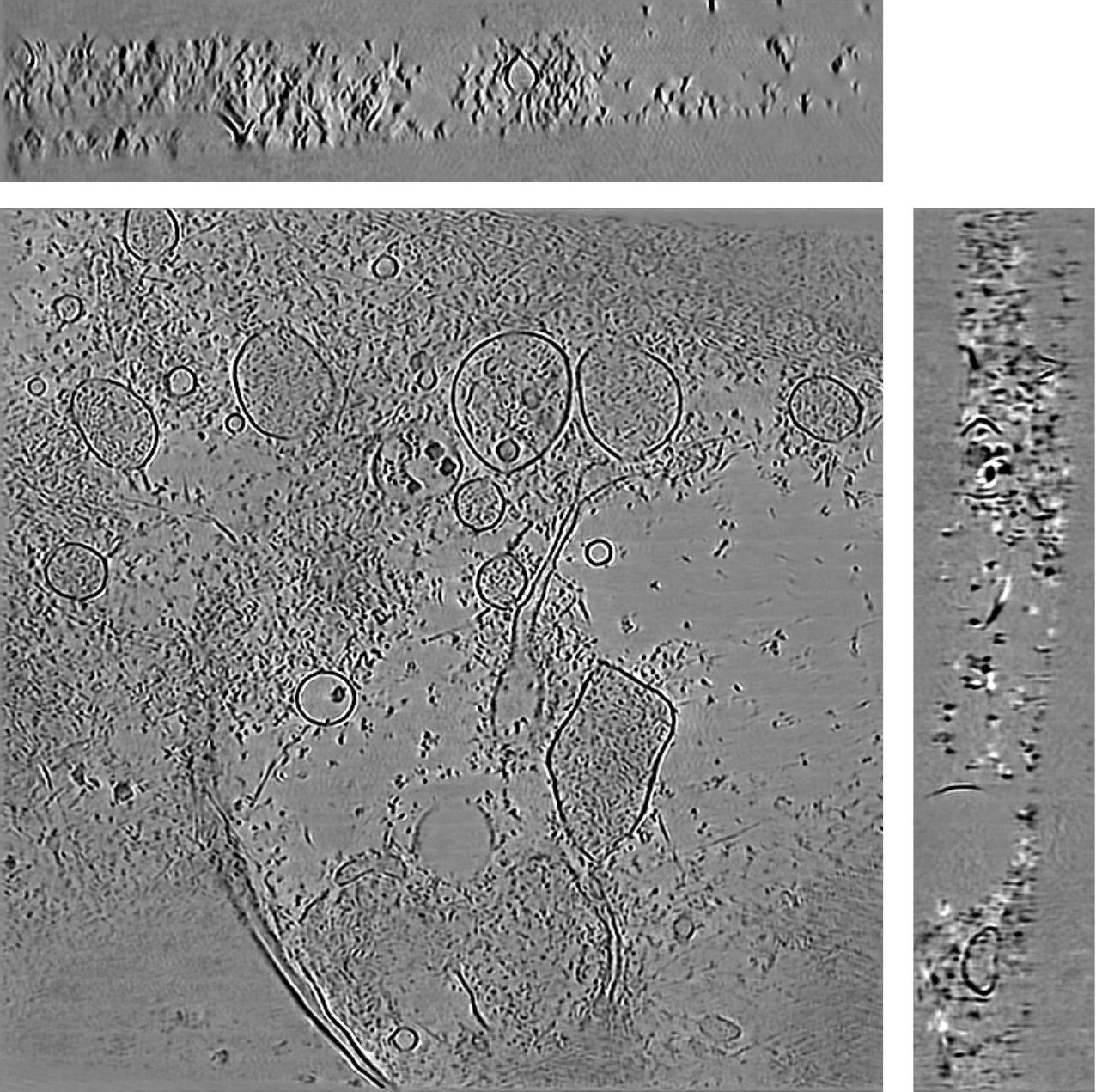}};
            		\spy on (-0.9,-0.2) in node [left] at (2.5,1.4);
        		\end{tikzpicture} 
        		\caption{FBP+\icecream{} ($\approx 12h$).}
        	\end{subfigure}\hfill
            \centering
	\begin{subfigure}[t]{\ps\textwidth}
	    \centering
	    \begin{tikzpicture}[spy using outlines={circle,orange,magnification=5,size=2.5cm, connect spies}]
    		\node[ rotate=90] at (0*\xx,-0*\yy) {}; outlines={circle,orange,magnification=2,size=3cm, connect spies}]
    		\node[rotate=0, line width=0.05mm, draw=white] at (0*\xx,-0.*\yy) { \includegraphics[height=\sz]{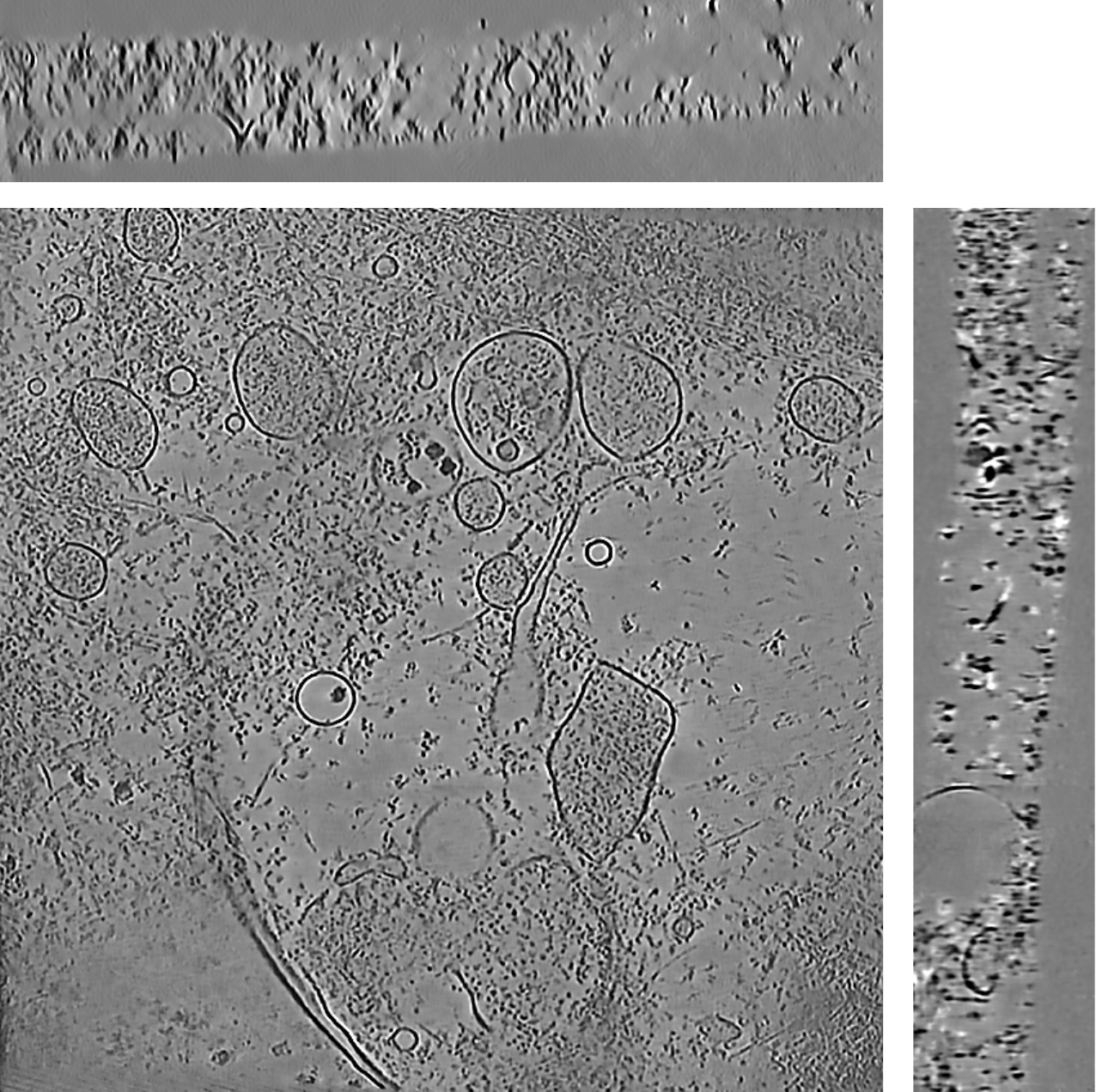}
            };
    		\spy on (-0.9,-0.2) in node [left] at (2.5,1.4);
		\end{tikzpicture}		\caption{\method{} ($\approx 4$min).}
    \end{subfigure}\hfill
\caption{
Tomogram reconstruction from EMPIAR-12262 \cite{ishemgulova2024endosome} tilt series containing non-infected cos-7 cells sampled at 5.525 \AA{}/pixel. Projections are downsampled by a factor of 3. The tilt series is acquired between -51 and 51 degrees with a tilt increment of 3 degrees and with projections of size $1279 \times 1236$. We use AreTomo2 to align the projections before reconstruction.
\method{} is able to process unseen dense biological structures on challenging datasets. Notice the processing time (up to $75$x faster) and the absence of expert intervention (no parameter to tune) for \method{}.
}\label{fig:challenge_data}
\end{figure*}

\subsection*{Facilitating template matching}
In contrast to single-particle cryo-EM, where proteins are isolated, cryo-ET allows the observation of proteins in their native cellular environment, offering a view that is closer to biological reality. While this complexity can be challenging—especially in crowded environments, as discussed earlier—it also presents an opportunity to learn more about biological systems \textit{in situ}.
When the same protein complex appears multiple times within a tomogram, averaging techniques similar to those used in single-particle cryo-EM can be applied. This process, known as subtomogram averaging (STA), relies on accurately identifying the position and orientation of each instance of the macromolecule.
The most reliable approach for this is template matching. Templates can be obtained in several ways: from databases such as the EMDB for experimental maps and the PDB (experimental) or AlphaFoldDB (predicted) for atomic models, from other imaging techniques like single-particle cryo-EM (which can be expensive and impractical to generate if a high-resolution structure is not the primary goal of the study), from subtomogram averaging attempts (for example based on manual or semi-automated picking), or directly from the tomogram itself. The latter method requires a high-quality tomographic reconstruction, which is hampered in practice especially by the missing wedge problem.

To quantitatively evaluate the quality of the reconstructed tomograms in the context of STA, we follow the procedure introduced by Chaillet \text{et. al.} \cite{chaillet2023extensive}, which measures how well a known protein structure correlates with its counterparts in a tomographic reconstruction, see Fig.~\ref{fig:template-match}.
Using this approach, we demonstrate that the human 80S ribosomes from the template matching tutorial dataset \cite{TLGJCM_2023} are more accurately recovered from \method{} than from standard FBP reconstructions. Ribosomes exhibit higher correlation with the clean reference structure from the PDB when the tomogram is reconstructed using \method{} or FBP+\icecream{}. Notice that FBP+Cryo-CARE+IsoNet performs similarly or worse than the proposed method.

\def\ps{0.25}
\begin{figure*}
\begin{subfigure}[t]{\ps\textwidth}
\centering
           \begin{tikzpicture}
            \begin{axis}[
            width=1\linewidth, 
            height=1.2\linewidth, 
            axis lines = left,
            xmin = 0, xmax= 205, ymin= 0.2, ymax = 0.55,
			grid=major, 
			grid style={dashed,gray!30}, 
			xlabel= {\notsotiny{Particle Index}},
			ylabel={\notsotiny{Cross-Correlation}},
            legend style={at={(1.45,1.)}, legend cell align=left, align=right, draw=none,font=\notsotiny}]
            \addplot[mark=\FBPmark, mark size=\ms, line width=\lw,  mark repeat=20, color=\FBPColor] table [x expr=\coordindex, y=fbp, col sep=comma] {images_3/pytom_exp/lcc_values_tomo_100.txt};
            \addplot[mark=\isoCaremark, mark size=\ms, line width=\lw,  mark repeat=20, color=\isoCareColor] table [x expr=\coordindex, y=crc_iso, col sep=comma] {images_3/pytom_exp/lcc_values_tomo_100.txt};
            \addplot[mark=\isomark, mark size=\ms, line width=\lw,  mark repeat=20, color=\ddwColor] table [x expr=\coordindex, y=icecream, col sep=comma] {images_3/pytom_exp/lcc_values_tomo_100.txt};
            \addplot[mark=\methodWmark, mark size=\ms, line width=\lw,  mark repeat=20, color=\methodWColor] table [x expr=\coordindex, y=ours_icr, col sep=comma] {images_3/pytom_exp/lcc_values_tomo_100.txt};
            \end{axis}
            \end{tikzpicture} 
            \caption{ Tomogram-100.}
            \end{subfigure}\hfill
\begin{subfigure}[t]{\ps\textwidth}
\centering
          \begin{tikzpicture}
            \begin{axis}[
            width=1.15\linewidth, 
            height=1.13\linewidth, 
            axis lines = left,
            xmin = 0, xmax= 205, ymin= 0.2, ymax = 0.55,
			grid=major, 
			grid style={dashed,gray!30}, 
			xlabel= {\notsotiny{Particle Index}},
			legend style={at={(1.,1.)}, legend cell align=left, align=right, draw=none,font=\notsotiny}]
            \addplot[mark=\FBPmark, mark size=\ms, line width=\lw,  mark repeat=20, color=\FBPColor] table [x expr=\coordindex, y=fbp, col sep=comma] {images_3/pytom_exp/lcc_values_tomo_101.txt};
            \addlegendentry{FBP}
            \addplot[mark=\isoCaremark, mark size=\ms, line width=\lw,  mark repeat=20, color=\isoCareColor] table [x expr=\coordindex, y=crc_iso, col sep=comma] {images_3/pytom_exp/lcc_values_tomo_101.txt};
            \addlegendentry{FBP+Cryo-CARE+IsoNet}
            \addplot[mark=\isomark, mark size=\ms, line width=\lw,  mark repeat=20, color=\ddwColor] table [x expr=\coordindex, y=icecream, col sep=comma] {images_3/pytom_exp/lcc_values_tomo_101.txt};
            \addlegendentry{FBP+\icecream{}}
            \addplot[mark=\methodWmark, mark size=\ms, line width=\lw,  mark repeat=20, color=\methodWColor] table [x expr=\coordindex, y=ours_icr, col sep=comma] {images_3/pytom_exp/lcc_values_tomo_101.txt};
            \addlegendentry{\method{}}
            \end{axis}
            \end{tikzpicture} 
            \caption{ Tomogram-101.}
        \end{subfigure}\hfill
\begin{subfigure}[t]{\ps\textwidth}
\centering
          \begin{tikzpicture}
            \begin{axis}[
            width=1.15\linewidth, 
            height=1.13\linewidth, 
            axis lines = left,
            xmin = 0, xmax= 205, ymin= 0.2, ymax = 0.55,
			grid=major, 
			grid style={dashed,gray!30}, 
			xlabel= {\notsotiny{Particle Index}},
            legend style={at={(1,1)}, legend cell align=right, align=right, draw=none,font=\notsotiny}]
            \addplot[mark=\FBPmark, mark size=\ms, line width=\lw,  mark repeat=20, color=\FBPColor] table [x expr=\coordindex, y=fbp, col sep=comma] {images_3/pytom_exp/lcc_values_tomo_107.txt};
            \addplot[mark=\isoCaremark, mark size=\ms, line width=\lw,  mark repeat=20, color=\isoCareColor] table [x expr=\coordindex, y=crc_iso, col sep=comma] {images_3/pytom_exp/lcc_values_tomo_107.txt};
            \addplot[mark=\isomark, mark size=\ms, line width=\lw,  mark repeat=20, color=\ddwColor] table [x expr=\coordindex, y=icecream, col sep=comma] 
            {images_3/pytom_exp/lcc_values_tomo_107.txt};
            \addplot[mark=\methodWmark, mark size=\ms, line width=\lw,  mark repeat=20, color=\methodWColor] table [x expr=\coordindex, y=ours_icr, col sep=comma] {images_3/pytom_exp/lcc_values_tomo_107.txt};

            \end{axis}
            \end{tikzpicture} 
            \caption{Tomogram-107.}
            \end{subfigure}\hfill
\begin{subfigure}[t]{\ps\textwidth}
\centering
          \begin{tikzpicture}
            \begin{axis}[
            width=1.15\linewidth, 
            height=1.13\linewidth, 
            axis lines = left,
            xmin = 0, xmax= 205, ymin= 0.2, ymax = 0.55,
			grid=major, 
			grid style={dashed,gray!30}, 
			xlabel= {\notsotiny{Particle Index}},
            legend style={at={(1,1)}, legend cell align=right, align=right, draw=none,font=\notsotiny}]
            \addplot[mark=\FBPmark, mark size=\ms, line width=\lw,  mark repeat=20, color=\FBPColor] table [x expr=\coordindex, y=fbp, col sep=comma] {images_3/pytom_exp/lcc_values_tomo_108.txt};
            \addplot[mark=\isoCaremark, mark size=\ms, line width=\lw,  mark repeat=20, color=\isoCareColor] table [x expr=\coordindex, y=crc_iso, col sep=comma] {images_3/pytom_exp/lcc_values_tomo_108.txt};
            \addplot[mark=\isomark, mark size=\ms, line width=\lw,  mark repeat=20, color=\ddwColor] table [x expr=\coordindex, y=icecream, col sep=comma] 
            {images_3/pytom_exp/lcc_values_tomo_108.txt};
            \addplot[mark=\methodWmark, mark size=\ms, line width=\lw,  mark repeat=20, color=\methodWColor] table [x expr=\coordindex, y=ours_icr, col sep=comma] {images_3/pytom_exp/lcc_values_tomo_108.txt};
            \end{axis}
            \end{tikzpicture} 
            \caption{ Tomogram-108.}
            \end{subfigure}
		\caption{
        Cross-correlation between the template of human 80S ribosome (EMD-2938 \cite{khatter2015structure}) and the tomograms provided in the template matching tutorial dataset \cite{TLGJCM_2023}. The tilt series are first pre-processed using Aretomo3 for alignment and CTF correction. The aligned tilt series serve as input to our method, while the corresponding FBP reconstruction from IMOD is used as input for FBP+Cryo-CARE+IsoNet and FBP+\icecream{}.  All volumes are reconstructed at Bin 4. We use pytom-match-pick \cite{chaillet2025pytom} to perform template matching and obtain the cross-correlation scores. We then select the 200 particles with the highest scores and plot the values in descending order. }
    \label{fig:template-match}
\end{figure*}

\subsection*{Effect of the CTF}
The CTF distorts the volume reconstruction, especially for thick samples or when data are acquired at high magnification (i.e., small pixel size). Ignoring the CTF can result in the loss of high-resolution details. 
To evaluate the impact of CTF correction on tomographic reconstruction, we use a tomogram of \textit{Chlamydomonas reinhardtii} thylakoid membranes obtained from EMPIAR-11830. 

As shown in Figure~\ref{fig:ctf-recons-hiv}, \method{} performs similarly when given the CTF corrected tilt series (Fig.~\ref{fig:ctf-cryolithe}) and the non-CTF corrected tilt series (Fig.~\ref{fig:noctf-cryolithe}). In particular, CTF correction leads to improved contrast and slightly attenuated the noise. 
The performance of the baseline reconstruction (FBP+Cryo-CARE+IsoNet), however, strongly depends on whether CTF correction is applied. Without CTF correction, the baseline reconstruction exhibits clear artifacts, such as oscillatory patterns between membranes (Figure~\ref{fig:baseline-no-ctf}). When CTF correction is not included, using Wiener filtering as described in IsoNet's original publication \cite{liu2022isotropic}, the baseline performance improves. These results suggest that \method{} is more robust to CTF artifacts, enabling high-quality reconstructions even in the absence of explicit CTF correction.

\def\ps{0.32}
\begin{figure*}
\begin{subfigure}[t]{\ps\textwidth}
	    \centering
    	\begin{tikzpicture}[spy using outlines={circle,orange,magnification=5,size=2.5cm, connect spies}]
    		\node[ rotate=90] at (0,0) {}; outlines={circle,orange,magnification=2,size=3cm, connect spies}]
    		\node[rotate=0, line width=0.05mm, draw=white] at (0,0) { \includegraphics[height=\sz]{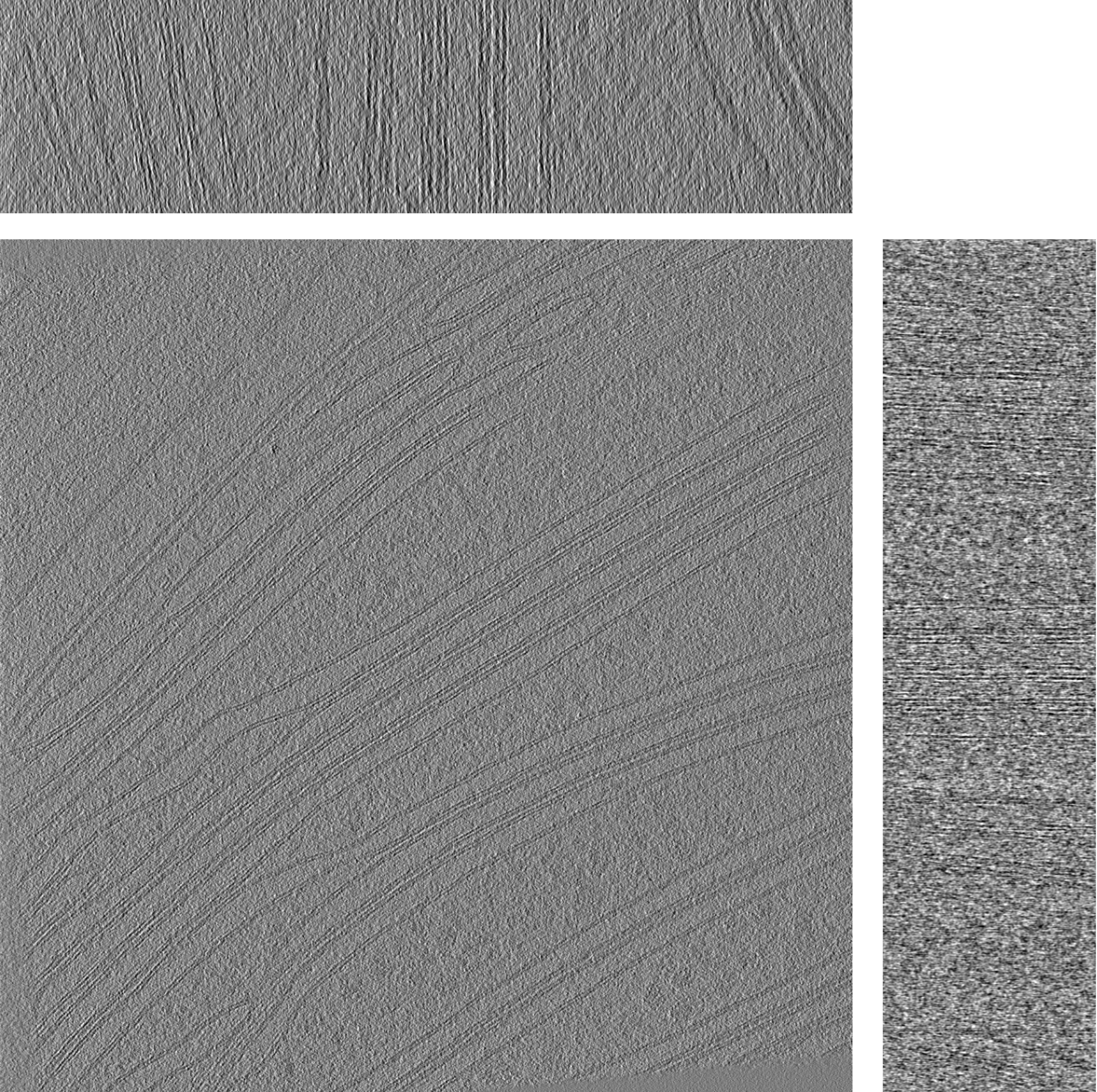}};
    		 \spy on (-0.9,-0.8) in node [left] at (2.5,1.2);
		\end{tikzpicture} 
		\caption{FBP.}
	\end{subfigure}\hfill    
     	\begin{subfigure}[t]{0.35\textwidth}
	     \centering
     	\begin{tikzpicture}[spy using outlines={circle,orange,magnification=5,size=2.5cm, connect spies}]
     		\node[ rotate=90] at (0,0) {}; outlines={circle,orange,magnification=2,size=3cm, connect spies}]
     		\node[rotate=0, line width=0.05mm, draw=white] at (0,0) { \includegraphics[height=\sz]{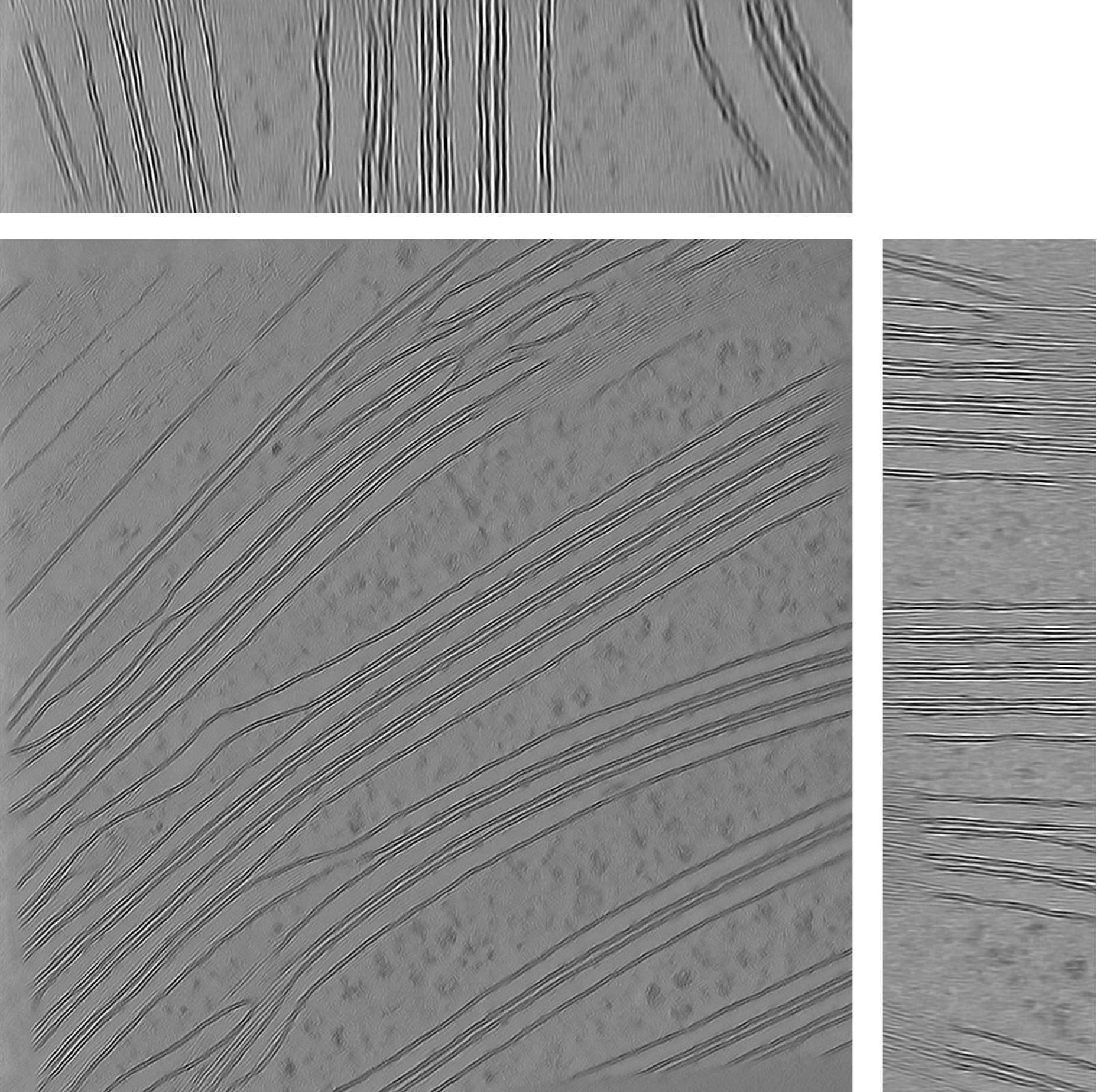}
             };
                 \spy on (-0.9,-0.8) in node [left] at (2.5,1.2);
	 	\end{tikzpicture} 
	 	\caption{FBP+Cryo-CARE+IsoNet. \label{fig:baseline-no-ctf}}
	 \end{subfigure}
     	\begin{subfigure}[t]{\ps\textwidth}
	     \centering
     	\begin{tikzpicture}[spy using outlines={circle,orange,magnification=5,size=2.5cm, connect spies}]
     		\node[ rotate=90] at (0,0) {}; outlines={circle,orange,magnification=2,size=3cm, connect spies}]
     		\node[rotate=0, line width=0.05mm, draw=white] at (0,0) { \includegraphics[height=\sz]{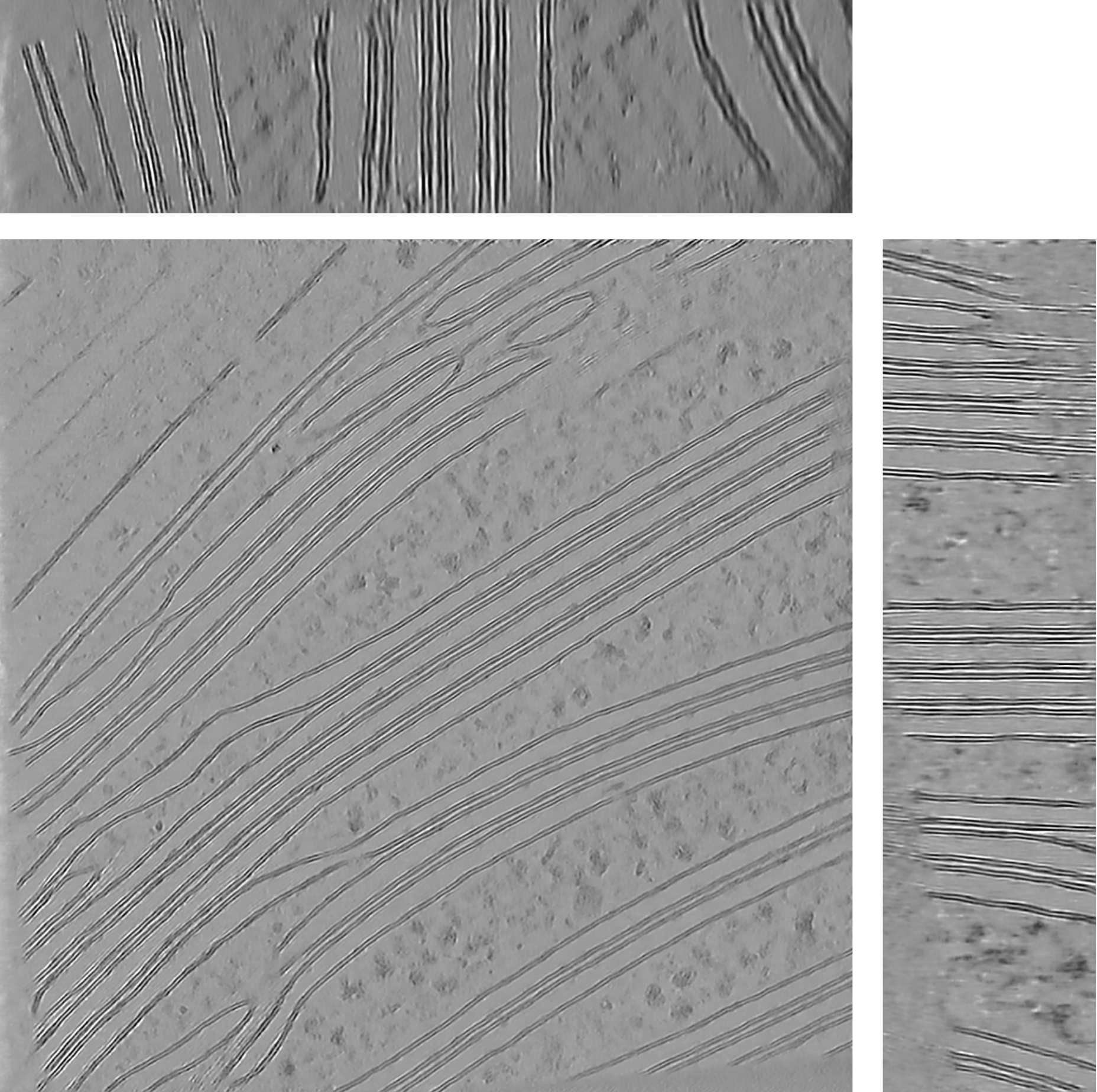}
             };
                 \spy on (-0.9,-0.8) in node [left] at (2.5,1.2);
	 	\end{tikzpicture} 
	 	\caption{\method{}. \label{fig:noctf-cryolithe}}
	 \end{subfigure}
\begin{subfigure}[t]{\ps\textwidth}
	    \centering
    	\begin{tikzpicture}[spy using outlines={circle,orange,magnification=5,size=2.5cm, connect spies}]
    		\node[ rotate=90] at (0,0) {}; outlines={circle,orange,magnification=2,size=3cm, connect spies}]
    		\node[rotate=0, line width=0.05mm, draw=white] at (0,0) { \includegraphics[height=\sz]{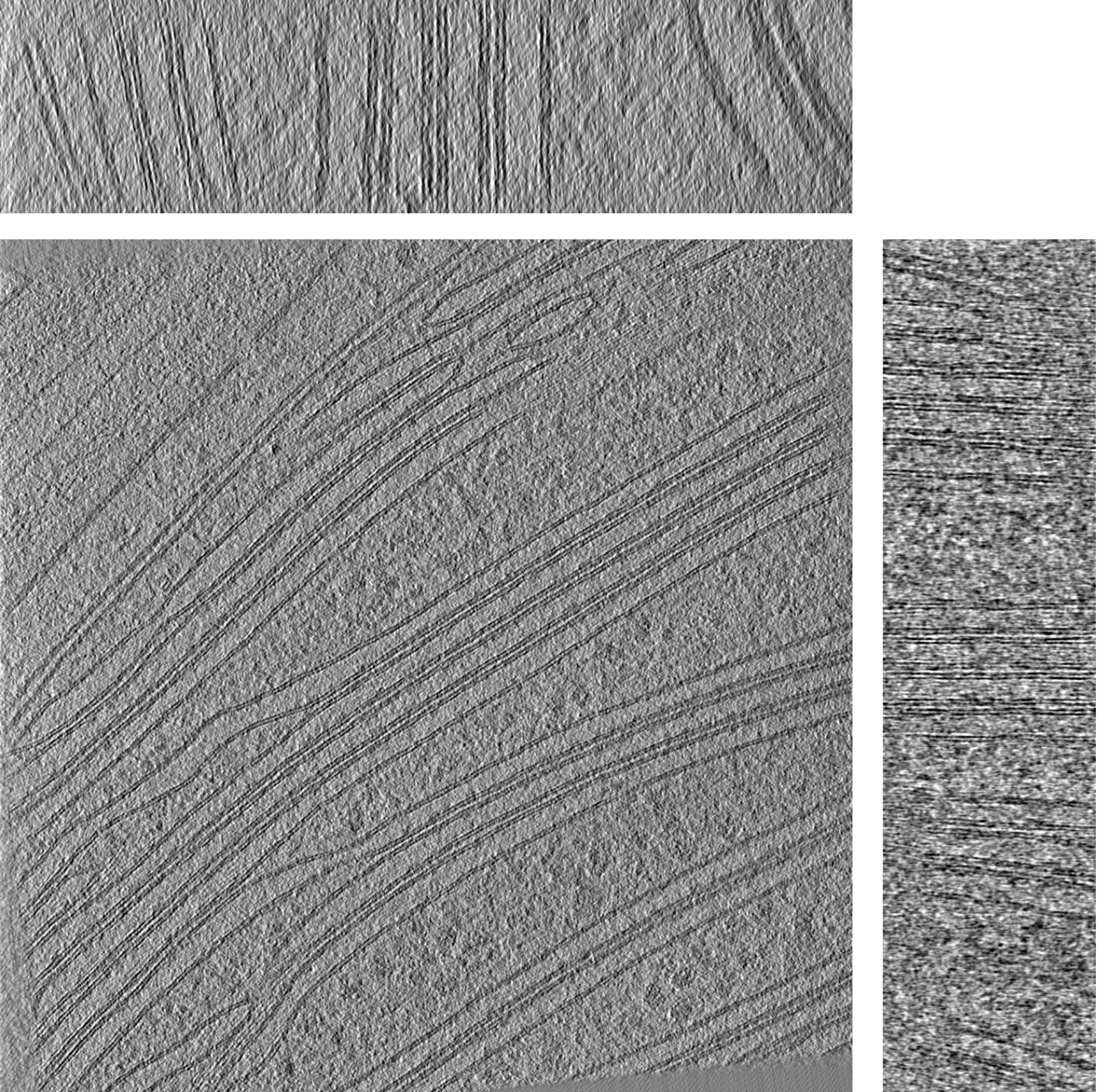}};
    		 \spy on (-0.9,-0.8) in node [left] at (2.5,1.2);
		\end{tikzpicture} 
		\caption{CTF corrected+FBP.}
	\end{subfigure}\hfill
\begin{subfigure}[t]{0.35\textwidth}
	    \centering
    	\begin{tikzpicture}[spy using outlines={circle,orange,magnification=5,size=2.5cm, connect spies}]
    		\node[ rotate=90] at (0,0) {}; outlines={circle,orange,magnification=2,size=3cm, connect spies}]
    		\node[rotate=0, line width=0.05mm, draw=white] at (0,0) { \includegraphics[height=\sz]{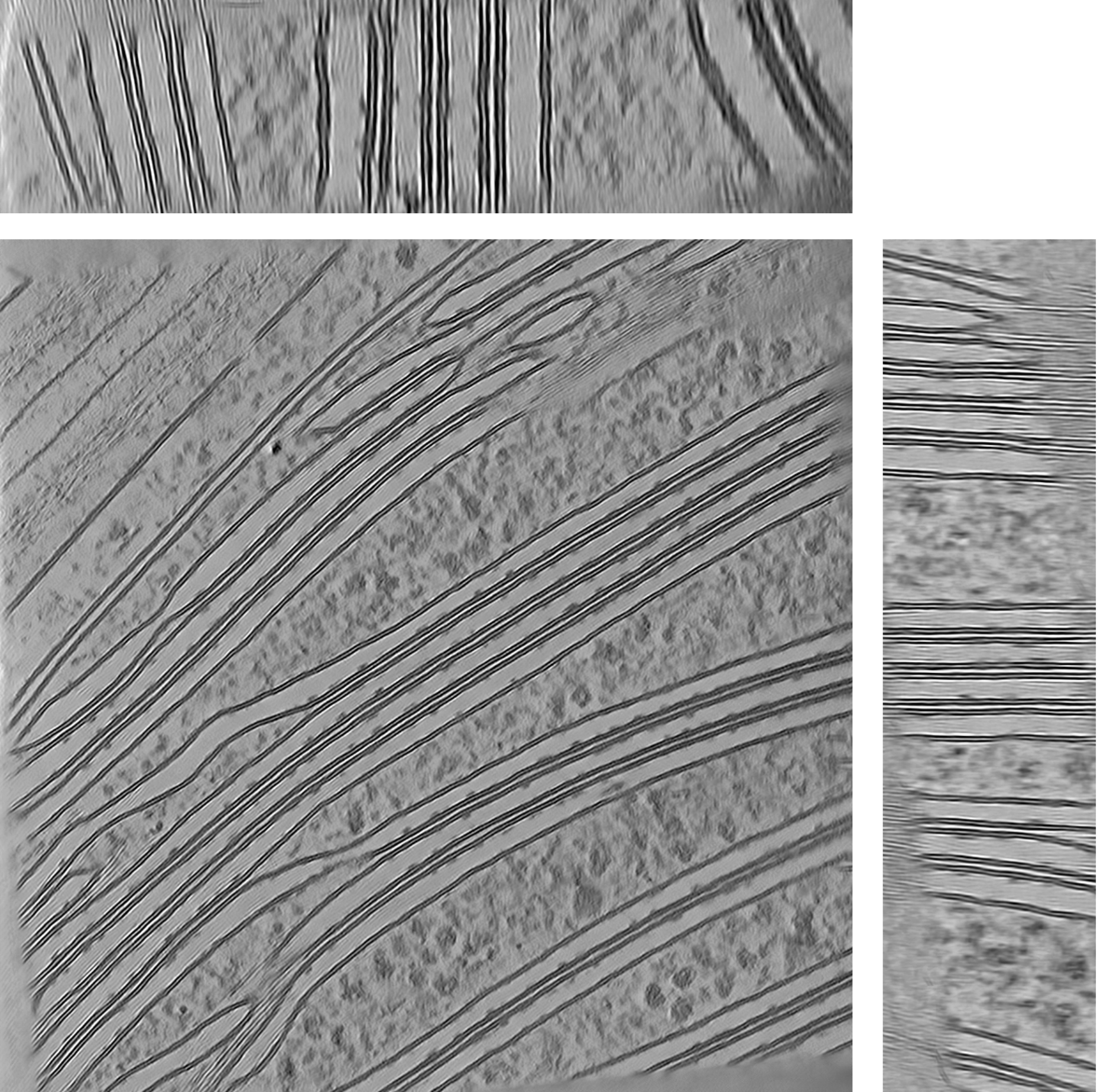}};
    		 \spy on (-0.9,-0.8) in node [left] at (2.5,1.2);
		\end{tikzpicture} 
		\caption{CTF corrected+FBP+ Cryo-CARE+IsoNet.}
	\end{subfigure}\hfill
\begin{subfigure}[t]{\ps\textwidth}
	    \centering
    	\begin{tikzpicture}[spy using outlines={circle,orange,magnification=5,size=2.5cm, connect spies}]
    		\node[ rotate=90] at (0,0) {}; outlines={circle,orange,magnification=2,size=3cm, connect spies}]
    		\node[rotate=0, line width=0.05mm, draw=white] at (0,0) { \includegraphics[height=\sz]{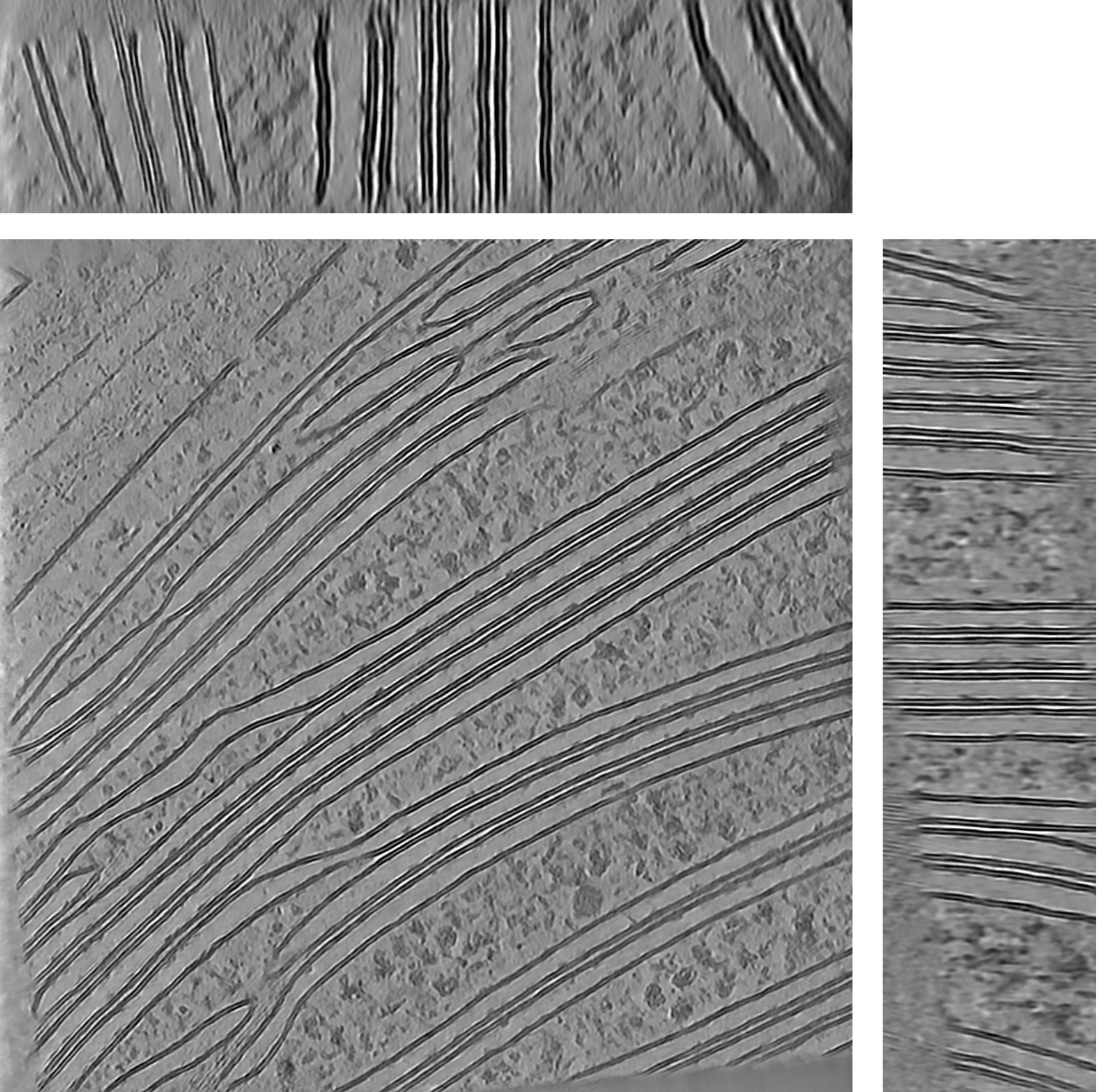}};
    		 \spy on (-0.9,-0.8) in node [left] at (2.5,1.2);
		\end{tikzpicture} 
		\caption{CTF corrected+\method{}. \label{fig:ctf-cryolithe}}
	\end{subfigure}\hfill    
\caption{Impact of correcting for the CTF on \textit{C. reinhardtii} thylakoid membranes membranes (\texttt{01122021\_BrnoKrios\_arctis\_lam2\_pos13}, EMPIAR-11830). 
The performance of \method{} remains similar with and without CTF correction, contrary to the FBP+Cryo-CARE+IsoNet for which performance are improved by CTF correction.
We adapted AreTomo3 \cite{peck2025aretomolive} to produce the CTF corrected aligned tilt series as \method{} requires aligned tilt series. AreTomo3 corrects the CTF in tile based manner with a mixture of Wiener filtering for low frequencies and phase flipping for high frequencies. This is done to avoid correcting the high frequency CTF parameters, as they have very low magnitude while maintaining reliable high frequency estimations.
}\label{fig:ctf-recons-hiv}
\end{figure*}

\addcontentsline{toc}{section}{Discussion}
\section*{Discussion}
The availability of cryo-ET datasets is growing fast; as a case in point, EMPIAR-11830 alone has almost 2,000 volumes \cite{kelley_toward_2026}. This expansion will likely accelerate due to the increasing adoption of modern instrumentation and ongoing advancements in the automation and throughput of cryo-ET sample preparation and data acquisition. There is thus a growing need for fast, reliable reconstruction methods.

\method{} can be used off-the-shelf to efficiently reconstruct arbitrary volumes. Training on new data is also possible---though, as shown here, hardly necessary---making it easy for both experienced users and novices to obtain high-quality reconstructions in a short time compared with current self-supervised methods. In principle, self-supervised approaches could also be accelerated by training on a larger dataset; DeepDeWedge suggests using up to ten similar volumes (e.g., from the same cell type and imaging session). However, extending self-supervised strategies to much larger datasets is generally challenging and the issues of robustness to distribution shift remain. In contrast, \method{} is specifically designed to be robust to such variations thanks to its localized architecture that prevent overfitting.

A particularly time-consuming aspect of developing \method{} was the manual curation of a high-quality training set from EMPIAR, which required repeatedly running existing self-supervised reconstruction methods, and visually inspecting the tomograms slice by slice. We believe this curated dataset will be valuable for others developing deep-learning-based approaches and anticipate the need for a more scalable curation strategy as we work to further improve results. Such data are also useful for building deep learning methods in other stages of the cryo-ET pipeline, including denoising, missing wedge compensation, tilt-series alignment, and CTF correction.

As resolution continues to improve, refinements to the forward model \eqref{eq:forward} may become necessary---for instance to account for CTF effects more explicitly (especially harmful in thick samples or lower electron energies \cite{russo_cryomicroscopy_2022}). The global filtering and the nonlinear local ``backprojection'' framework implemented in \method{} can be adapted to handle these more complex physics. For the current state of cryo-ET, where resolutions typically remain around 8–10~Å, the forward model in \eqref{eq:forward} is standard and effective.

Our approach bypasses the training of deep neural network necessary in the current state-of-the-art methods, while still leveraging their high performance for training. This leads to a method with several advantages including ease of use with no need for parameter tuning, low resource requirements and robustness to unseen data without introducing hallucinated artifacts.
Our best-trained model is available by downloading the network weights and running a Python script \href{https://github.com/swing-research/CryoLithe}{https://github.com/swing-research/CryoLithe}.

\addcontentsline{toc}{section}{Methods}
\section*{Methods}
\subsection*{Definition of the cryo-ET acquisition model}

The unknown volume density $V: \mathbb{R}^3 \to \mathbb{R}$ has to be estimated from its noisy tilt series $\{y_\theta\}_{\theta=\theta_{\text{min}}}^{\theta_{\text{max}}}$. Letting $\vr = (r_x,r_y,r_z)$, the noiseless tilt series can be modeled as 
\begin{equation}\label{eq:forward}
    y_{\theta}(r_x,r_y) = \int V(\vR_{\theta}\vr)dr_z, 
\end{equation}
where $\vR_{\theta}(\cdot)$ is the 3D rotation matrix around the second axis,
\begin{equation*}
    \vR_{\theta} = \begin{bmatrix}
\cos(\theta) & 0 & \sin(\theta)\\
0 & 1 & 0 \\
-\sin(\theta) & 0&  \cos(\theta)
\end{bmatrix},
\end{equation*}
with $\theta$ the rotation angle.
We observe discretized, noisy projections $\yb_{\theta}\in\R^{n_1 \times n_2}$, given by
\begin{equation}\label{eq:forwad_noise}
    \yb_{\theta}[i,j] \approx \Pc\left( {[b \ast y_\theta]}(\Delta_x i,\Delta_y j) \right),
\end{equation}
where $b$ is a sampling kernel modeling the finite pixel size, $\Pc$ is the (random) degradation process, typically modeled as multiplicative Poisson noise or additive white Gaussian noise, and $\Delta_x, \Delta_y$ are the pixel dimensions on the sensor.

\subsection*{Local property of the cryo-ET acquisition model}
The projection operator defined in Equation \eqref{eq:forward} has a local impulse response: a point in the volume influences only one point in each tilt. This has been observed previously in 2D CT \cite{khorashadizadeh2024glimpse}. 
More formally, let $V(\vr) = \delta(\vr - \vr^0)$  a point source, with $\delta$ being the Dirac delta.
In this case, for any $(r_x, r_y) \in\R^2$ and any projection angle $\theta\in[-\pi,\pi)$, the noiseless projections are given by

\begin{eqnarray}
\label{eq:delta_support}
     y_{\theta}(r_x,r_y) = \delta(r_x - (r^0_x\cos(\theta) - r^0_z\sin(\theta)), r_y-r^0_y).
\end{eqnarray}
This means that a point source influences the intensity of the tilt series along the curve
\begin{equation}\label{eq:line}
    \theta \mapsto \vr_{\theta}(\vr^0) = (r^0_x\cos(\theta) - r^0_z\sin(\theta), r^0_y),
\end{equation}
which is a sinusoid in $\theta$.
Thus, in principle, all the information of the volume from location $\vr^0$ is present at the points $\vr_{\theta}$ in the projections. We note that the points on the curve $\vr_\theta$ are also influenced by  other points in the volume. This mixing can be partially undone using a simple filter along the $x$  analogous to the ``filtering'' step of the FBP.

In this paper, we exploit the local support of the impulse response of the cryo-ET projection operator to build a reconstruction method that uses only a small subset of the tilt series around $\vr_\theta(\vr^0)$ to estimate $V(\vr^0)$. Our goal is thus, loosely speaking, to invert the mapping $V(\vr) = \delta(\vr - \vr^{0}) \mapsto ( y_{\theta}(\vr_\theta)) _\theta$.
This mapping is generally not invertible due to the missing wedge and the sparse sampling of the tilt angles. To overcome this problem, we propose a local neural network to approximate the inverse. 

Using only $(y_{\theta}(\vr_\theta))_\theta$ as input to the inversion network has the advantage of drastically reducing the dimension of the inversion problem. However, it amounts to assume that there is no correlation between pixels of the tilt series and becomes suboptimal in the presence of noise. For this reason, we consider measurements in a neighborhood which is a ``thickening'' of $\vr_\theta$. This reflects the fact that the strong correlations in the data are primarily local, and only these local correlations can be robustly estimated from finite data, while still greatly reducing the computational and statistical complexity of learning the inverse. 

\subsection*{Transform-localized training}

Let $N$ denote the number of projections in the tilt series, let $M \times M$ denote the size of the projections and let $P \times P$ denote the size of the patch of measurements extracted from the tilt series. For simplicity, we only consider square projections, but non-square tilt series are easily handled. The released model was obtained with $P=21$. Notice that the number of tilts $N$ can vary from one tilt series to another and \method{} can process an arbitrary number of tilts. Of importance, the tilt range doesn't need to be uniform either as long as the tilt value is provided.
In order to estimate the volume density at a given position $\vr\in\R^3$, \method{} processes an $N\times P \times P$ dimensional input through a neural network. 
We let $f_\gammab: \R^{N\times P \times P} \to \R$ denote the neural network with $\gammab \in \R^K$ its $K$ learnable parameters. The specific details about the architecture can be found in Appendix \ref{app:details-method}.
As mentioned above, a simple 1D convolutional filtering along the $x$ coordinate localizes information in a similar way that the filtering step of the FBP does. Intuitively, it is a way to incorporate global information in the reconstruction in a way which does not lead to overfitting. In practice, we apply the cosine ramp filter to the input tilt series, a high-pass filter with a large receptive field similar to the ramp filter used in FBP. We let $\hat{\yb}$ denote the filtered projections.

The patch extraction is formally defined by the cropping operator $C(\cdot,\cdot)$, which takes a filtered projection $\hat \yb_\theta$ and the locations of the projections $\vr^0$ and crops a patch around these locations,
\begin{equation}\label{eq:crop}
	\begin{split}
			C(\hat{\yb}_{\theta},\vr^0)[i,j] =\hat{\yb}_{\theta} \bigg(r_{\theta}&(\vr^0)- \Delta \begin{bmatrix}
			i \\ j 
		\end{bmatrix}\bigg) ,\\
		& i,j \in \bigg(-\bigg\lfloor \frac{P}{2} \bigg\rfloor, \bigg\lfloor \frac{P}{2} \bigg\rfloor \bigg),
	\end{split}
\end{equation}
where $\Delta$ depends on the resolution of the projection and $\vr_\theta(\vr^0)$ is given by Equation \eqref{eq:line}. We use bilinear interpolation to estimate the projection values if the sampling location is not on the discrete grid.

The neural network $f_\gammab$ is trained using $L$ paired training examples composed of $L$ filtered tilt series and their corresponding volume densities $\{(\hat{\yb}_l, V_l)\}_{l=1}^L$, by approximately solving
\begin{equation}\label{eq:optim_super}
	\begin{split}
		\min_{\gammab}\mathbb{E}_{V,\yb,\vr}&\| V(\vr)  - f_\gammab(\vp)\|^2,\\ 
		&\vp = [C(\hat{\yb}_{\theta_1}, r_{\theta_1}(\vr)), \hdots, C(\hat{\yb}_{\theta_N},  r_{\theta_N}(\vr)) ].
	\end{split}
\end{equation}
We use Adam \cite{kingma2014adam} optimizer to minimize the cost function. In particular, we sample the expectation by taking batches over the finite number of volumes density available, but also pixels, which allows to trade-off training time for GPU memory and enables training on large volumes. 

More precisely, for each volume, we first sample target locations in the volume at random. Our network outputs the value of the estimated volume at arbitrary real coordinates, so the sampled locations need not lie on a fixed grid. Then, we extract the corresponding patches from the tilt series to serve as input of \method{}'s neural network. Finally, we evaluate the cost function defined in Equation \eqref{eq:optim_super} and use automatic differentiation to update the learnable parameters.

\subsection*{Link between \method{} and Filtered Back-Projection}
The FBP is arguably the most used reconstruction algorithm in cryo-ET. It is simple, fast, and can handle moderate noise. However, in the limited-angle setting (and due to incomplete sampling of accessible angles) of cryo-ET, the FBP generates streak-like artifacts and renders details at certain orientations invisible.
In practice, the filter is chosen empirically among several high-pass filters, such as the ramp filter, although "exact" filters can be derived under certain assumptions \cite{harauz1986exact}.
Then, the filtered projection is backprojected to reconstruct the 3D volume. Let $\hat{\yb}_{\theta}(r_x,r_y)$ denote the filtered projection. The FBP volume reconstruction is 
\begin{equation}
	\label{eq:FBP_recon}
	\begin{split}
		\hat{V}_{\text{FBP}}(\vr^0_\theta ) &= \int_{
			\theta} \hat{\yb}_{\theta}(r_x^0\cos(\theta) - r_z^0\sin(\theta) , r_y^0)d\theta \\
			&= \int_{\theta} \hat{\yb}_{\theta}(\vr^0_\theta)d\theta.
	\end{split}
\end{equation}
Our proposed approach can then be seen as a nonlinear generalization of the FBP, where the integral (the average) along the sinusoidal support is replaced by a learnable function parameterized by a neural network. In contrast to FBP, \method{} uses a neural network in order to optimally process the local measurements. 
This also underlines that \method{} is a deep learning approach for cryo-ET which is substantially different from the existing postprocessing methods.

\subsection*{Training \method{}}
We selected tilt series from the EMPIAR-11830 dataset \cite{kelley_toward_2026} which contains \textit{Chlamydomonas reinhardtii} samples prepared using cryo-plasma FIB milling. The dataset comprises 1829 tilt series, with the number of projections per series ranging between 16 and 61. The projections are of size $4096 \times 4096$, sampled at 1.96 \AA{}/pixel.
The dataset is one of the few, if not the only publicly available dataset where full raw data and corresponding even-odd reconstructions are available.
Tilt series were aligned using AreTomo2 on the entire stack, which is then used to reconstruct tomogram pairs from the even-odd split. Then, denoising is performed using Cryo-CARE on both even and odd $4\times$ downsampled projections. 
Among tilt series with 41 projections or more, we visually inspected and ranked Cryo-CARE reconstructions based on the presence of artifacts and the quality of denoising, retaining the top 114.

These 114 tomograms were then further refined using two independent approaches. IsoNet was applied to the Cryo-CARE reconstructions to mitigate missing-wedge artifacts and improve denoising. In parallel, ICECREAM was applied directly to tomograms reconstructed from odd–even splits, without using Cryo-CARE outputs.
Reconstructions from IsoNet and ICECREAM were visually compared for each tomogram, and the best visual quality was retained. Reconstructions exhibiting prominent artifacts were excluded from the training data.

Finally, we downsampled the projections by a factor of 4 to match the reference reconstructed volumes, resulting in a sampling rate of 7.84 \AA{}/pixels. 
In order to train \method{} to process an arbitrary number of tilts, during the training phase, we randomly remove between 0 and 30 projections from the tilt series. 
In order to be robust to the effect of CTF, we also CTF corrected 64 aligned tilt series obtained using AreTomo3 \cite{peck2025aretomolive} and their corresponding reconstruction.

\bibliography{refs}

\backmatter

\addcontentsline{toc}{section}{Supplementary information}
\bmhead{Supplementary information}

\bmhead{Acknowledgements}
This work was supported by the European Research Council Starting Grant 852821---SWING. 
Calculations were performed at sciCORE (\href{http://scicore.unibas.ch/}{http://scicore.unibas.ch/}) scientific computing center at University of Basel.

\addcontentsline{toc}{section}{Data availability}
\section*{Data availability}
We release the training data to facilitate the development of new deep learning algorithms for cryo-ET: \dataref{}.

\addcontentsline{toc}{section}{Code availability}
\section*{Code availability}
The latest pretrained models are openly available following the instructions provided in the Github page: \git. 
\method{} allows straightforward reconstructions directly from aligned tilt series by running a few lines of Python code.
We also released a modified version of AreTomo3 that allows to obtain CTF corrected tilt series \href{https://github.com/vinith2/AreTomo3}{https://github.com/vinith2/AreTomo3}.

\section*{Acknowledgment}
The authors would like to thank Christophe Chaumet, Christiane Riedel, for valuable discussions.
This project was supported by the European Research Council Starting under Grant 852821—SWING.
V.D. is supported by the Agence National de la Recherche (ANR) and the Ministère de l'Enseignement Supérieur et de la Recherche.
Calculations were partially performed at sciCORE (\href{https://scicore.unibas.ch/}{https://scicore.unibas.ch/}) scientific computing center at University of Basel.


\appendixpageoff
\appendixtitleoff
\renewcommand{\appendixtocname}{Supplementary material}
\begin{appendices}
\section*{Supplementary material}
\section{Numerical implementation}\label{app:details-method}
\subsection*{Architecture}
An important advantage of our transform-domain local approach is that we can build a true 3D reconstructor which optimally exploits the 3D correlation structure of the measurement data and the volumes. Doing this naively, however, still results in high compute and memory complexity. We thus exploit the structure of the cryo-ET forward model in Equation \eqref{eq:delta_support} and combine MLPs which operate on constant-$y$ slices with an MLP acting along the $y$ dimension.

We split the patches along the rotation axis to obtain $P$ vectors of dimension $N \times P$. Each of these $P$ vectors is processed by a set-MLP block; we denote by $\smlp_s$ the set-MLP that processes the $s$-th vector. The architecture of the set-MLP block follows that of the Deep Sets \cite{zaheer2017deep} architecture as the number of projections in the tilt series can vary due to microscope setup and random corruptions present in the projections. The set-MLP block contains two MLPs: one acting along each of the projections sliced along the rotation axis, mapping $P$ vectors of size $P\times N$ to $P$ feature vectors of size $F$; and then the second MLP acting on the $PF$ features. Additionally, we incorporate positional encoding by converting the angle information to a feature vector that is used inside the architecture. 

More formally, let $\smlp_{s}: \mathbb{R}^{P \times N}\times \R^N \to \mathbb{R}^F$ for  $1\leq s \leq P$ and $\overline{\mlp}: \mathbb{R}^{PF} \to \mathbb{R}$. The overall neural network $f_\gammab$ can be written as 
\begin{equation*}
	\begin{split}
			f_\gammab(\vp) = \overline{\mlp}([\smlp_{1}(\vp_1,& \thetab),\smlp_{2}(\vp_2, \thetab), \dots,\\
			&\smlp_{P}({\vp}_{P}, \thetab)]),
	\end{split}
\end{equation*}
where $\vp_s$ denotes the $s$-th slice of the tilt series patches obtained at angles $\thetab\in \R^{P}$ and the learnable weights $\gammab$ are the concatenation of all the learnable parameters of the different MLPs. 
The base block of the neural network architecture consists of standard MLPs, defined as 
\begin{equation*}
	\begin{split}
		\mlp: \R^{n} \ni x \mapsto W_{\text{out}}\relu(W_{H}&\relu(W_{H-1}\dots \\
		&\relu(W_{\text{in}}(x)) )),
	\end{split}
\end{equation*}
where $\relu$ is the rectified linear unit, $n$ is the dimension of the input, $W_{\text{in}} \in \mathbb{R}^{n \times \text{hidden}}$, $W_{i} \in \mathbb{R}^{\text{hidden} \times \text{hidden}}$, $W_{\text{out}} \in \mathbb{R}^{\text{1} \times \text{hidden}}$, and we selected $\text{hidden}=128$ and $H=5$ for the released model.
The set-MLPs, $\smlp_{s}$ are simply MLPs to which the tilt angle information and the projection are previously encoded. First, a single linear layer is applied to map the slice of the patch of size $P\times N$ to $128\times N$, the $N$ coefficients being processed by the same weights. The angle coordinates $\thetab\in\R^{N\times 1}$ is processed by standard positional encoding to obtain a vector of size $N\times 128$. The two resulting vectors are then multiplied element wise and the result is processed by the $\mlp$ described above.
Note that we use linear layers without bias. This has been shown to improve robustness to noise \cite{mohan2019robust}. Moreover, it results in scaling equivariant (1-homogeneous) networks which greatly facilitates handling different normalizations of volumes and data. 

\begin{figure*}
    \centering
    \includegraphics[scale=0.3]{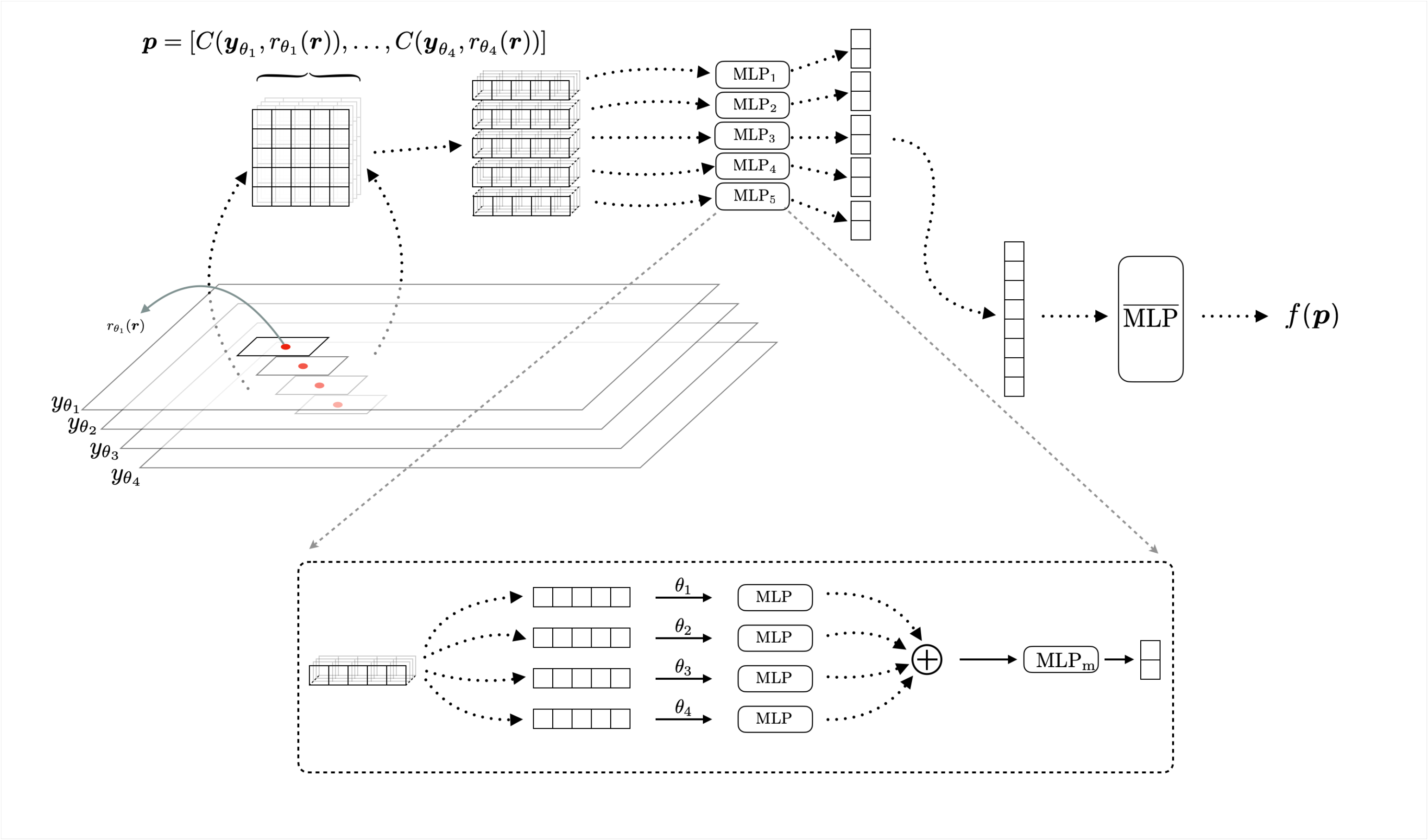}
    \caption{Schematic diagram of the SliceMLP architecture. The patches are grouped along the slices according to the axis of the tilt series. These slices are fed into the corresponding MLPs ($\text{MLP}_{i}$) to obtain a feature vector, which is used by a combination MLP ($\overline{\mlp}$) to estimate the volume at a particular location $\vr$. While the architecture appears slicewise for computational reasons, it is fundamentally a 3D architecture that exploits the 3D correlation structure of the measurement data to recover the volume. }
    \label{fig:slicemlp}
\end{figure*}

Finally, the averaging network $\overline{\mlp}$ transform the $PF$-dimensional feature vector by a second standard MLP as described above. The schematic diagram of the network architecture is provided in the Fig.~\ref{fig:slicemlp}.

\subsection*{Data augmentation for better denoising}
Modern direct electron detectors collect multiple frames at each tilt angle, allowing compensation of beam-induced motion \cite{Brilot2012}. By aligning and averaging these frames, small deformations can be compensated and sharper images  are obtained. Self-supervised denoising frameworks such as Cryo-CARE and Topaz-Denoise group the frames into two  subsets to generate independent noisy realizations of the measurements, enabling training in a Noise2Noise framework.

While training \method{}, we can also benefit from the Noise2Noise framework using the two independent and identically distributed tilt series available. More precisely, we randomly use one of the two dose-fractionated tilt series available with the tomogram reconstructed with the full tilt series.
We applied this strategy and compared the network's performance on the ribosome dataset (EMPIAR-10045).
Fig.~\ref{fig:noise_input_test} shows the effect of this data augmentation technique on reconstruction quality compared to using standard averaged projections, referred to as 'averaged'. We observe that the network trained with the proposed data augmentation step performs better visually, compared to the network trained with fully averaged projections. This can be seen with the increase in contrast between the background and the particles in the figure.

\def\xx{0}
\def\yy{0}
\def\zx{-0.2}
\def\zy{1.5}
\def\ps{0.24}
\def\sztmp{3.9cm}
\begin{figure}
	\centering
	\begin{subfigure}[t]{\ps\textwidth}
		\centering
		\begin{tikzpicture}[spy using outlines={circle,orange,magnification=5,size=1.7cm, connect spies}]
			\node[ rotate=90] at (0*\xx,-0*\yy) {}; outlines={circle,orange,magnification=2,size=3cm, connect spies}]
			\node[rotate=0, line width=0.05mm, draw=white] at (0*\xx,-0.*\yy) { \includegraphics[height=\sztmp]{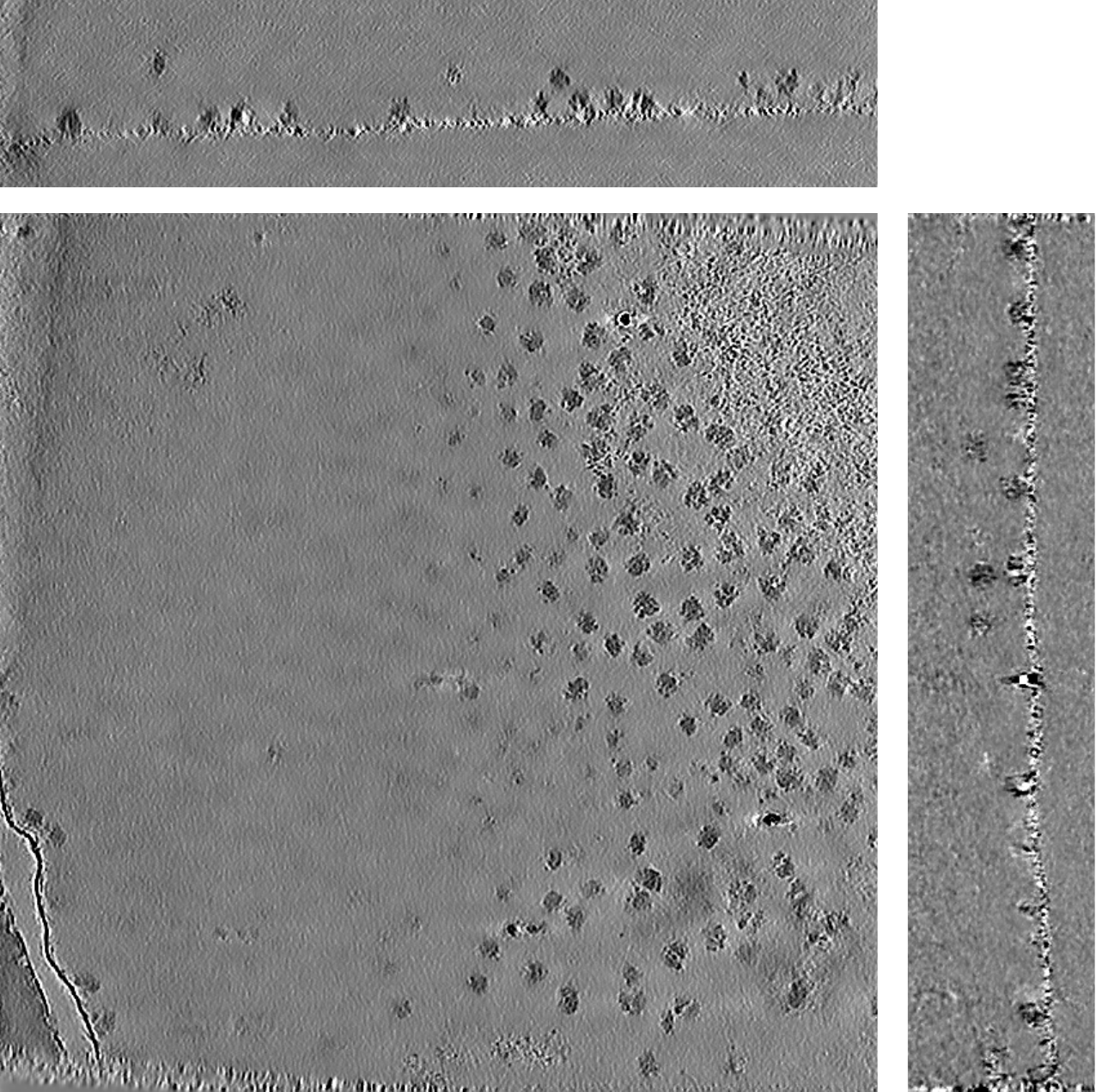}};
			\spy on (\zx,\zy) in node [left] at (2,-0.95);
		\end{tikzpicture} 
		\caption{\method{} trained only with raw averaged tilt series.}
	\end{subfigure}\hfill
	\begin{subfigure}[t]{\ps\textwidth}
		\centering
		\begin{tikzpicture}[spy using outlines={circle,orange,magnification=4,size=1.7cm, connect spies}]
			\node[ rotate=90] at (0*\xx,-0*\yy) {}; outlines={circle,orange,magnification=2,size=3cm, connect spies}]
			\node[rotate=0, line width=0.05mm, draw=white] at (0*\xx,-0.*\yy) { \includegraphics[height=\sztmp]{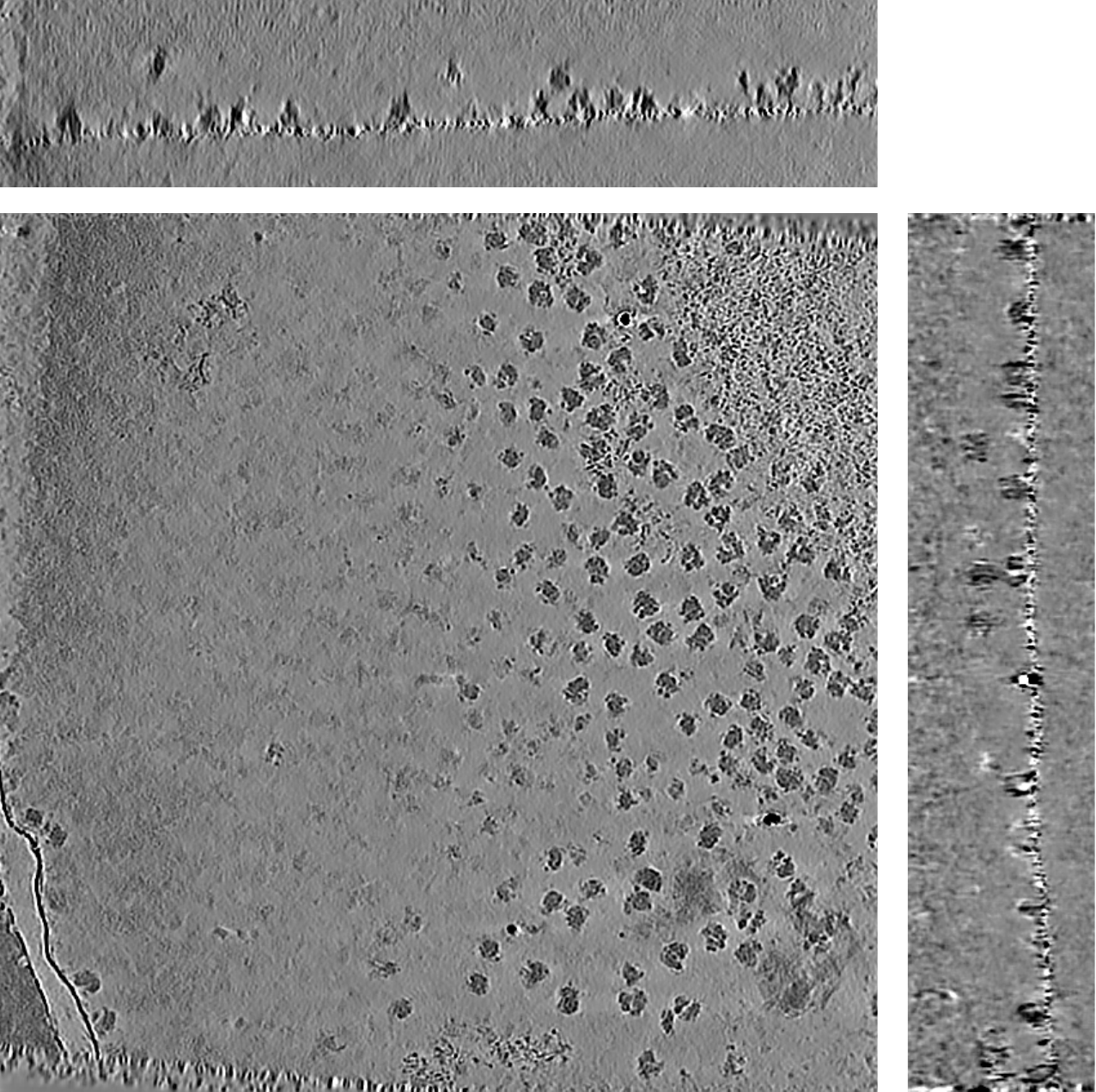}};
			\spy on (\zx,\zy) in node [left] at (2,-0.95);
		\end{tikzpicture} 
		\caption{\method{} trained with Noise2Noise like data augmentation.}
	\end{subfigure}
	\caption{Influence of data augmentation strategy on the reconstruction performance which improves the denoising capacity of \method{}. }\label{fig:noise_input_test}
\end{figure}

\subsection*{Fast reconstruction}
\begin{figure}
    \centering
    \includegraphics[scale=0.2]{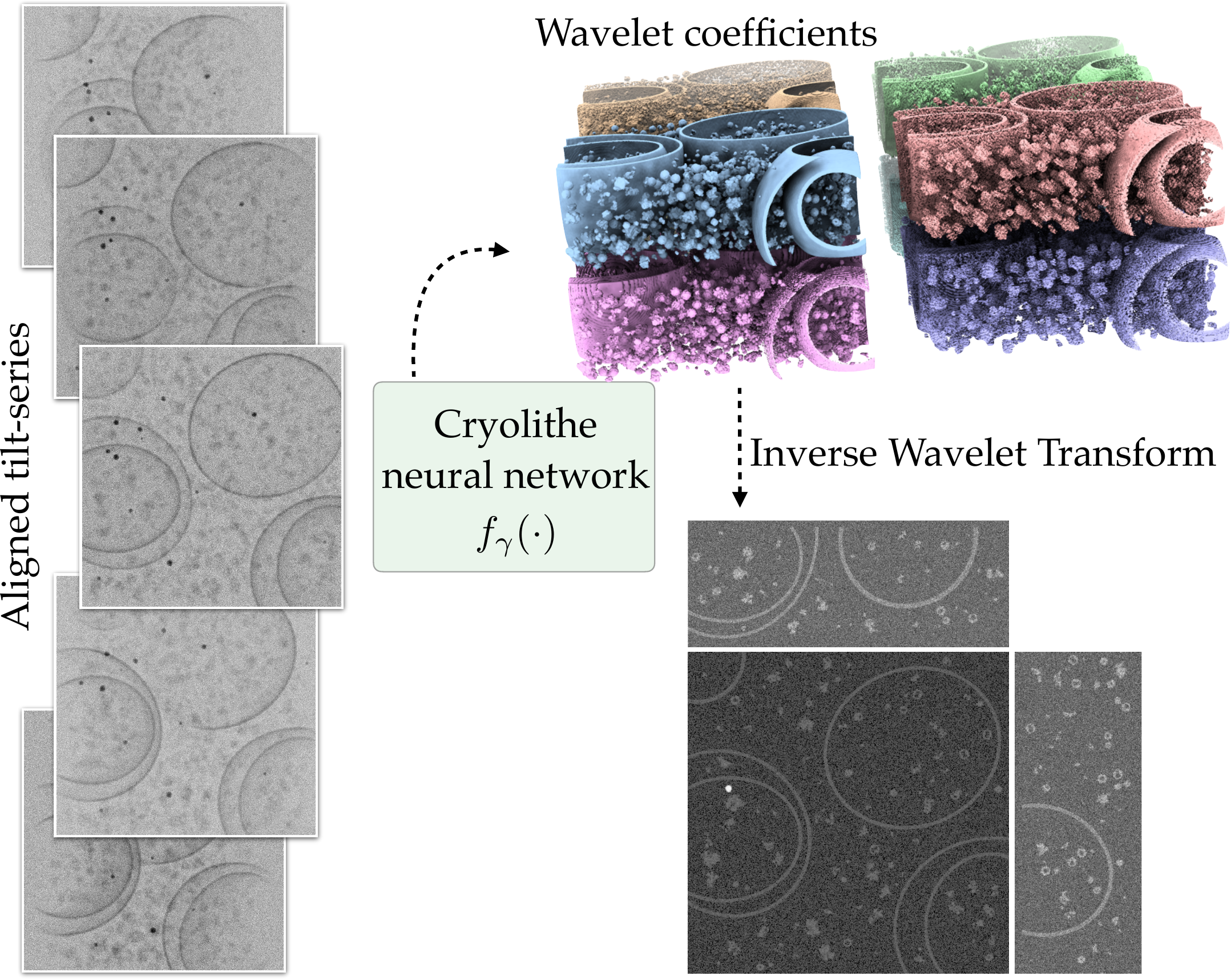}
    \caption{\method{} with the wavelet based reconstruction method. The network uses local information from the tilt series to estimate 8 wavelet coefficients simultaneously, which are combined to recover the volume using the standard inverse wavelet transform. This amounts to at least a 8$\times$ speed increase at inference. }
    \label{fig:set-wvelet-inference}
\end{figure}

The default model recovers the volumes in a voxel-wise manner, which makes the recovery time slow for large volumes. We propose to fasten the reconstruction process by reducing the number of pixels being sampled. To prevent patching artefacts, we trained the network to output wavelet coefficients instead. The wavelet coefficients estimates are combined and the inverse wavelet transform is used to recover the final volume. 
Note that each level of the wavelet transform splits the volume into high-pass and low-pass components in three orthogonal directions and further downsample it by a factor of 2. This amounts to splitting the volume into 8 different subvolumes of half the size in each direction, see Fig.~\ref{fig:set-wvelet-inference}. 
We extend our network architecture to output the 8 wavelet coefficients on a grid that is $\frac{1}{8}^{\text{th}}$ smaller than the target one, providing roughly a $8\times$ speed-up in reconstruction. The wavelet needs to be carefully chosen such that the support of the wavelet filter lies withing the support of the network input in order to accurately estimate the coefficients. We choose biorthogonal wavelets in our experiments. The network that outputs wavelet coefficients, \method{}, is trained as described previously except for the voxel sampling stage that is replaced by the wavelet coefficient sampling of the volume. The network is trained to predict these coefficients. 

\subsection*{Real-space vs Wavelet implementation}
The proposed approach comes with two different implementations: \method{} that predicts 8 wavelet coefficients on a coarse grid and predicts the final tomogram by applying the inverse Wavelet transform, and \method{}-Pixel that reconstructs a single pixel of the tomogram at a time.
We show that \method{} comes with only a small drop in accuracy compared to the standard model but with significant speed-up in the reconstruction. Fig.~\ref{fig:wavelet-est} shows the orthogonal slices of the reconstruction and the self-FSC curve computed on two tilt series obtained with dose fractionation, which have not been used for training. The FSC curves agree for most of the spectrum. However, the inference time is significantly faster ($\times 14$) for \method{}  on a tomogram of size $928 \times 928 \times 464$. 
In Fig.~\ref{fig:wavelet-comparison}, we provide additional visual comparison when tested on a dataset coming from a microscope not seen during training.
For these reasons, we chose to highlight \method{}'s reconstructions in the core of the paper. We let the possibility for users to select the model they want to use in practice, as it might change depending on applications.

\def\sztmp{5cm}
\def\ps{0.33}
\def\scc{512*0.5}
\begin{figure*}
	\begin{subfigure}[t]{\ps\textwidth}
		\centering
		\begin{tikzpicture}[spy using outlines={circle,orange,magnification=4,size=2cm, connect spies}]
			\node[ rotate=90] at (0,0) {}; outlines={circle,orange,magnification=2,size=3cm, connect spies}]
			\node[rotate=0, line width=0.05mm, draw=white] at (0,0) { \includegraphics[height=\sztmp]{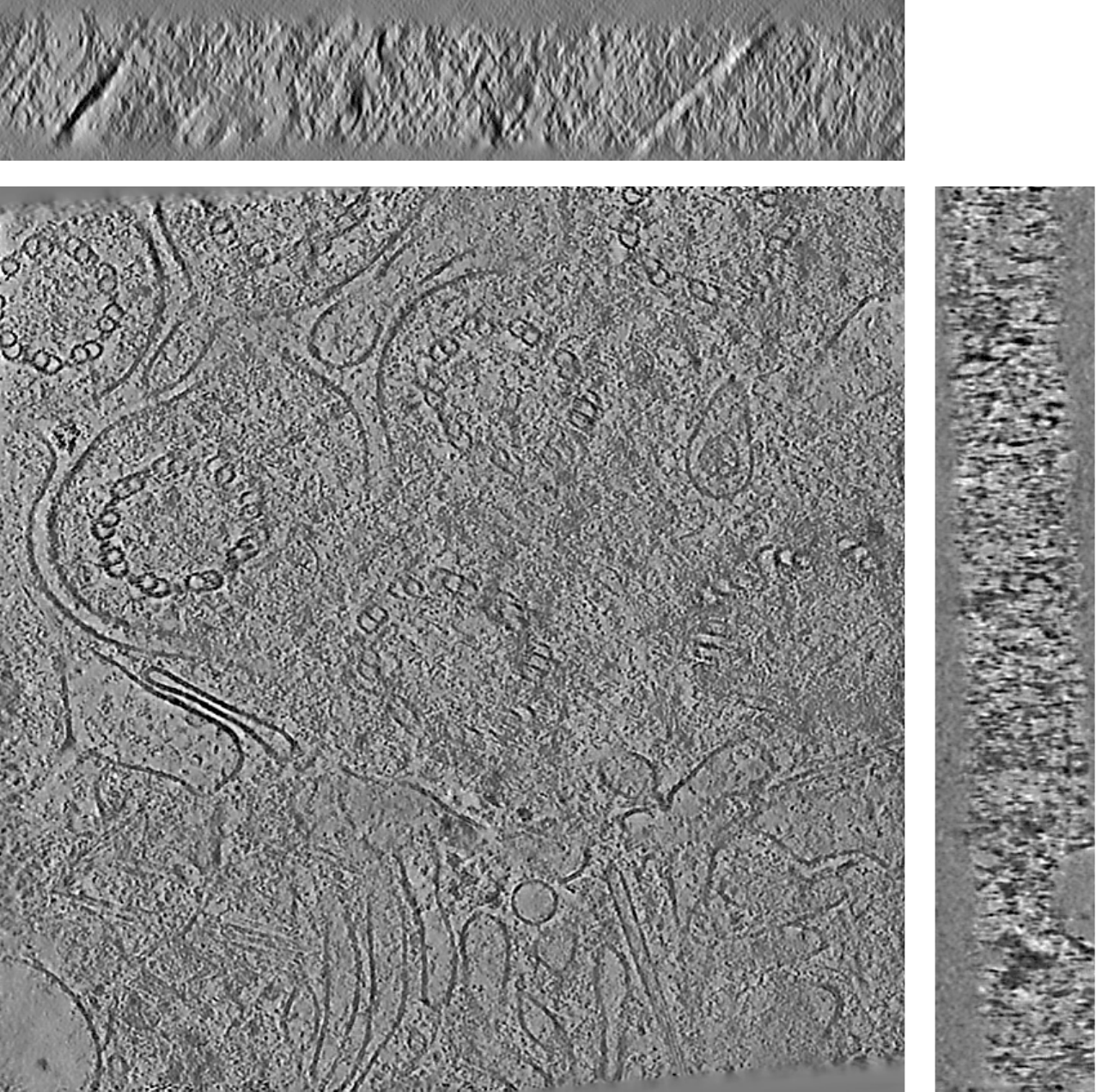}};
			\spy on (-1.5,-0.2) in node [left] at (2.2,1.2);
		\end{tikzpicture} 
		\caption{\method{}-Pixel (37min 6sec).}
	\end{subfigure}\hfill
	\begin{subfigure}[t]{\ps\textwidth}
		\centering
		\begin{tikzpicture}[spy using outlines={circle,orange,magnification=4,size=2cm, connect spies}]
			\node[ rotate=90] at (0,0) {}; outlines={circle,orange,magnification=2,size=3cm, connect spies}]
			\node[rotate=0, line width=0.05mm, draw=white] at (0,0) { \includegraphics[height=\sztmp]{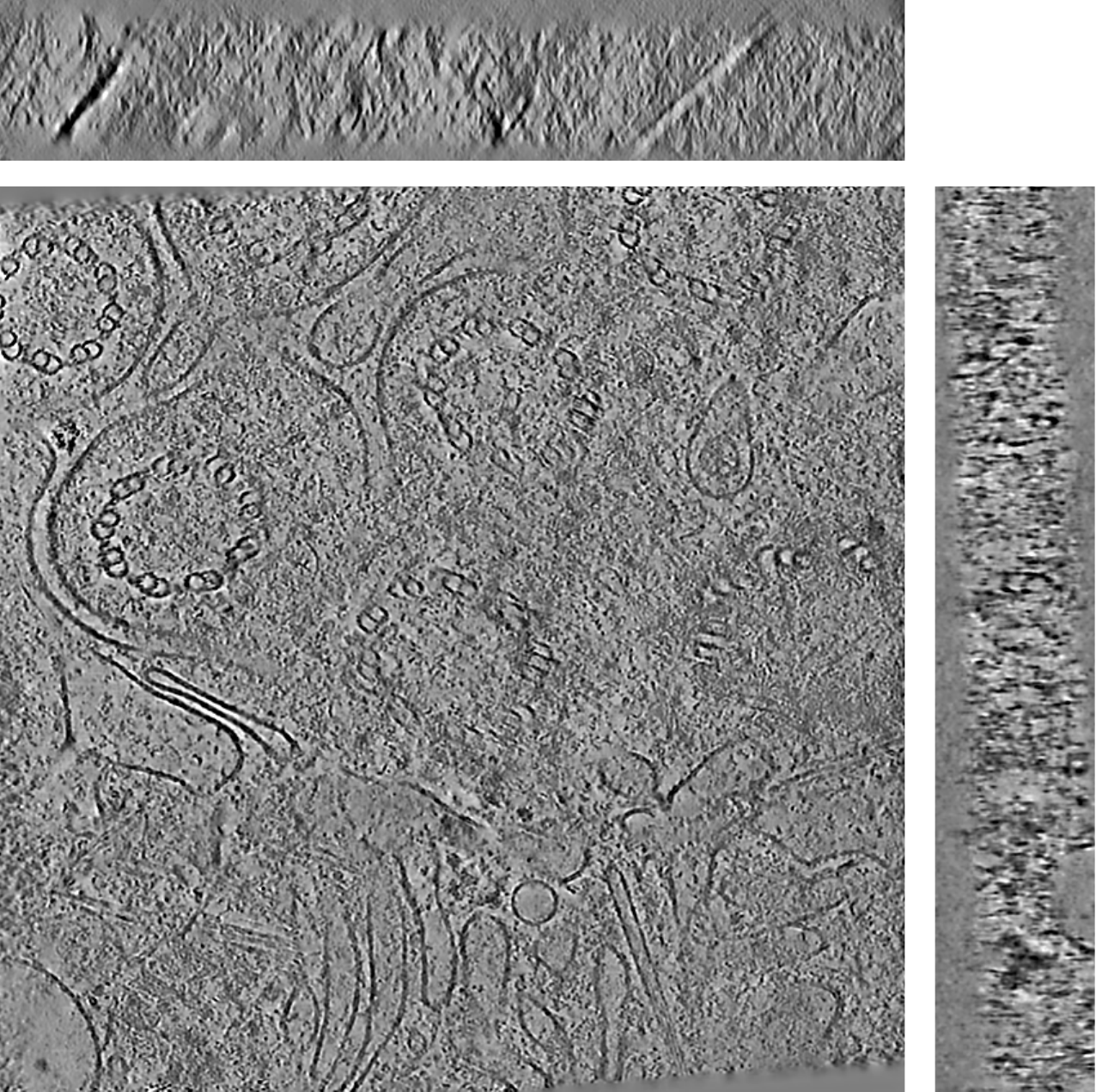}};
			\spy on (-1.5,-0.2) in node [left] at (2.2,1.2);
		\end{tikzpicture} 
		\caption{\method{} (4min 47sec).}
	\end{subfigure}
	\begin{subfigure}[t]{\ps\textwidth}
		\centering
		\begin{tikzpicture}
			\begin{axis}[
				axis lines = left,
                xmin = 0, xmax = 0.51,
                width=1.\linewidth, 
                height=1.13\linewidth, 
                height=5.6cm,
				grid=major, 
				grid style={dashed,gray!30}, 
				xlabel= {\notsotiny{Resolution (\AA)}},
				xtick={0.1,0.2,0.3,0.4, 0.5},
				xticklabels={78.4, 39.2, 26.1, 19.6, 15.7},
				ylabel={\notsotiny{FSC}},legend style={at={(1,1)}, legend cell align=left, align=left, draw=none,font=\notsotiny}]
				\addplot[mark=\methodmark, mark size=\ms, line width=\lw,  mark repeat=20, color=\methodColor] table [x expr=\coordindex/\scc, y=Voxel, col sep=comma] {FigureA4/self_fsc_odd_eve.txt};
			     \addlegendentry{\method{}-Pixel};
                 \addplot[mark=\isoCaremark, mark size=\ms, line width=\lw,  mark repeat=20, color=\isoCareColor] table [x expr=\coordindex/\scc, y=Wavelet, col sep=comma] {FigureA4/self_fsc_odd_eve.txt};
			      \addlegendentry{\method{}}
                \addplot[mark=\FBPmark, mark size=\ms, line width=\lw,  mark repeat=60, color=\FBPColor] table [x expr=\coordindex/\scc, y=fbp, col sep=comma] {FigureA4/self_fsc_odd_eve.txt};
                \addlegendentry{FBP}
			\end{axis}
		\end{tikzpicture}
		\caption{Self-FSC curve.}
	\end{subfigure}\hfill
	\caption{Comparison of the two proposed implementation: \method{}-Pixel and \method{} on EMPIAR-13055\cite{bertiaux_luminal_2025} dataset. The reconstructed tomograms are of size $928 \times 928 \times 464$. 
    The FSC curves are obtained via dose fractionation.  }\label{fig:wavelet-est}
\end{figure*}

\def\sztmp{3.9cm}
\def\xx{0}
\def\yy{0}
\def\zx{-0.1}
\def\zy{1.2}
\def\ps{0.25}
\begin{figure*}
	\centering
        \begin{subfigure}[t]{\ps\textwidth}
        	    \centering
        	    \begin{tikzpicture}[spy using outlines={circle,orange,magnification=5,size=2.cm, connect spies}]
            		\node[ rotate=90] at (-1,-1) {}; outlines={circle,orange,magnification=2,size=3cm, connect spies}]
            		\node[rotate=0, line width=0.05mm, draw=white] at (0*\xx,-0.*\yy) { \includegraphics[height=\sztmp]{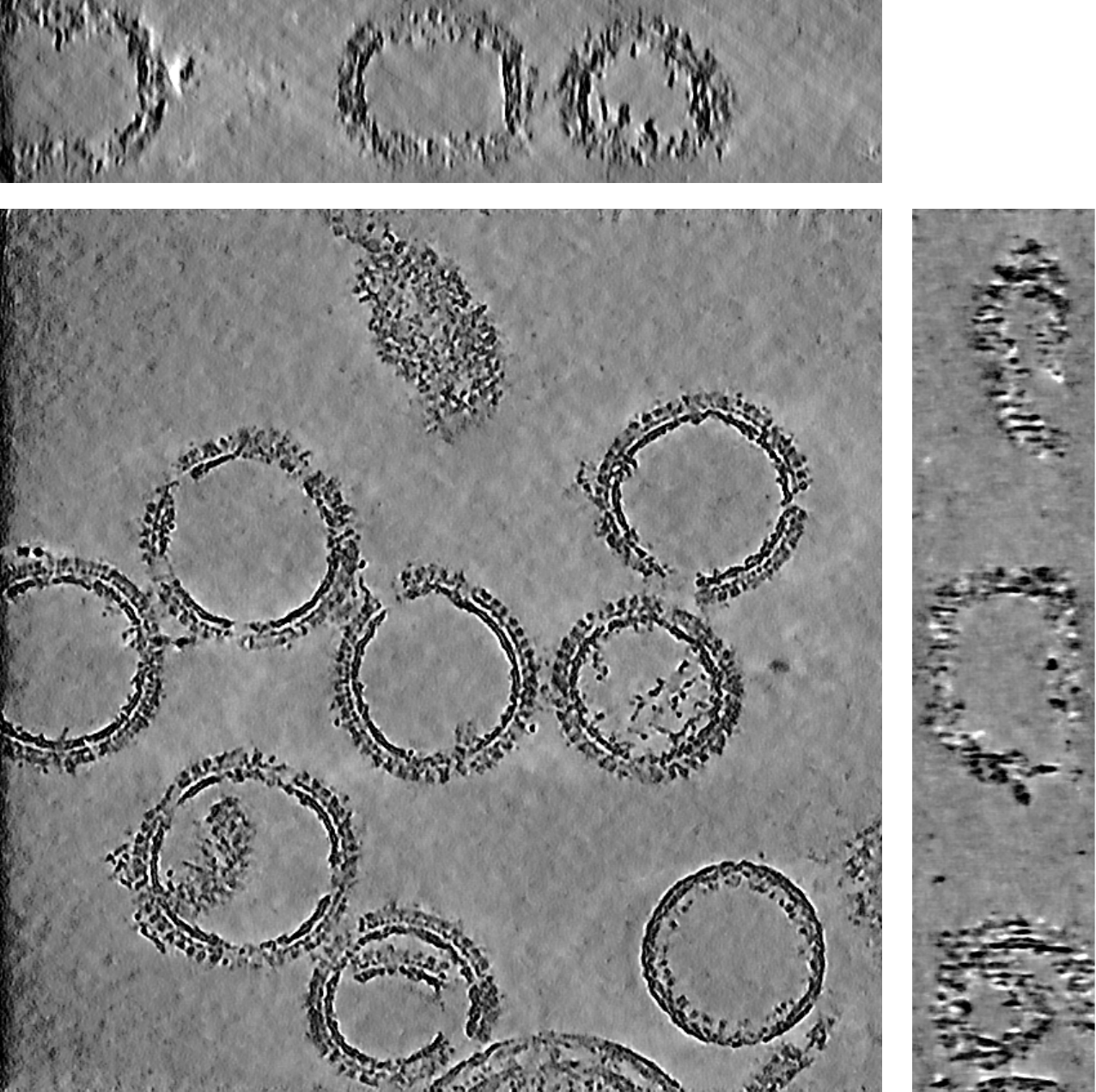}};
            		\spy on (-0.7,-0.2) in node [left] at (2,1.);
        		\end{tikzpicture} 
        		\caption{\method{}-Pixel ($\approx$40min)}
        	\end{subfigure}\hfill
        \begin{subfigure}[t]{\ps\textwidth}
        	    \centering
        	    \begin{tikzpicture}[spy using outlines={circle,orange,magnification=5,size=2cm, connect spies}]
            		\node[ rotate=90] at (0*\xx,-0*\yy) {}; outlines={circle,orange,magnification=2,size=3cm, connect spies}]
            		\node[rotate=0, line width=0.05mm, draw=white] at (0*\xx,-0.*\yy) { \includegraphics[height=\sztmp]{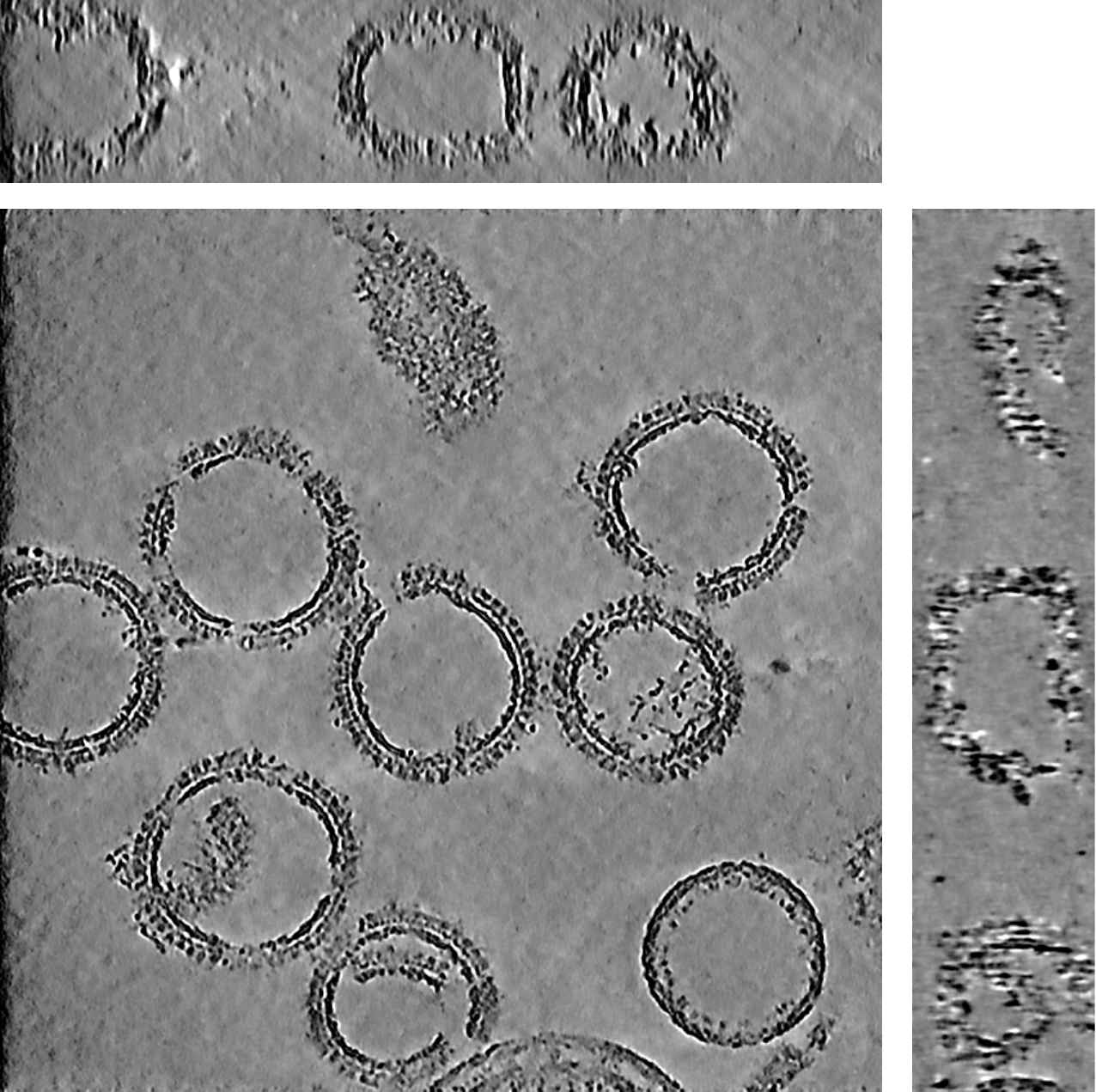}};
            		\spy on (-0.7,-0.2) in node [left] at (2,1);
        		\end{tikzpicture} 
        		\caption{\method{} ($5$min 12sec).}
        	\end{subfigure}\hfill
        \begin{subfigure}[t]{\ps\textwidth}
        	    \centering
        	    \begin{tikzpicture}[spy using outlines={circle,orange,magnification=5,size=2cm, connect spies}]
            		\node[ rotate=90] at (0*\xx,-0*\yy) {}; outlines={circle,orange,magnification=2,size=3cm, connect spies}]
            		\node[rotate=0, line width=0.05mm, draw=white] at (0*\xx,-0.*\yy) { \includegraphics[height=\sztmp]{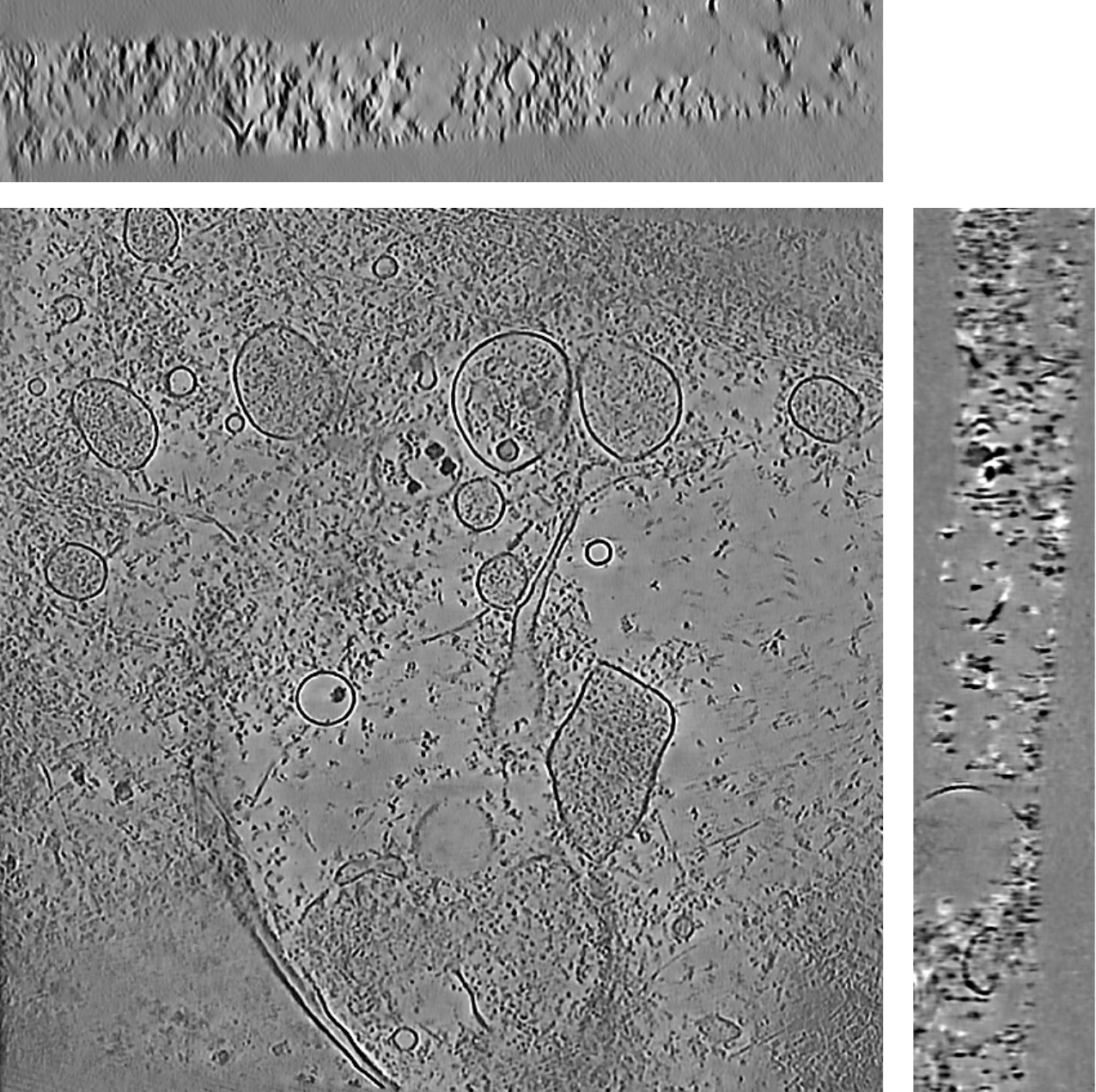}};
            		\spy on (-0.9,-0.2) in node [left] at (2,1);
        		\end{tikzpicture} 
        		\caption{\method{}-Pixel ($\approx$31min).}
        	\end{subfigure}\hfill
            \centering
	\begin{subfigure}[t]{\ps\textwidth}
	    \centering
	    \begin{tikzpicture}[spy using outlines={circle,orange,magnification=5,size=2cm, connect spies}]
    		\node[ rotate=90] at (0*\xx,-0*\yy) {}; outlines={circle,orange,magnification=2,size=3cm, connect spies}]
    		\node[rotate=0, line width=0.05mm, draw=white] at (0*\xx,-0.*\yy) { \includegraphics[height=\sztmp]{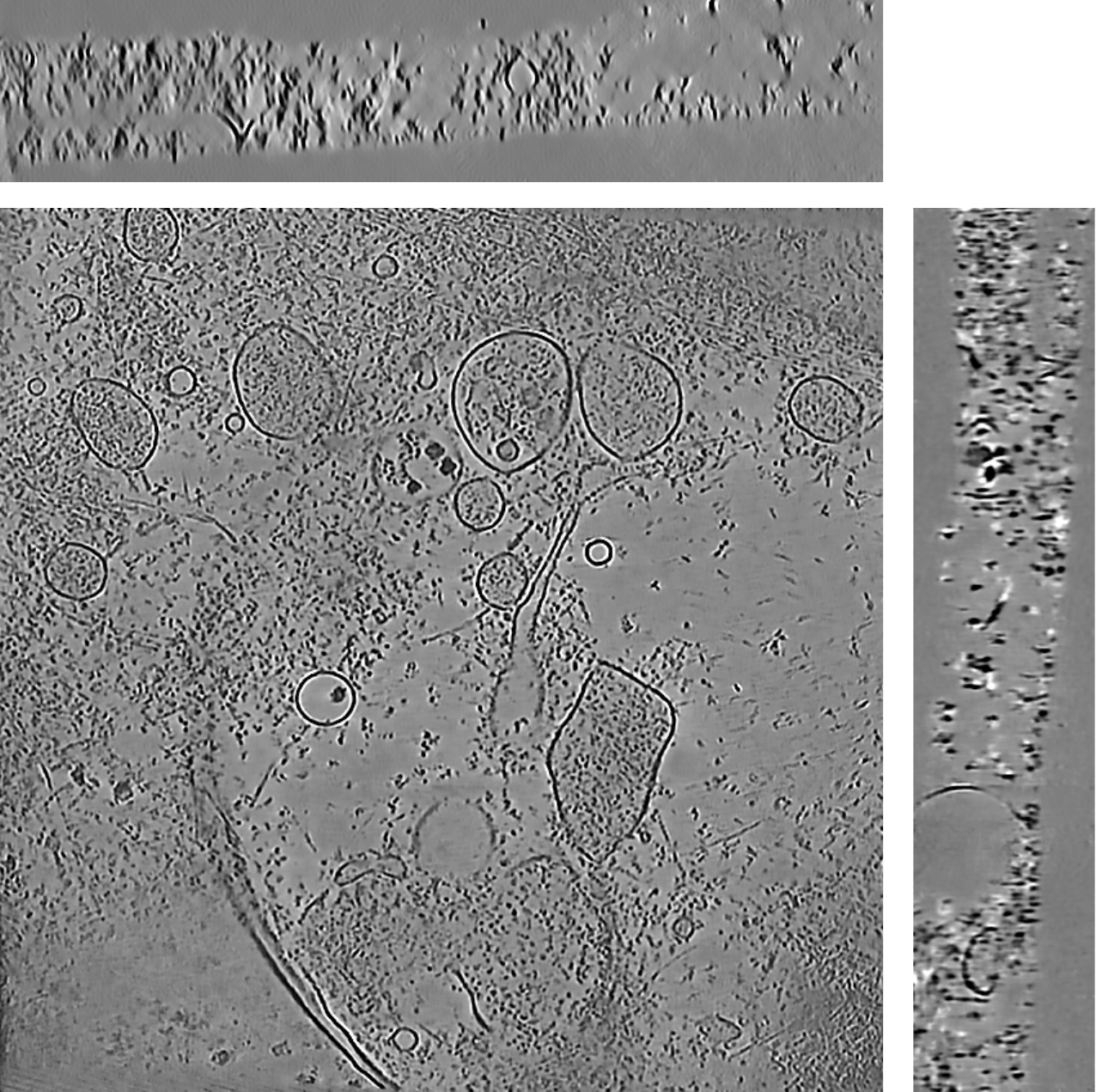}
            };
    		\spy on (-0.9,-0.2) in node [left] at (2,1);
		\end{tikzpicture}		\caption{\method{} (4min 2sec). }
    \end{subfigure}\hfill
\caption{Comparison of \method{}-Pixel and \method{} on data biologically different and not included in the training set, HIV-1 data from EMPIAR-10643 and non-infected cos-7 from EMPIAR-12262.  }\label{fig:wavelet-comparison}
\end{figure*}

\section{Additional reconstructions}

\subsection*{FBP reconstruction}
In Fig.~\ref{fig:mosaic-2-fbp}, we display the FBP reconstructions corresponding to to the tomograms reconstructed in Fig.~\ref{fig:mosaic-2}.

\begin{table*}
	\centering
    \resizebox{1.5\columnwidth}{!}{%
	\begin{tabular}{@{}cccccc@{}}
			\textbf{Image} &
			\textbf{EMPIAR ID}&
			\textbf{Tomogram ID}&
			\textbf{Pixel size (Å)}& \textbf{Dimensions (px)} & \textbf{Run time (s)} \\
			\hline
			A  &11058&9&14.08 & 928 × 928 × 464  & 248.5 \\
			     B &11078&2&13.68 & 928 × 928 × 464 & 251.9\\
			C &11538&113 &4.0 &1024 × 1024 × 512 & 339.3 \\
			D &11538&1435 &4.0 &1024 × 1024 × 512 & 338.4  \\
			E &11658&7 &7.84 & 1024 × 1024 × 512 & 198.3 \\
			F &11658&255 &7.84 & 1024 × 1024 × 512 & 244.1 \\
			G &11577& 98 &18.22 & 1022 × 1440 × 512 & 440.1 \\
			H &11830&2912 &7.84 & 1024 × 1024 × 512 & 301.9 \\
			I &12262 & 1 &5.0 & 1278 × 1236 × 512 & 461.1 \\			
		\end{tabular}
        }
	\caption{Information about the tilt series used to generate Fig.~\ref{fig:mosaic-2} and Fig.~\ref{fig:mosaic-2-fbp}.\label{table:mosaic-2}}
\end{table*}

%
 \begin{table*}
 \centering
\begin{tabular}{@{}c@{}c@{}c@{}}
			\overlayerimage{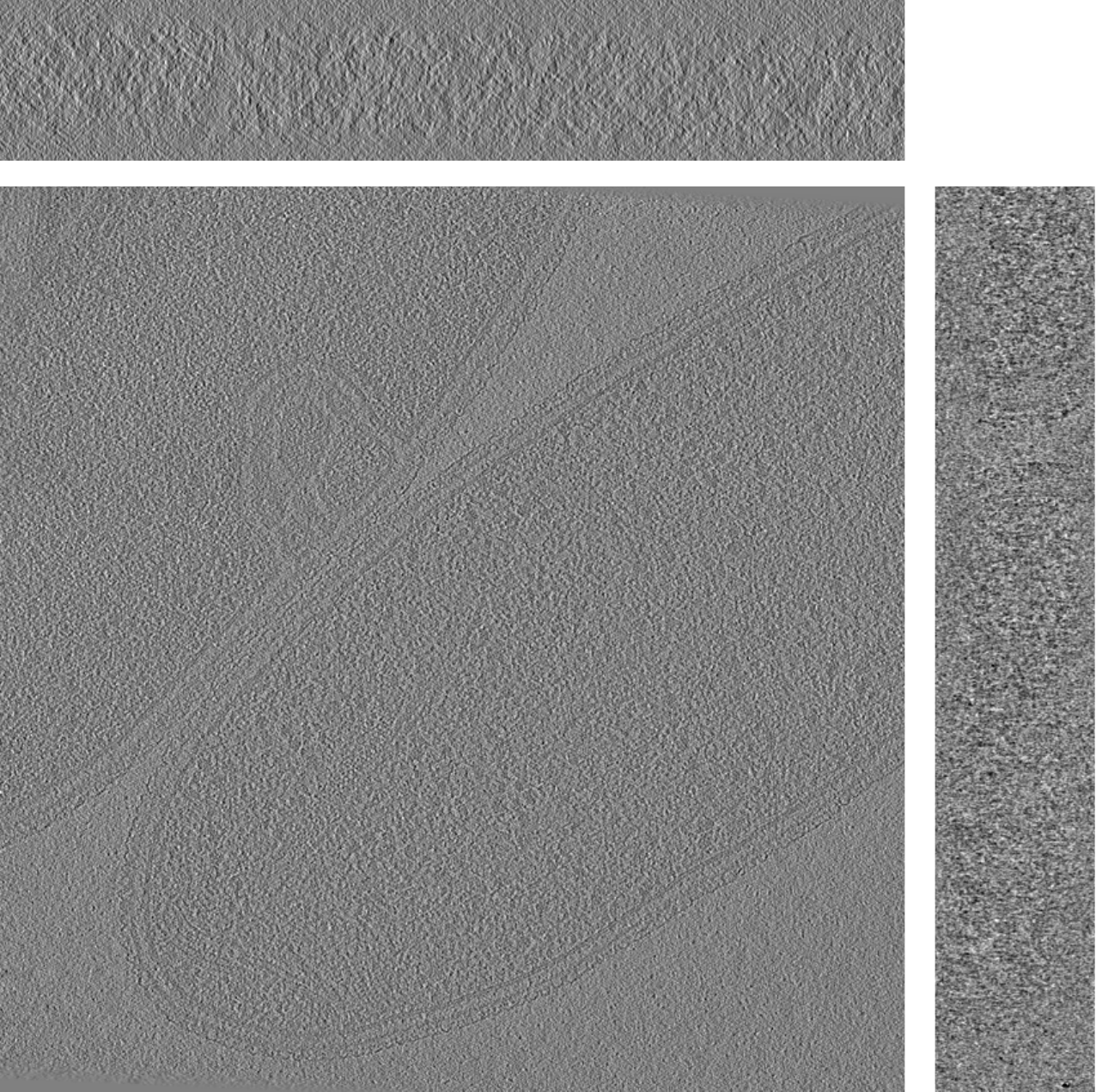}{\sz}{0.247}{100 nm}{A.} &
			\hspace{0.5mm} 
            \overlayerimage{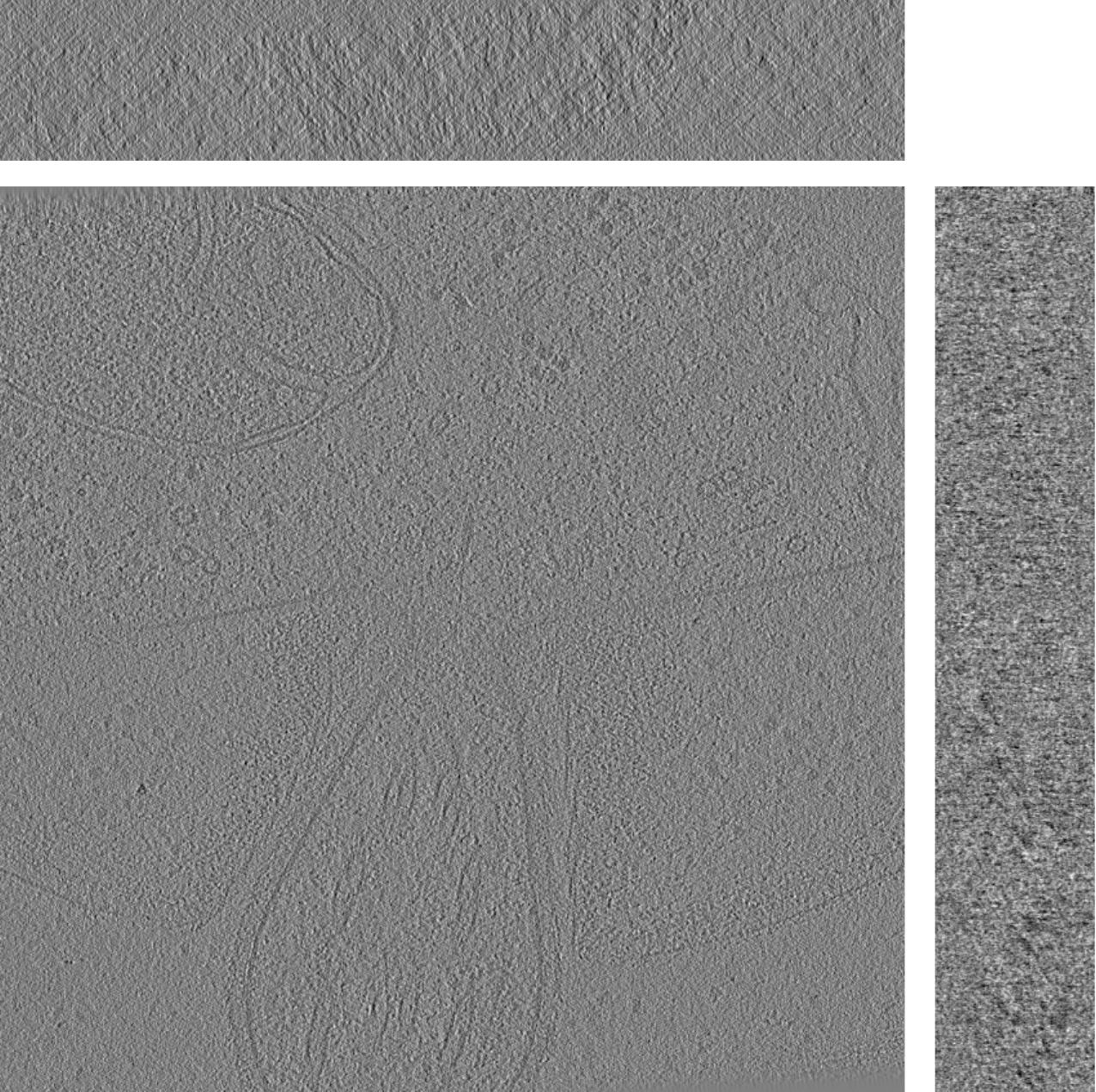}{\sz}{0.255}{100 nm}{B.} &
			\hspace{0.5mm} 
			\overlayerimage{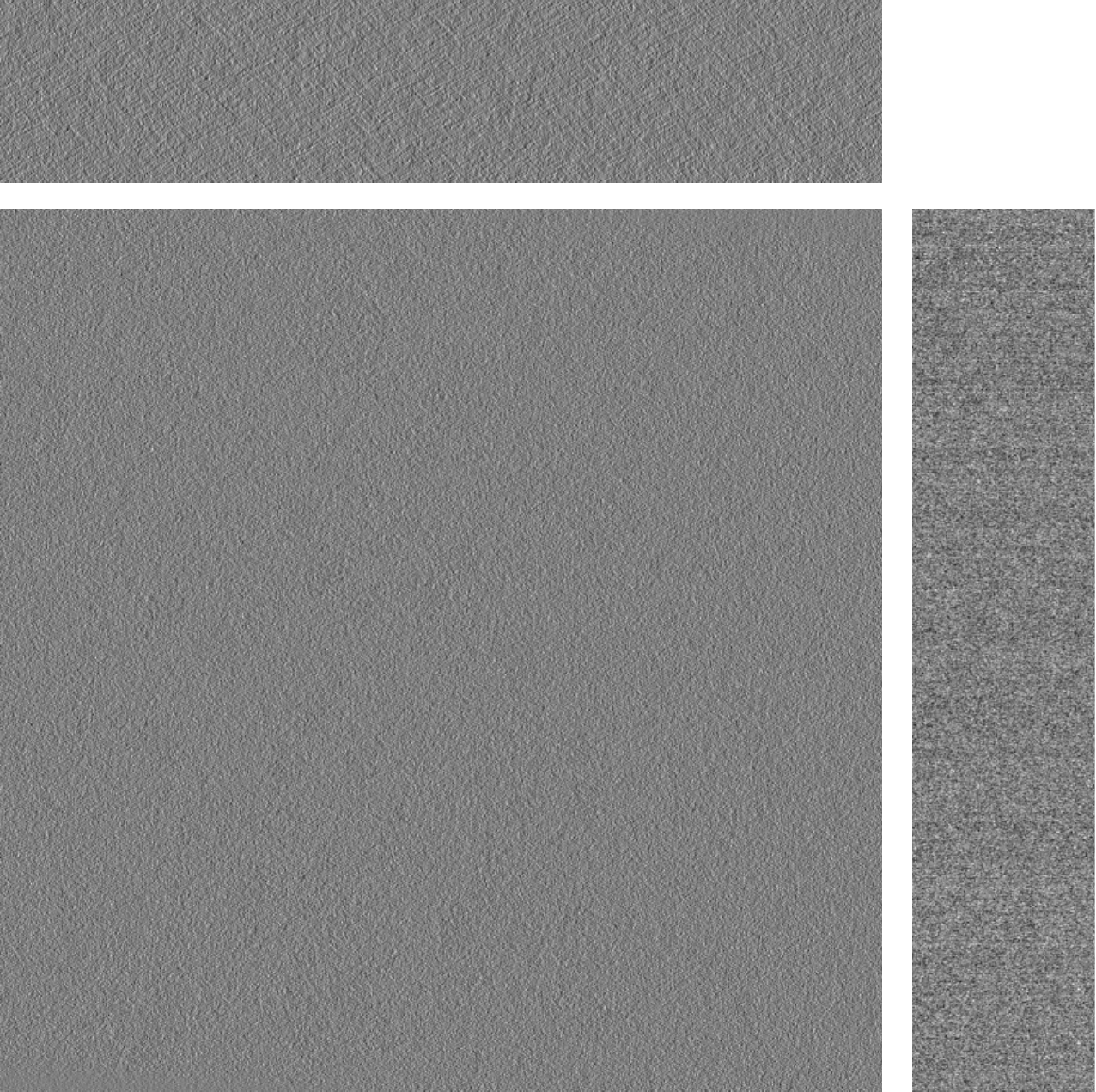}{\sz}{0.4177}{50 nm}{C.}  \\
			\overlayerimage{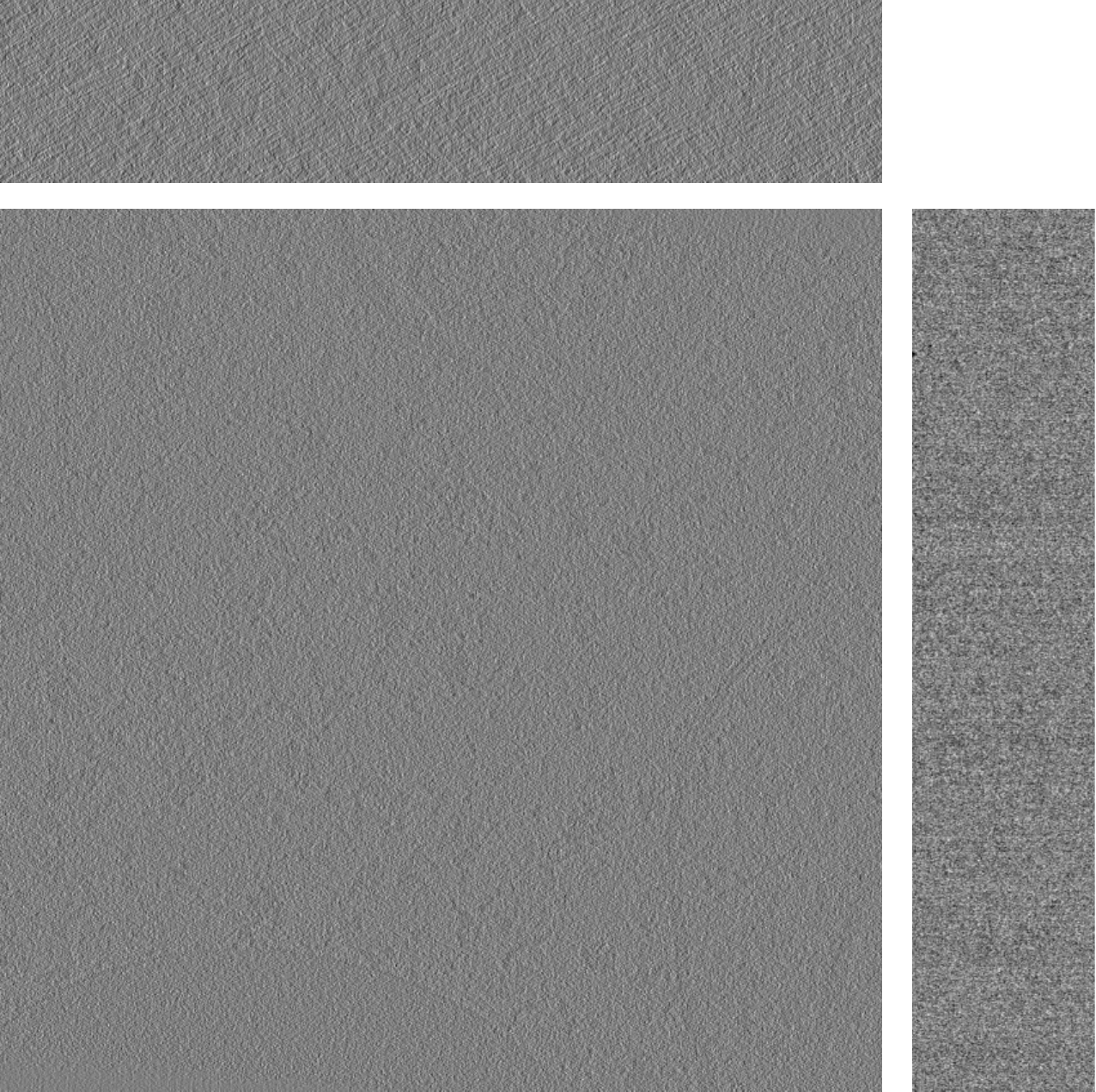}{\sz}{0.2956}{100 nm}{D.}  &
			\hspace{0.5mm} 
			\overlayerimage{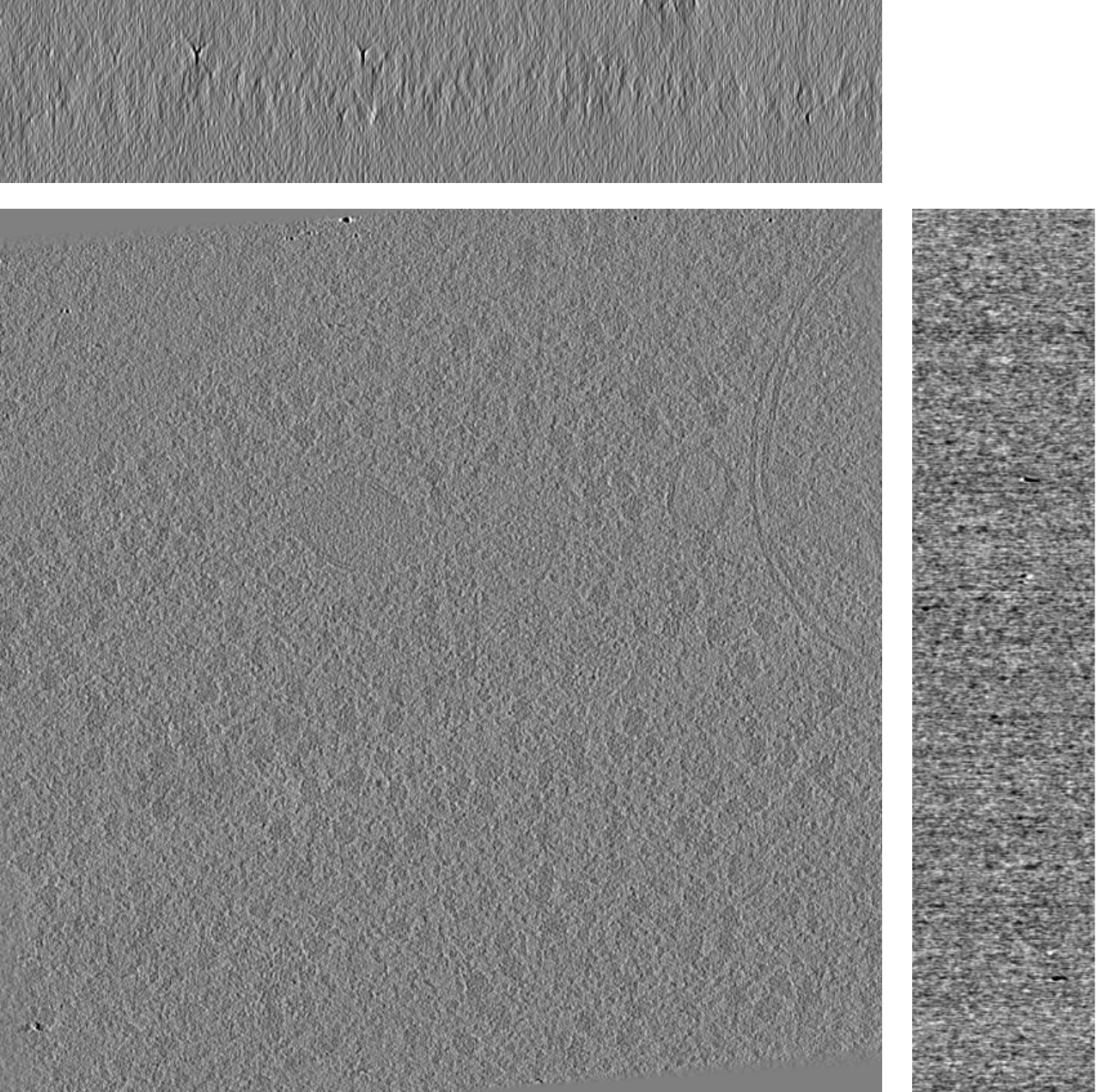}{\sz}{0.488}{50 nm}{E.}  &
			\hspace{0.5mm} 
			\overlayerimage{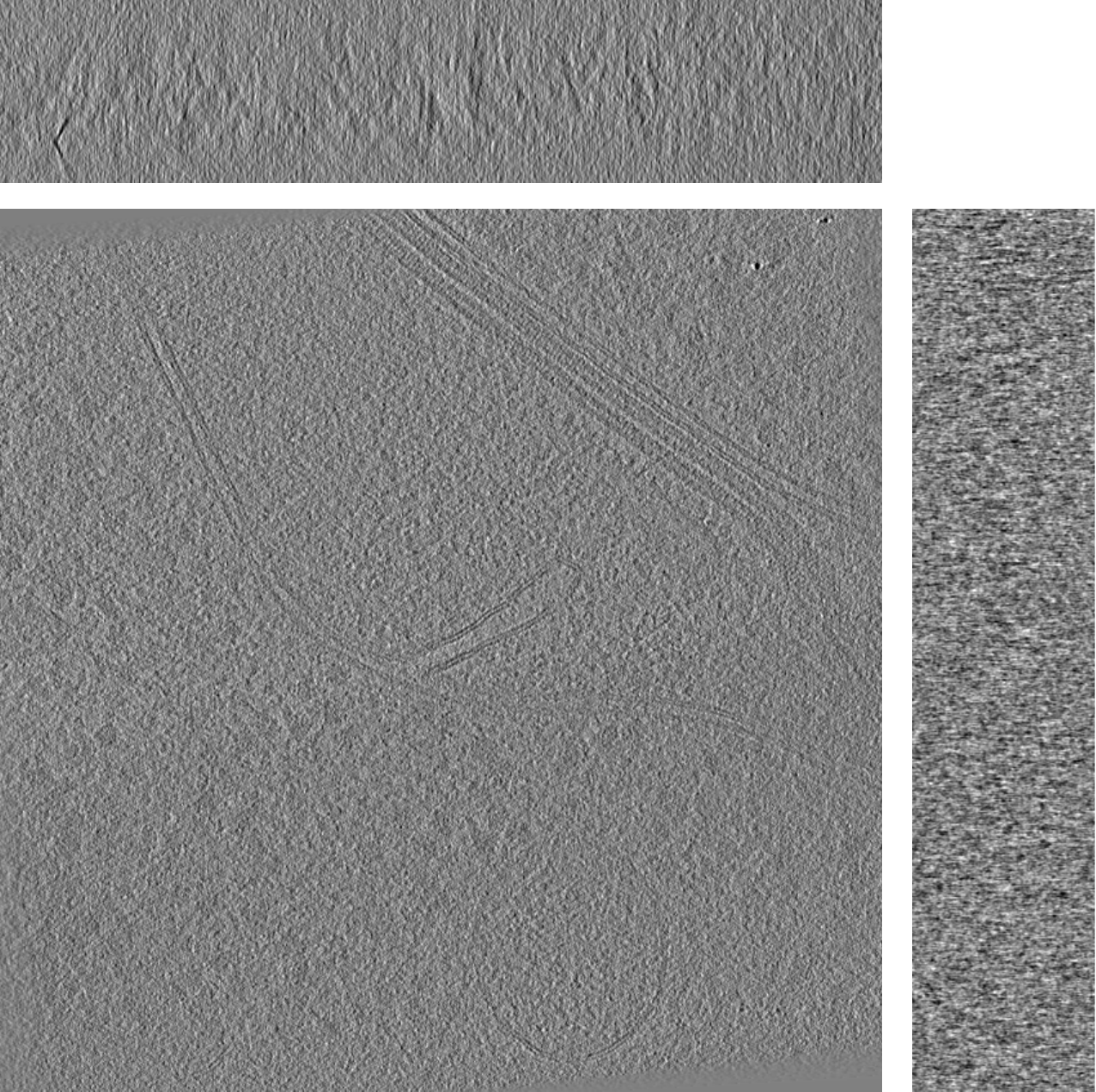}{\sz}{0.306}{100 nm}{F.}  \\
                \overlayerimage{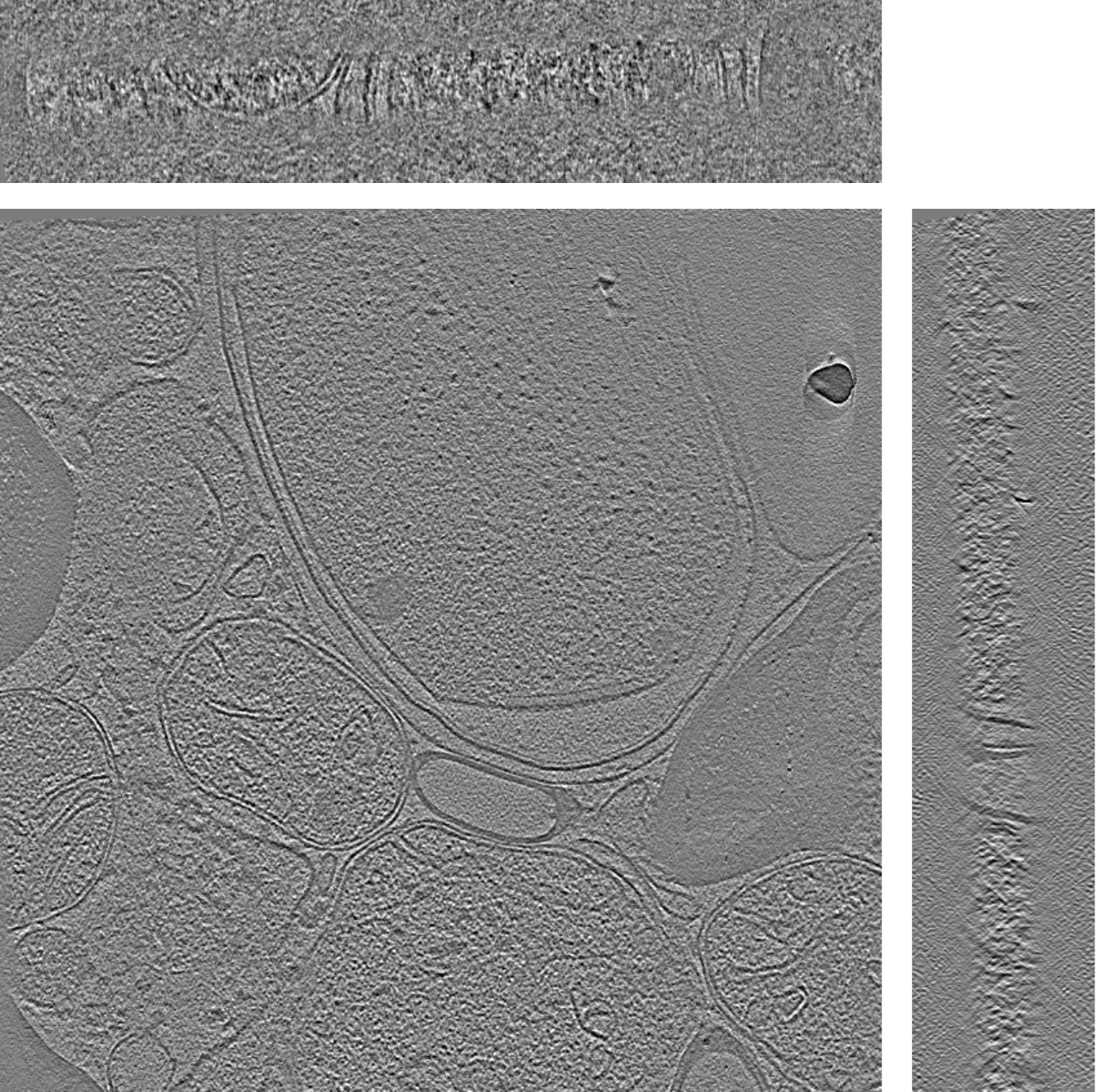}{\sz}{0.178}{100 nm}{G.} &
			\hspace{0.5mm} 
			\overlayerimage{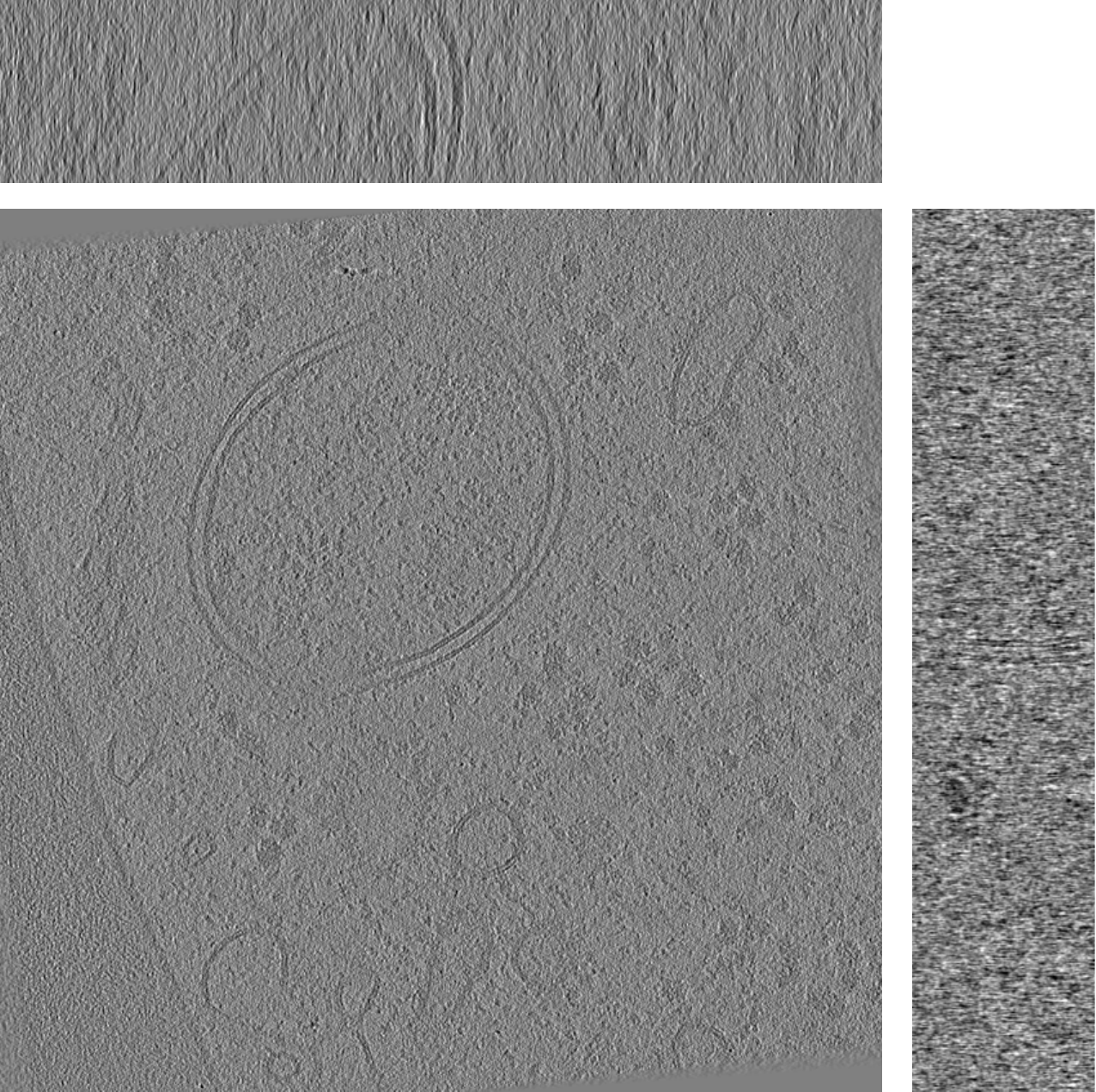}{\sz}{0.4102}{100 nm}{H.}  &
			\hspace{0.5mm} 
			\overlayerimage{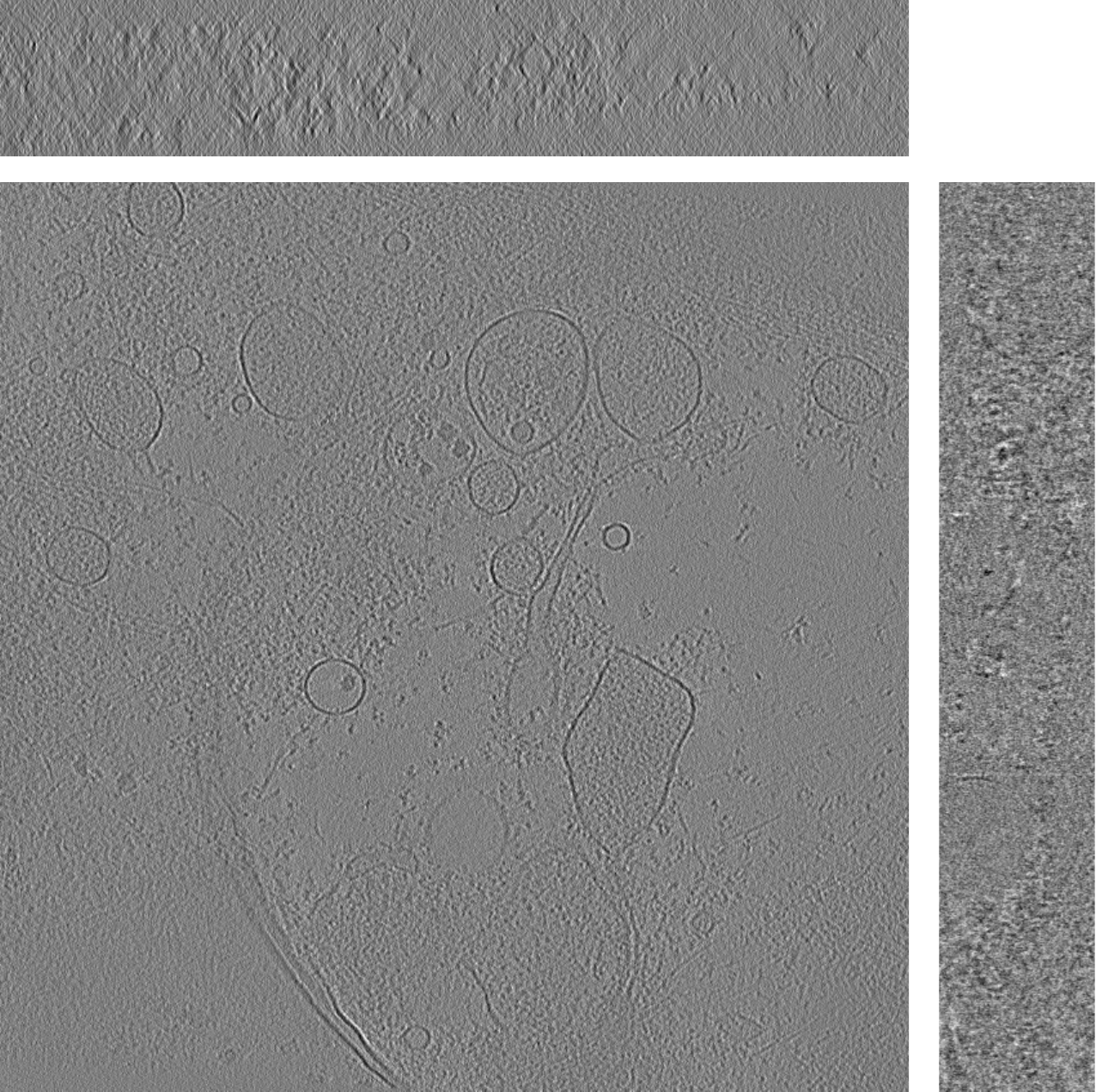}{\sz}{0.316}{50 nm}{I.}  \\
\end{tabular}
\captionof{figure}{Orthogonal slices through a variety of tomograms representing different biological context and imaging conditions using FBP corresponding to Fig.~\ref{fig:mosaic-2}. 
\label{fig:mosaic-2-fbp}}
\end{table*}

\subsection*{Reconstruction of Ribosomes}
We evaluate \method{} on the EMPIAR-10045 dataset, which contains 7 tilt series of purified \textit{S. cerevisiae} 80S Ribosomes.  The tilt series contains 41 aligned projections between -60 to 60 degrees sampled at $2.17$ \AA/px. We apply the different reconstruction approaches on the tilt series from 'tomogram 5' after downsampling it by a factor of 4. Results are shown in Fig.~\ref{fig:real-recons-ribo}.

\def\sztmp{3.9cm}
\def\ps{0.32}
\begin{figure*}
	\begin{subfigure}[t]{\ps\textwidth}
		\centering
		\begin{tikzpicture}[spy using outlines={circle,orange,magnification=4,size=2cm, connect spies}]
			\node[ rotate=90] at (0,0) {}; outlines={circle,orange,magnification=2,size=3cm, connect spies}]
			\node[rotate=0, line width=0.05mm, draw=white] at (0,0) { \includegraphics[height=\sz]{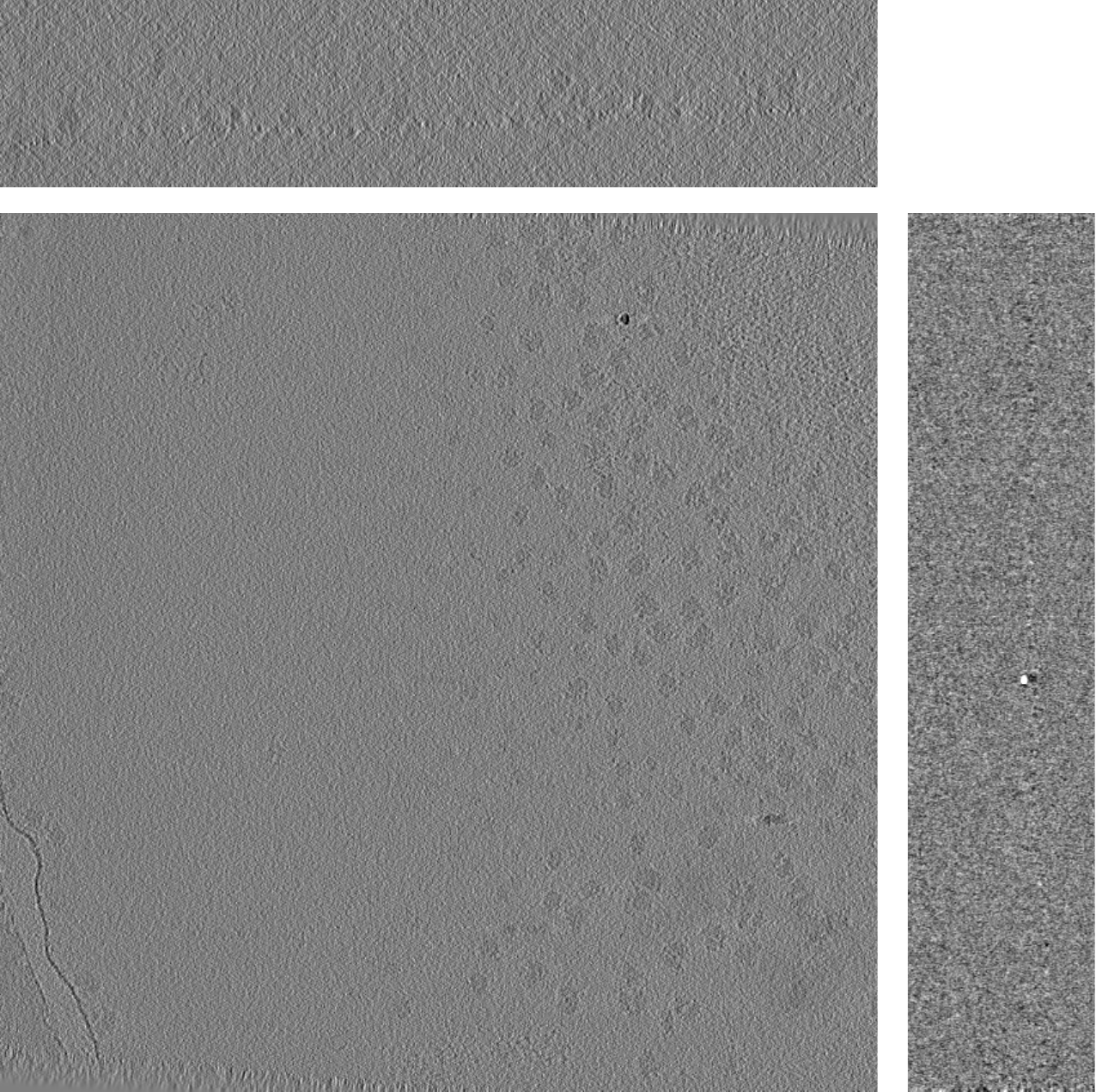}};
			\spy on (-0.1,0.4) in node [left] at (-0.7,0.5);
		\end{tikzpicture} 
		\caption{FBP ($<60$sec).}
	\end{subfigure}\hfill
	\begin{subfigure}[t]{\ps\textwidth}
		\centering
		\begin{tikzpicture}[spy using outlines={circle,orange,magnification=4,size=2cm, connect spies}]
			\node[ rotate=90] at (0,0) {}; outlines={circle,orange,magnification=2,size=3cm, connect spies}]
			\node[rotate=0, line width=0.05mm, draw=white] at (0,0) { \includegraphics[height=\sz]{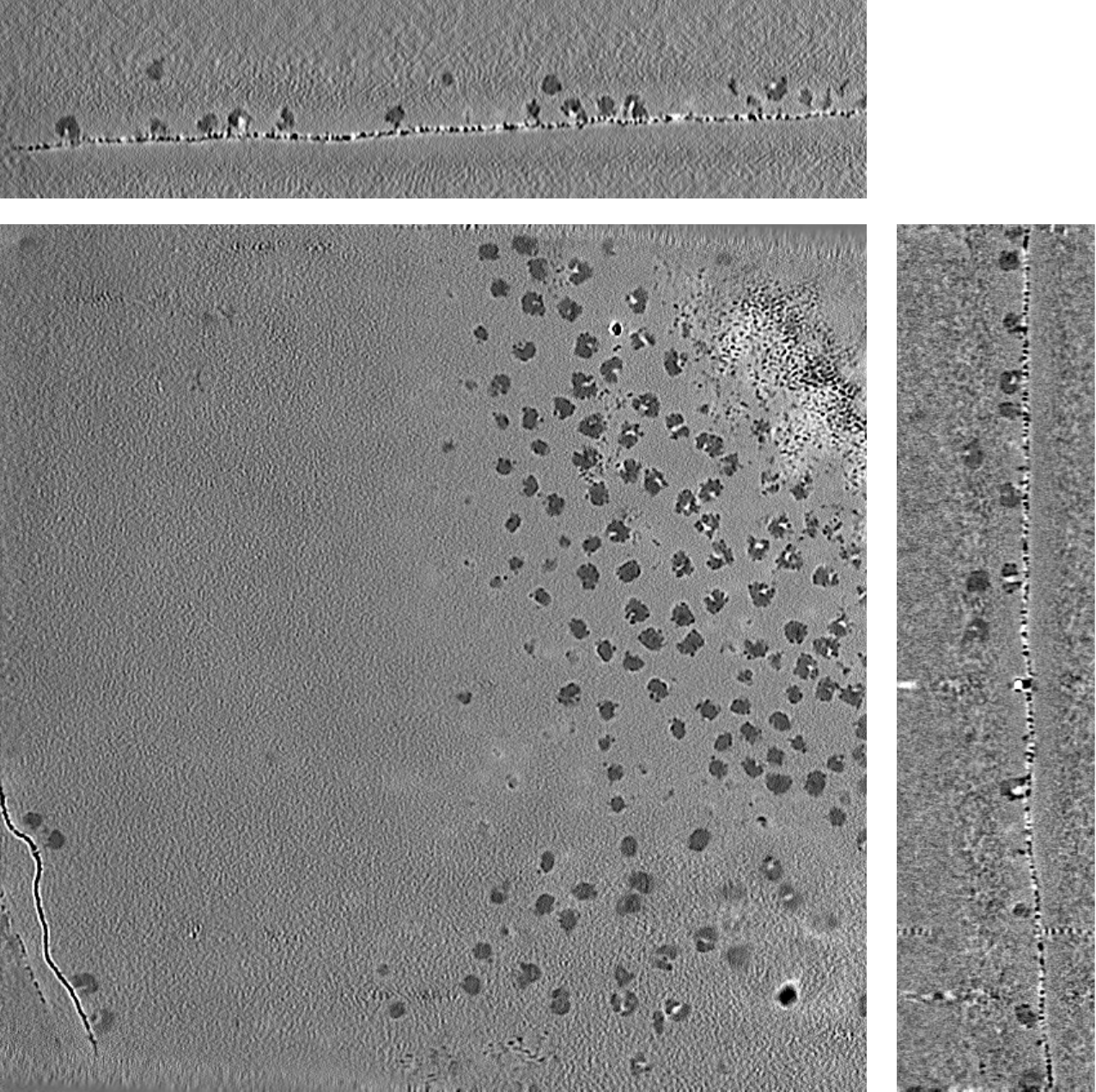}};
			\spy on (-0.1,0.4) in node [left] at (-0.7,0.5);
		\end{tikzpicture} 
		\caption{FBP+DeepDeWedge ($\approx$17h).}
	\end{subfigure}\hfill
	\begin{subfigure}[t]{\ps\textwidth}
		\centering
		\begin{tikzpicture}[spy using outlines={circle,orange,magnification=4,size=2cm, connect spies}]
			\node[ rotate=90] at (0,0) {}; outlines={circle,orange,magnification=2,size=3cm, connect spies}]
			\node[rotate=0, line width=0.05mm, draw=white] at (0,0) { \includegraphics[height=\sz]{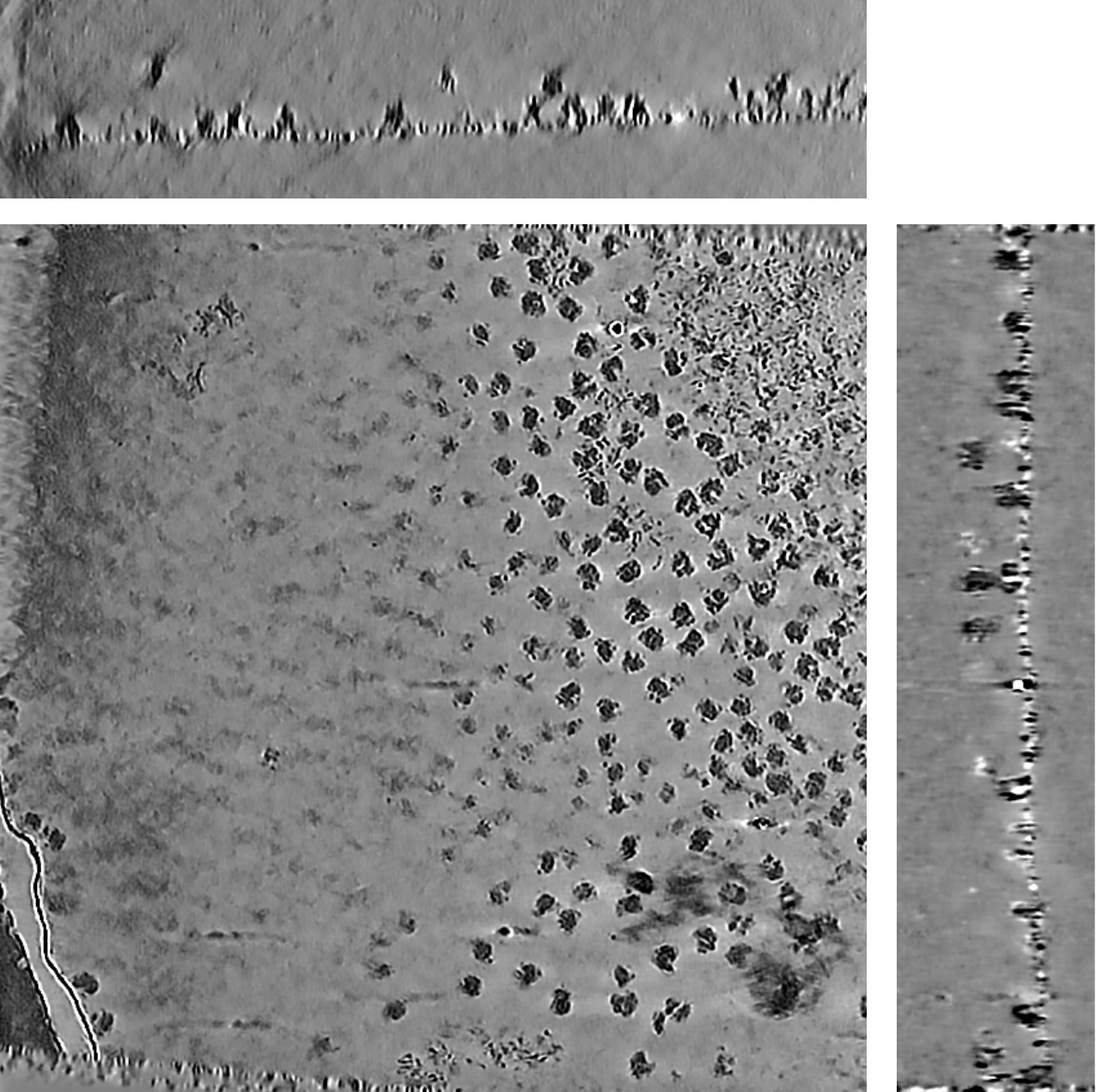}};           
			\spy on (-0.1,0.4) in node [left] at (-0.7,0.5);
		\end{tikzpicture} 
		\caption{\method ($\approx$4min).}
	\end{subfigure}
	\caption{\method{} reconstructs tomograms of similar quality to those obtained from state-of-the-art algorithms, which were not included in the training set, in just a few minutes.}\label{fig:real-recons-ribo}
\end{figure*}

\section{Quantifying denoising performance}
To assess the denoising capabilities of each reconstruction method, we compared their ability to recover empty regions in the tomogram using the HIV-1 dataset (EMPIAR-10643), as shown in Figure~\ref{fig:hiv-background}. All reconstructions were first normalized to have unit standard deviation. We then extracted a subtomogram from within the cell boundary in a region assumed to be empty, subtracted its mean, and plotted the voxel intensity histogram (Fig.~\ref{fig:histo_bg}).
The FBP reconstruction exhibits a large intensity spread, which is indicative of large residual noise. In contrast, \method{} and its wavelet variant yield histograms comparable to FBP+Cryo-CARE+IsoNet, with pixel intensities sharply concentrated around zero, reflecting effective noise suppression. FBP+Cryo-CARE alone performs slightly worse, showing a less pronounced peak at zero. These results suggest that \method{} provides better denoising compared to state-of-the-art pipelines, while avoiding their complexity.

\def\sztmp{3.9cm}
\def\ps{0.242}
\def\scc{1274}
\begin{figure*}
	\begin{subfigure}[t]{\ps\textwidth}
	    \centering
    	\begin{tikzpicture}[spy using outlines={circle,orange,magnification=5,size=2cm, connect spies}]
    		\node[ rotate=90] at (0,0) {}; outlines={circle,orange,magnification=2,size=3cm, connect spies}]
    		\node[rotate=0, line width=0.05mm, draw=white] at (0,0) { \includegraphics[height=\sztmp]{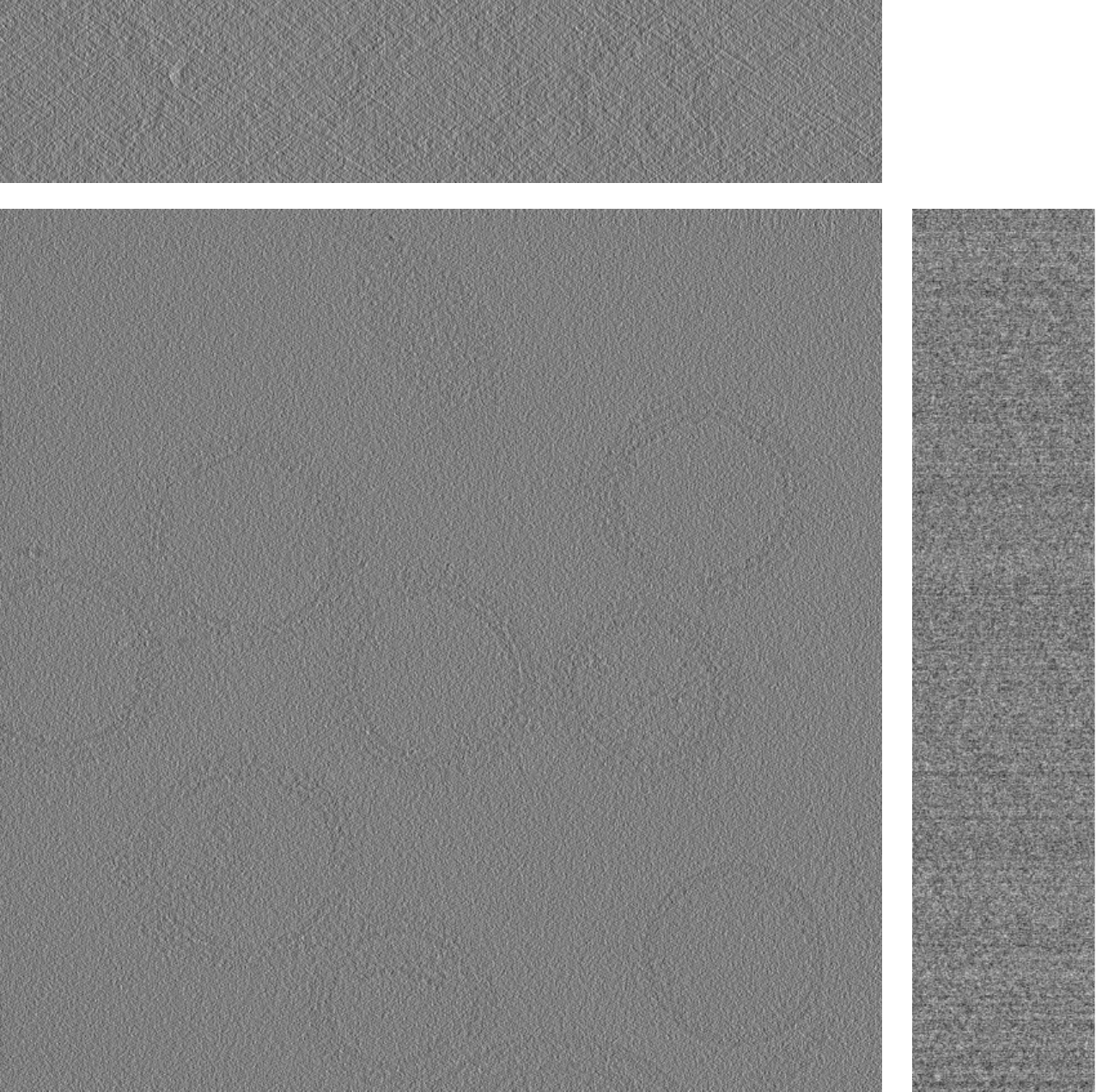}};
    		\spy on (-0.8,-0.5) in node [left] at (2,1);
		\end{tikzpicture} 
		\caption{FBP ($< 60$s).}
	\end{subfigure}\hfill
	\begin{subfigure}[t]{0.27\textwidth}
	    \centering
    	\begin{tikzpicture}[spy using outlines={circle,orange,magnification=5,size=2cm, connect spies}]
    		\node[ rotate=90] at (0,0) {}; outlines={circle,orange,magnification=2,size=3cm, connect spies}]
    		\node[rotate=0, line width=0.05mm, draw=white] at (0,0) { \includegraphics[height=\sztmp]{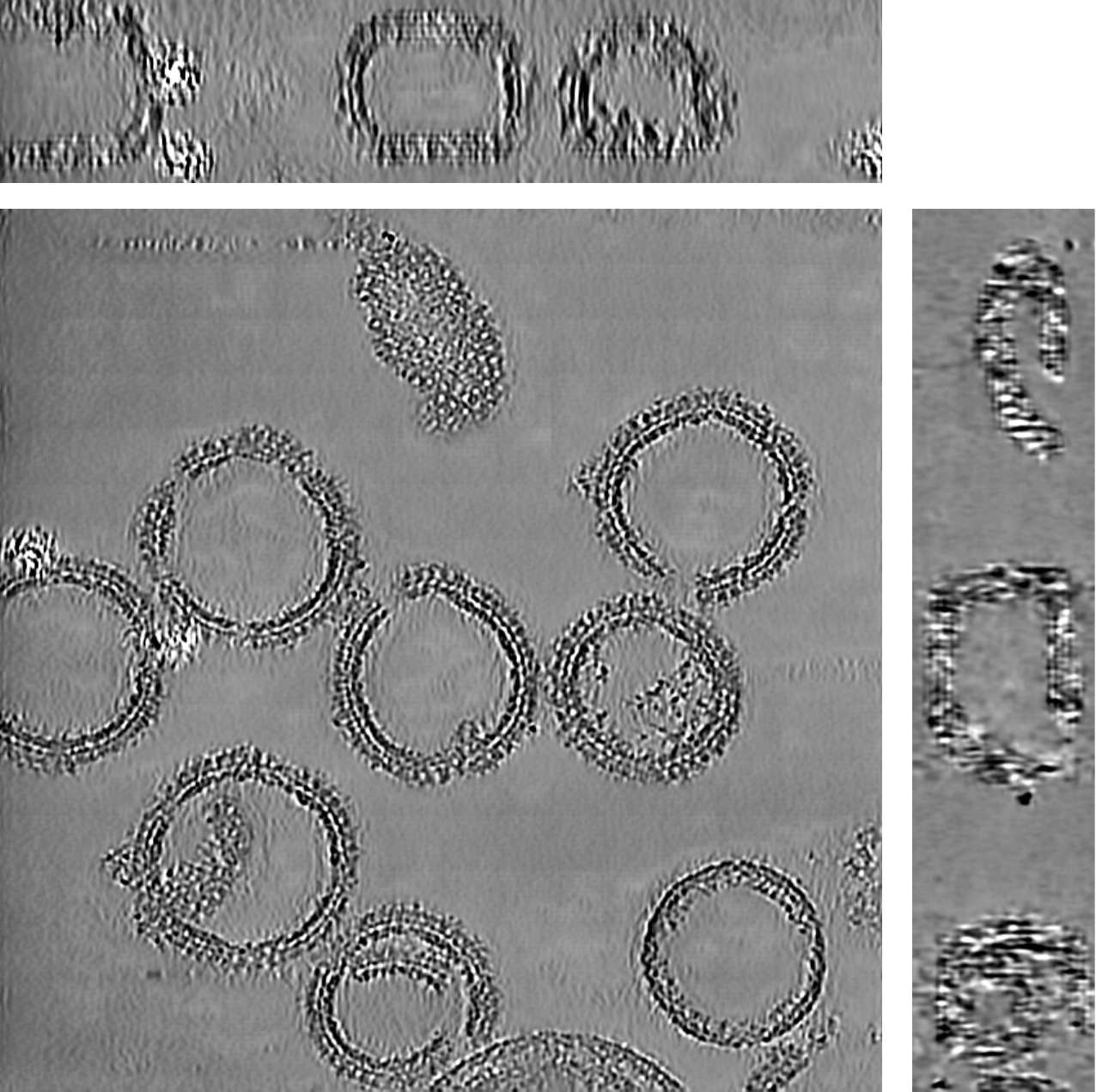}};
    		\spy on (-0.8,-0.5) in node [left] at (2,1);
		\end{tikzpicture} 
		\caption{FBP+Cryo-CARE+IsoNet ($5$h).}
	\end{subfigure}\hfill
	\begin{subfigure}[t]{\ps\textwidth}
	    \centering
    	\begin{tikzpicture}[spy using outlines={circle,orange,magnification=5,size=2cm, connect spies}]
    		\node[rotate=0, line width=0.05mm, draw=white] at (0,0) { \includegraphics[height=\sztmp]{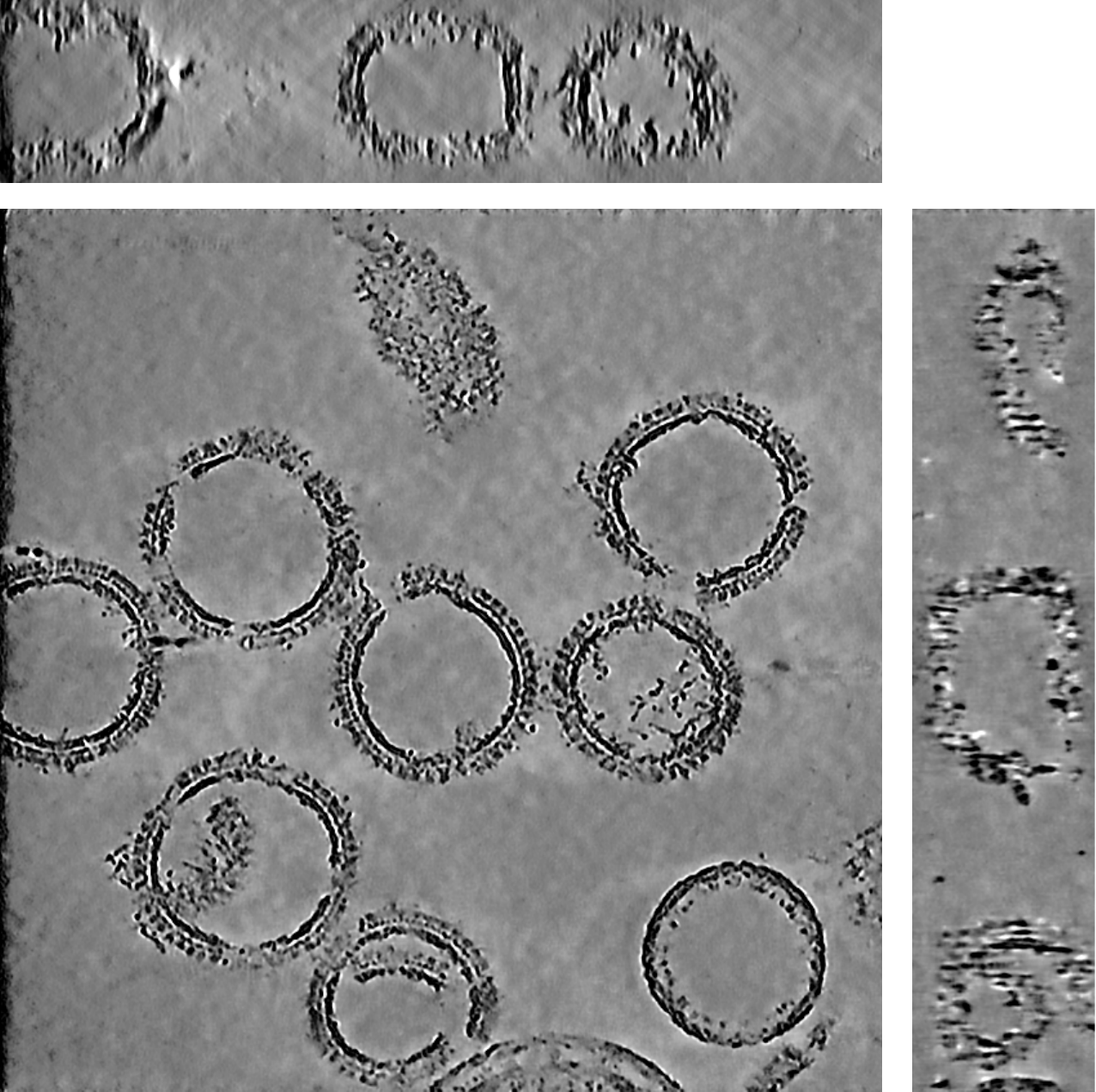}};
    		\spy on (-0.8,-0.5) in node [left] at (2,1);
		\end{tikzpicture} 
		\caption{\method (5min).}
	\end{subfigure}\hfill
        \begin{subfigure}[t]{\ps\textwidth}
	    \centering
          \begin{tikzpicture}
           \begin{axis}[
           xmin=-2, xmax=2,
           ymin=0, ymax=5.8,
           axis lines = left,
           width=1.15\linewidth, 
           height=1.13\linewidth, 
			grid=major, 
			grid style={dashed,gray!30}, 
			ytick={0,0.5,1,2,4},
			xlabel={\notsotiny{Voxel values}},
			ylabel={\notsotiny{Density}},legend style={at={(1,1.1)}, legend cell align=right, align=right, draw=none,font=\tiny}]

            \addplot[mark=\FBPmark, mark size=\ms, line width=\lw,  mark repeat=5, color=\FBPColor] table [x=y, y=x, col sep=comma] {FigureC8/fbp_hist.txt};
            \addlegendentry{FBP}

            \addplot[mark=\Ccaremark, mark size=\ms, line width=\lw,  mark repeat=5, color=\CcareColor] table [x=y, y=x, col sep=comma] {FigureC8/crc_hist.txt};
            \addlegendentry{FBP+Cryo-CARE}
            \addplot[mark=\isoCaremark, mark size=\ms, line width=\lw,  mark repeat=5, color=\isoCareColor] table [x=y, y=x, col sep=comma] {FigureC8/crc_isonet_hist.txt};
            \addlegendentry{FBP+Cryo-CARE+IsoNet}

            \addplot[mark=\isomark, mark size=\ms, line width=\lw,  mark repeat=5, color=\methodWColor] table [x=y, y=x, col sep=comma] {FigureC8/ours_wavelet_icecream_hist.txt};
            \addlegendentry{\method{}}
            \end{axis}
            \end{tikzpicture}
		\caption{Histogram.} \label{fig:histo_bg}
	\end{subfigure}\hfill
\caption{(a-c) Orthogonal slices of the reconstruction of a HIV-1 virions from EMPIAR-10643 dataset using different methods. (d) Histograms of the voxel values at empty locations of the normalized tomograms.
\method{} and FBP+Cryo-CARE+IsoNet produce quantitatively best denoising performance in the sense that the pixel intensity of empty areas are close to their mean and with a small variance.
}\label{fig:hiv-background}
\end{figure*}

\section{Training samples}
We display a random selection of the training tomograms, see Fig.~\ref{fig:mosaic-training}.

 \begin{table*}
 \centering
\begin{tabular}{@{}c@{}c@{}c@{}}
			\overlayerimage{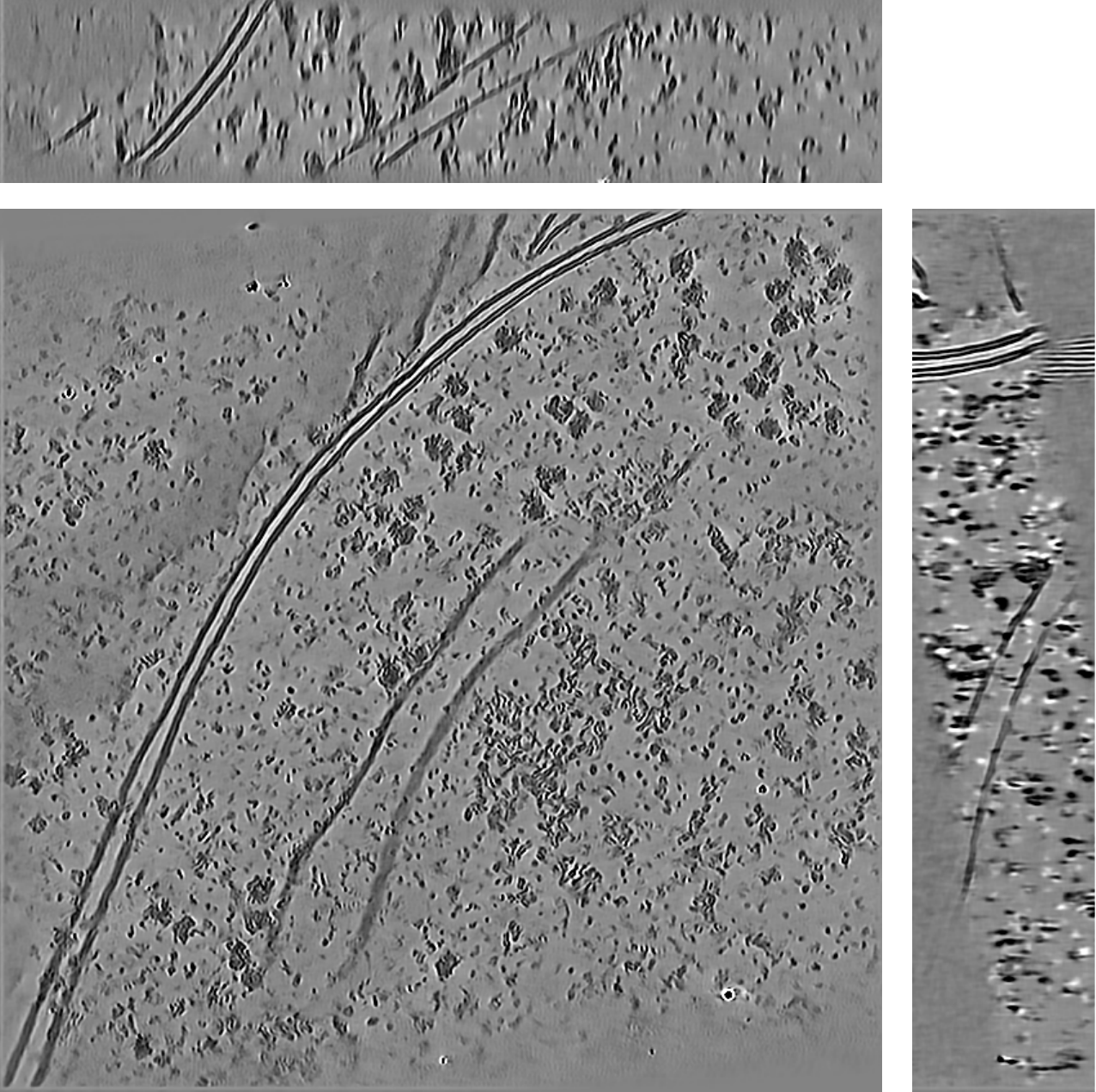}{\sz}{0.4035}{100 nm}{A.} &
			\hspace{0.5mm} 
            \overlayerimage{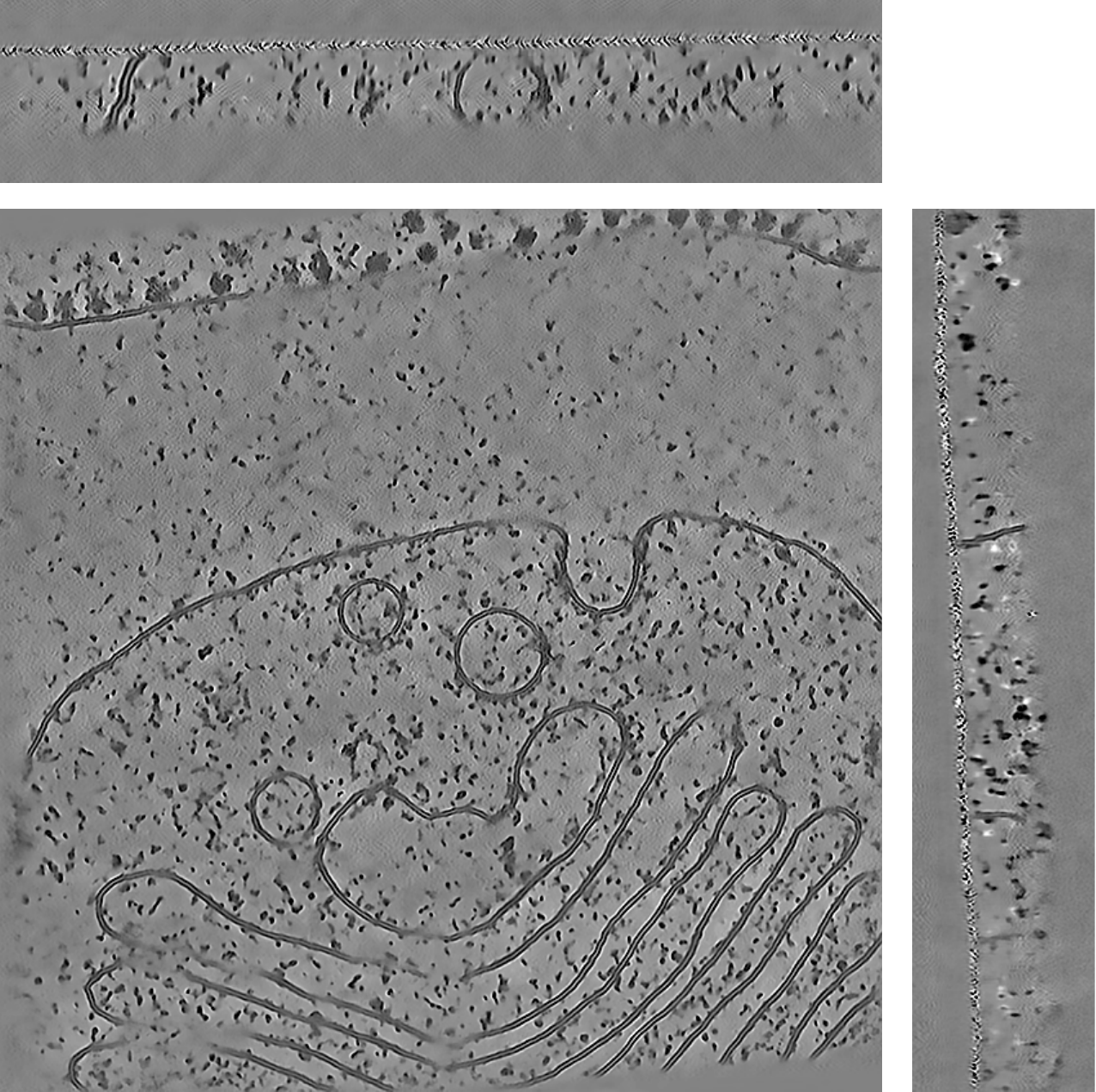}{\sz}{0.4346}{100 nm}{B.} &
			\hspace{0.5mm} 
			\overlayerimage{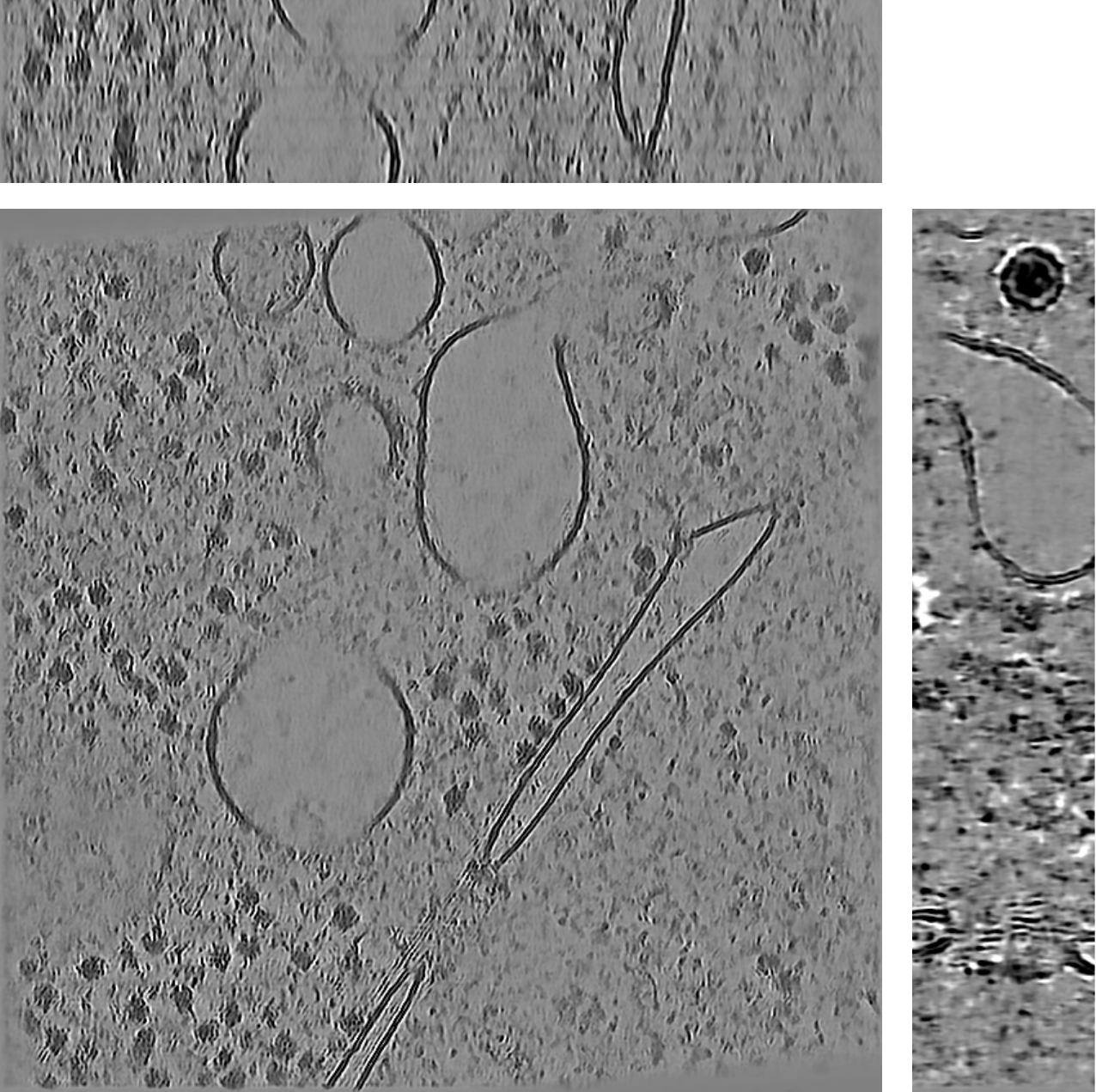}{\sz}{0.4035}{100 nm}{C.}  \\
			\overlayerimage{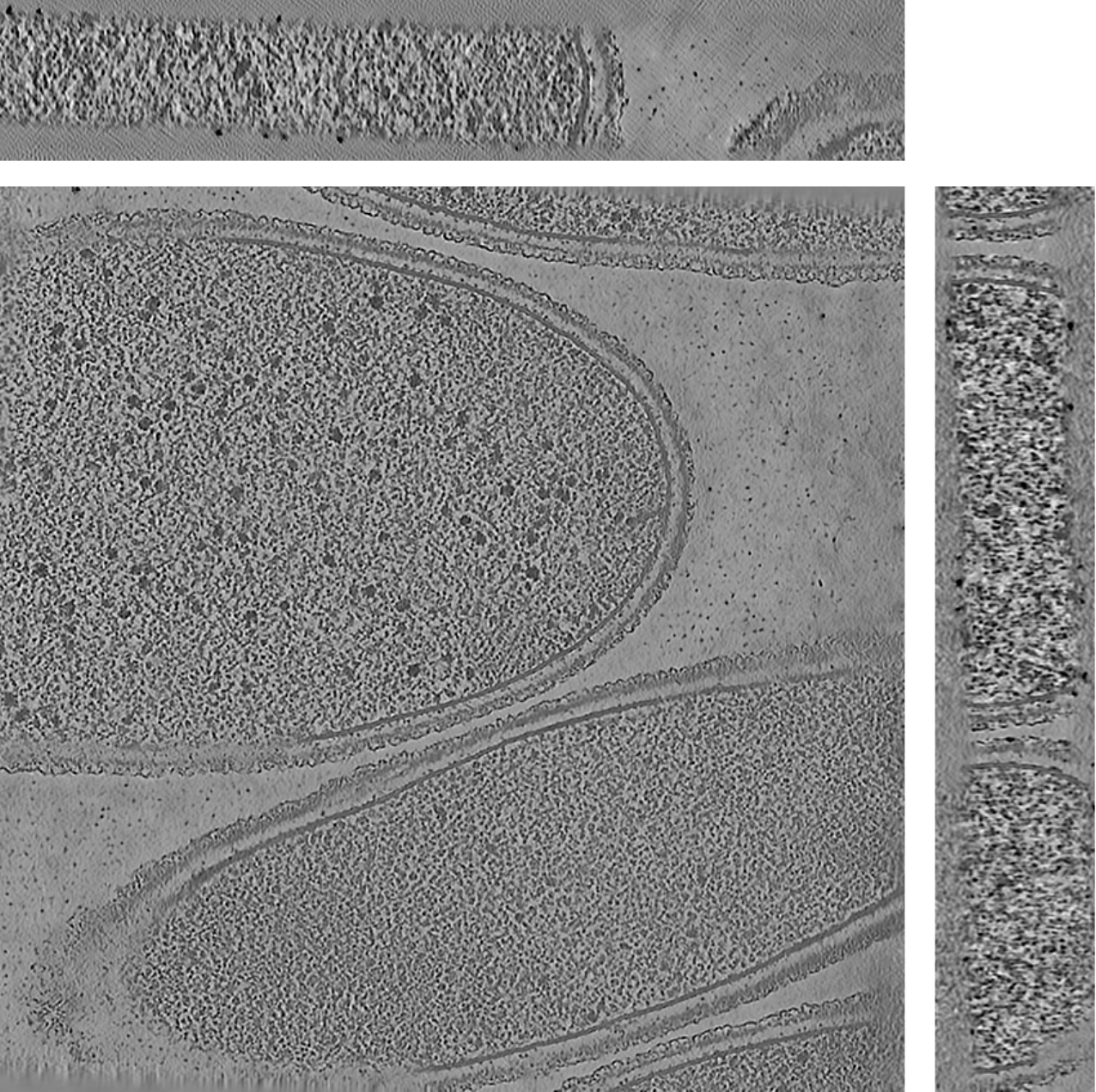}{\sz}{0.2480}{100 nm}{D.}  &
			\hspace{0.5mm} 
			\overlayerimage{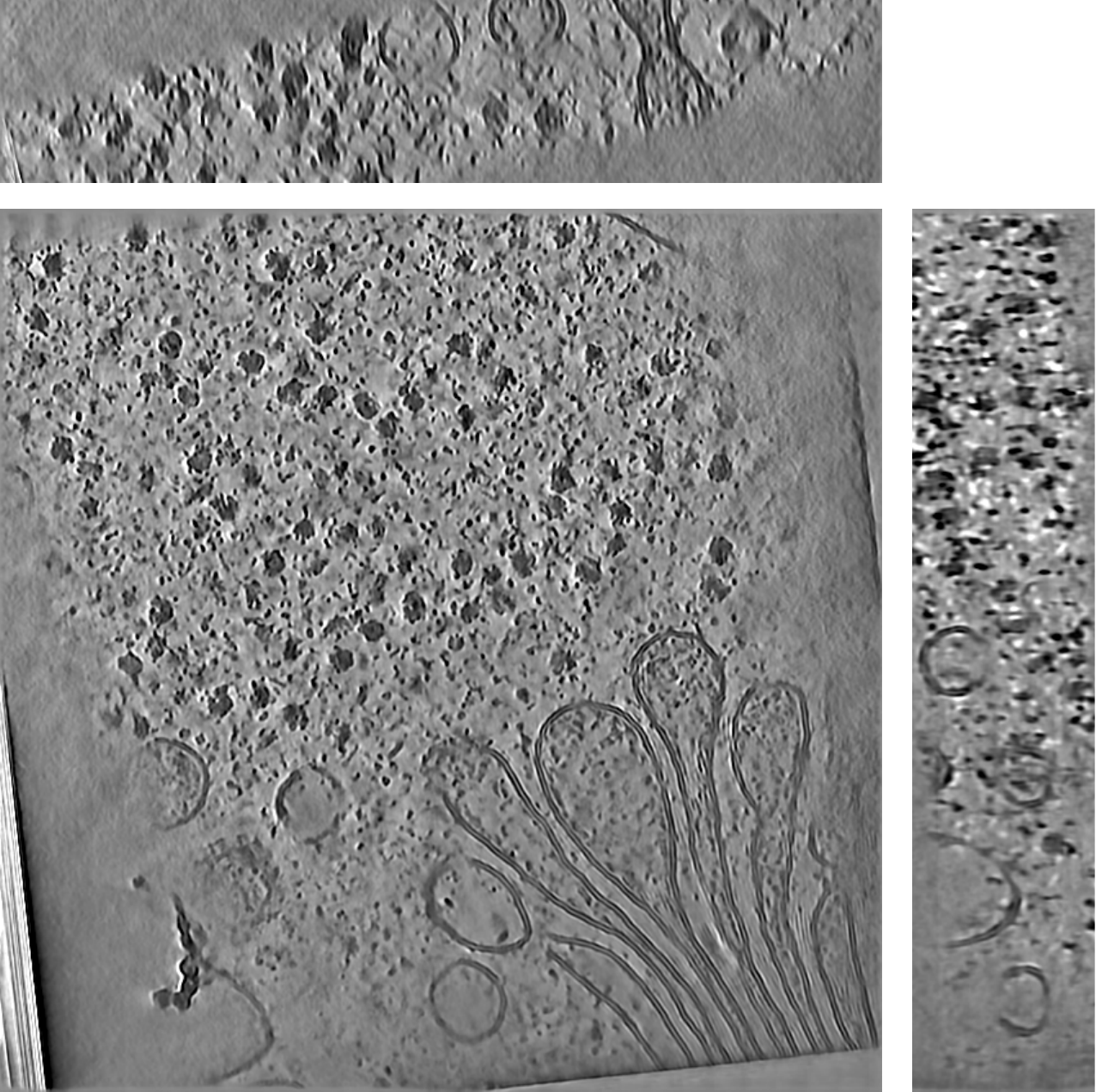}{\sz}{0.4163}{100 nm}{E.}  &
			\hspace{0.5mm} 
			\overlayerimage{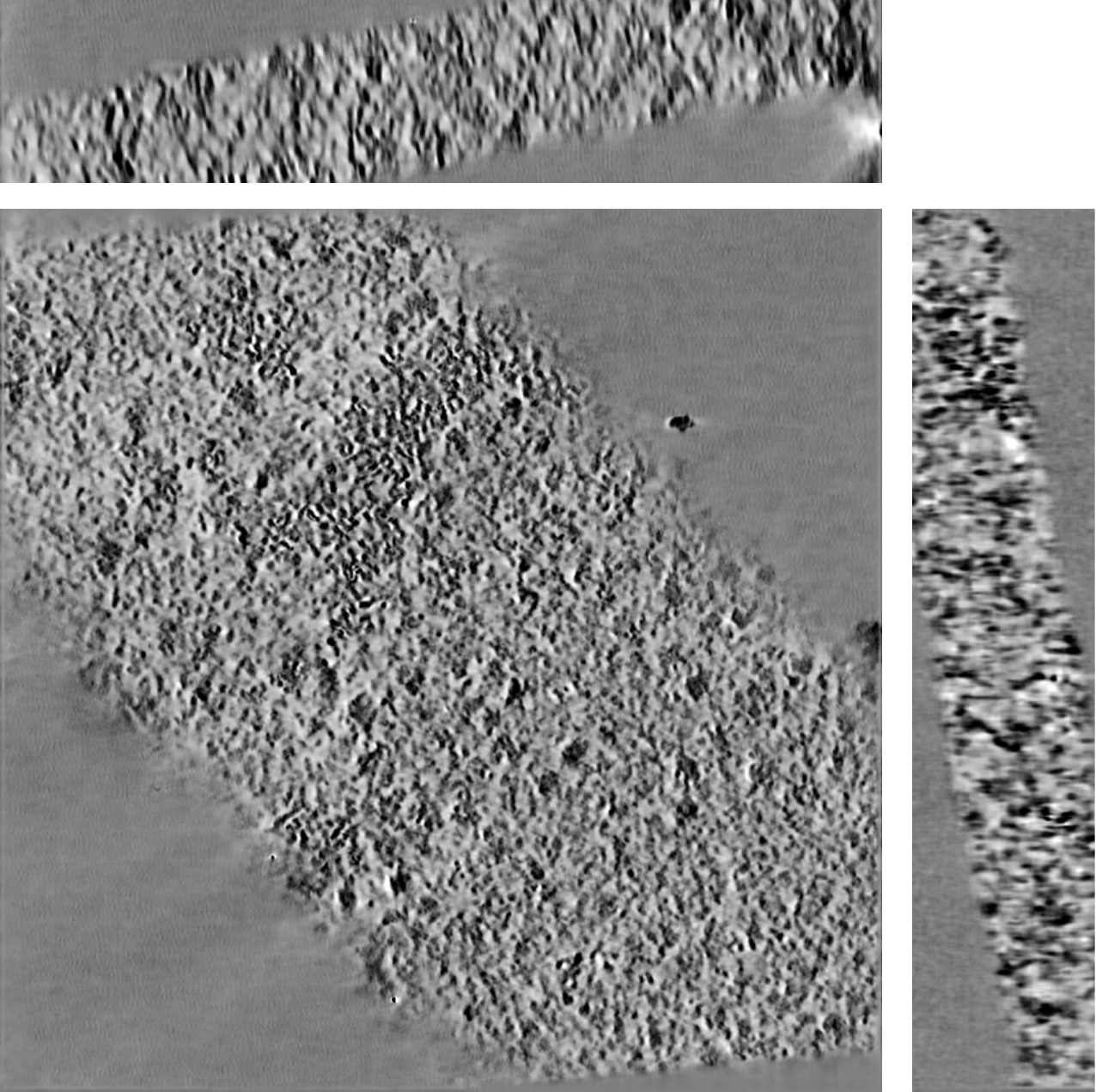}{\sz}{0.4035}{100 nm}{F.}  \\
            \overlayerimage{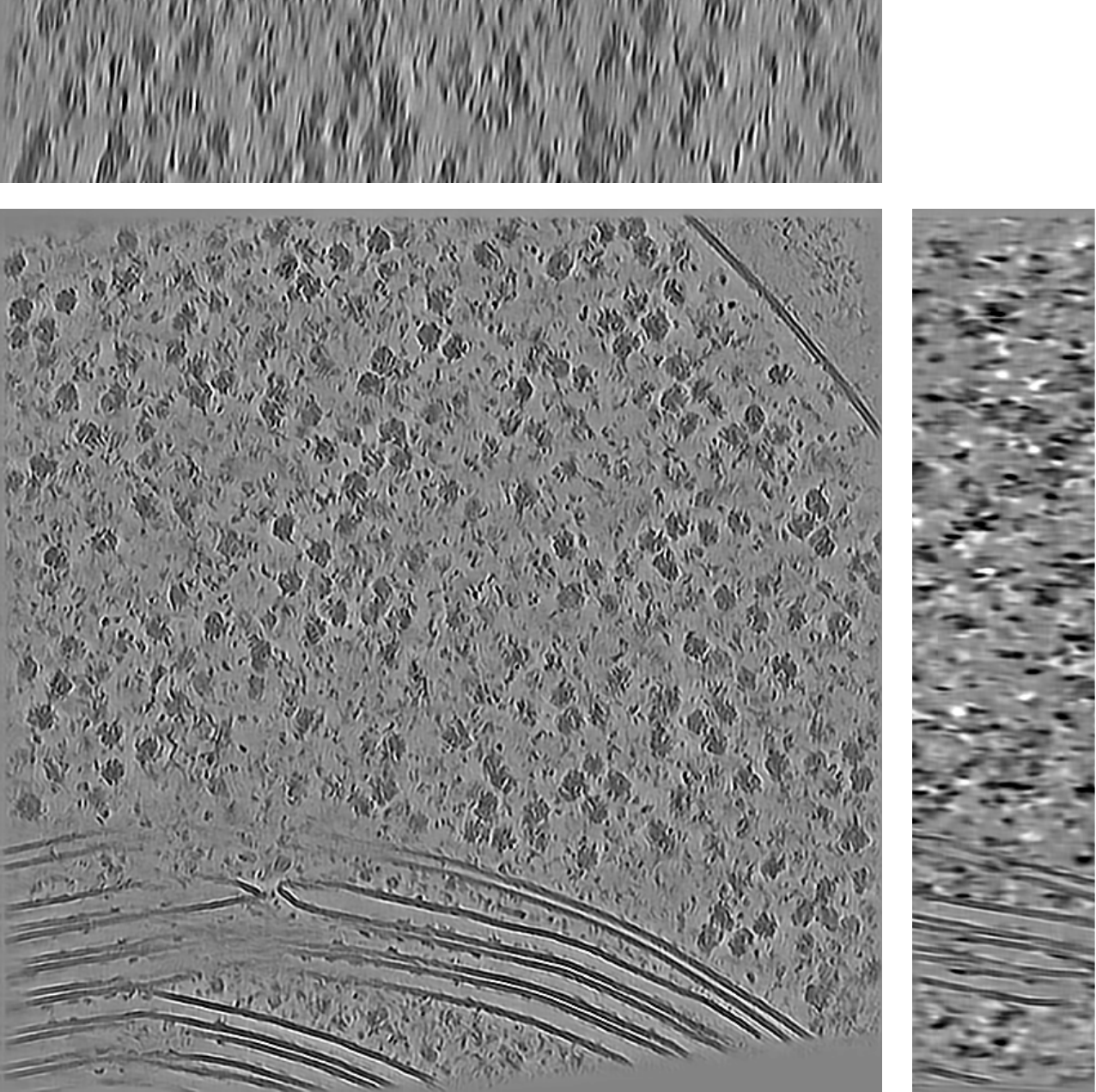}{\sz}{0.4035}{100 nm}{G.} &
			\hspace{0.5mm} 
			\overlayerimage{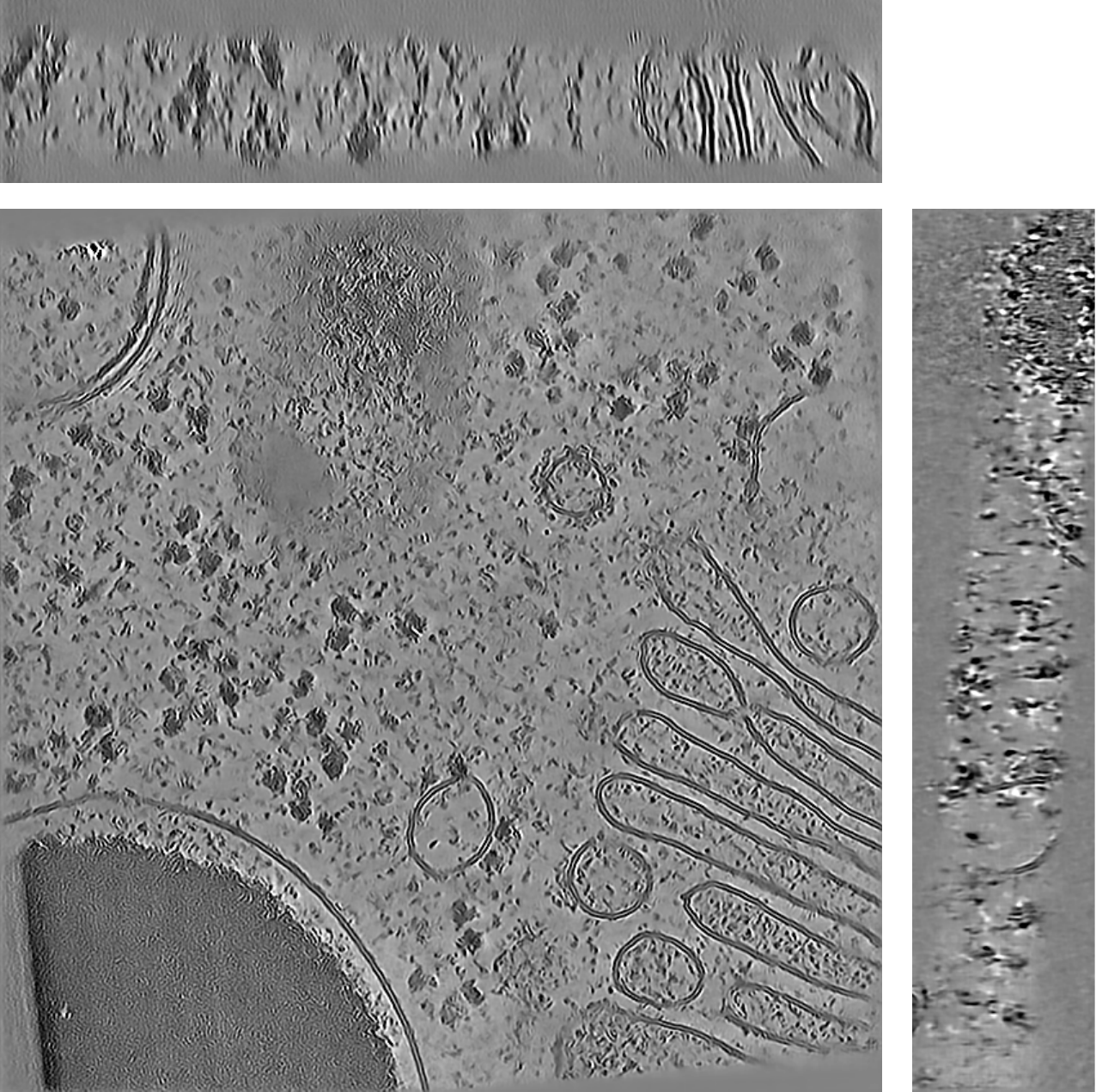}{\sz}{0.4035}{100 nm}{H.}  &
			\hspace{0.5mm} 
			\overlayerimage{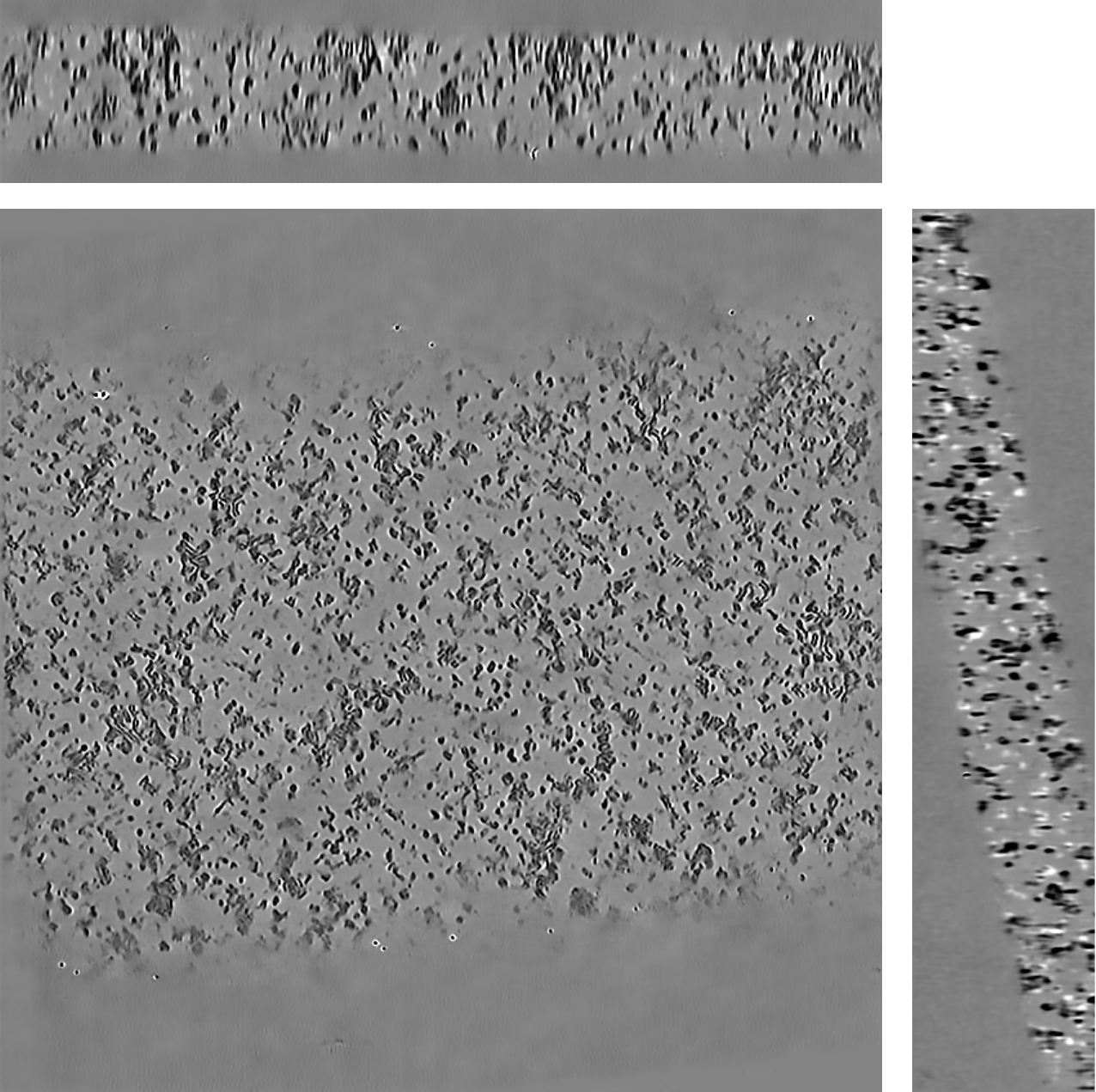}{\sz}{0.4035}{100 nm}{I.}  \\
\end{tabular}
\captionof{figure}{Orthogonal slices through a variety of tomograms used for training. Tomograms are reconstructed from aligned tilt series provided in EMPIAR-11830 using FBP+\icecream{}, except for panel D, which is obtained with FBP+Cryo-CARE+IsoNet. Among the 114 training volumes, 26 are obtained with FBP+Cryo-CARE+IsoNet and 88 with FBP+\icecream{}.  }
\label{fig:mosaic-training}
\end{table*}

\section{Self-FSC on the test data}
We evaluate quantitatively the performance of \method{} by computing the mean FSC between the odd and even split of the tilt series in the test set, see Fig.~\ref{fig:selffsc-test}. 
We used four tilt series from EMPIAR-11830 as test data.
The shaded regions in the plot represent the standard deviation of the FSC curves. Notably, for all volumes, \method{} outperforms the FBP reconstructions.
Notice that we don't compute the FSC for the baseline FBP+Cryo-CARE+IsoNet or DeepDeWedge as they explicitly use the odd and even frames to train the neural network. Hence, the independence assumption, crucial for the self-FSC to be meaningful \cite{verbeke2024self}, is lost.
\def\scc{1024}
\begin{figure}[!htbp]
\centering
		\begin{tikzpicture}
			\begin{axis}[
				axis lines = left,
                xmin = 0, xmax = 0.475,
                ymin = 0, ymax=1,
                width=1.05\linewidth, 
                height=1.13\linewidth, 
                ytick={0, 0.143, 0.5,1},
                yticklabels={0, 0.143, 0.5,1},
                height=5.6cm,
				grid=major, 
				grid style={dashed,gray!30}, 
				xlabel= {\notsotiny{Resolution ($1/ \text{pixel size}$)}},
				ylabel={\notsotiny{FSC}},legend style={at={(1,1)}, legend cell align=left, align=left, draw=none,font=\notsotiny}]
\addplot+[mark=\FBPmark, mark size=\ms, line width=\lw, mark repeat=50, color=\FBPColor]
    table[x expr=\coordindex/\scc, y=mean, col sep=comma] {images_3/test-odd-eve/fsc_fbp_stats_ice.txt};
\addlegendentry{FBP}

\addplot[name path=fbp_upper, draw=none,forget plot]
    table[x expr=\coordindex/\scc, y expr=\thisrow{mean}+\thisrow{std}, col sep=comma]
    {images_3/test-odd-eve/fsc_fbp_stats_ice.txt};

\addplot[name path=fbp_lower, draw=none, forget plot]
    table[x expr=\coordindex/\scc, y expr=\thisrow{mean}-\thisrow{std}, col sep=comma]
    {images_3/test-odd-eve/fsc_fbp_stats_ice.txt};

\addplot[\FBPColor!20, forget plot] fill between[of=fbp_upper and fbp_lower];

\addplot+[mark=\isoCaremark, mark size=\ms, line width=\lw, mark repeat=50, color=\isoCareColor]
    table[x expr=\coordindex/\scc, y=mean, col sep=comma] {images_3/test-odd-eve/fsc_wavelet_stats_ice.txt};
\addlegendentry{\method{}}

\addplot[name path=wavelet_upper, draw=none]
    table[x expr=\coordindex/\scc, y expr=\thisrow{mean}+\thisrow{std}, col sep=comma]
    {images_3/test-odd-eve/fsc_wavelet_stats_ice.txt};

\addplot[name path=wavelet_lower, draw=none]
    table[x expr=\coordindex/\scc, y expr=\thisrow{mean}-\thisrow{std}, col sep=comma]
    {images_3/test-odd-eve/fsc_wavelet_stats_ice.txt};
\addplot[mark=none, black, samples=2] {0.143};

\addplot[\isoCareColor!20, forget plot] fill between[of=wavelet_upper and wavelet_lower];

			\end{axis}
		\end{tikzpicture}
		\caption{The FSC curve is computed using the reconstruction obtained from the odd and even split of the tilt series in the test set. \method{} consistently improves over FBP across all frequencies.}
    \label{fig:selffsc-test}
\end{figure}
\end{appendices}

\end{document}

%% file: preambule.tex
\usepackage{amsmath}

\usepackage{amsthm}
\usepackage{amssymb}
\usepackage{amsfonts}
\usepackage{dsfont}
\usepackage{accents}
\usepackage{bm}
\usepackage{hyperref}

\usepackage{tabularx} 
\usepackage{caption}
\usepackage{multirow}
\usepackage{booktabs}
\usepackage{graphicx}
\graphicspath{{Figures/}}
 \usepackage{subcaption}

 \usepackage{tcolorbox}
 \usepackage{wasysym}
  \usepackage{longtable}

\usepackage{tikz,pgfplots}
\usepgfplotslibrary{groupplots}
\usepgfplotslibrary{fillbetween}
\usetikzlibrary{intersections,calc,arrows,matrix,spy}
\usepackage{color}
\definecolor{darkred}{rgb}{0.6,0,0}
\definecolor{darkgreen}{rgb}{0,0.5,0}
\definecolor{darkblue}{rgb}{0,0,0.5}
\definecolor{SkyBlue}{rgb}{0.53, 0.81, 0.92}
\pgfplotsset{compat=1.5.1}

\newcommand\notsotiny{\@setfontsize\notsotiny\@vipt\@viipt}
\pgfplotsset{every tick label/.append style={font=\notsotiny}}
\tikzset{fontscale/.style = {font=\notsotiny}
    }

\usepackage{algorithm}
\usepackage{algpseudocode}
\usepackage[squaren,Gray]{SIunits}

\newcommand{\vr}{\ensuremath{\boldsymbol{r}}}
\newcommand{\vp}{\ensuremath{\boldsymbol{p}}}

\newcommand{\vR}{\ensuremath{\boldsymbol{R}}}

\newcommand{\relu}{\ensuremath{\text{ReLU}}}
\newcommand{\mlp}{\ensuremath{\text{MLP}}}
\newcommand{\smlp}{\ensuremath{\text{SMLP}}}
\usetikzlibrary{plotmarks}
\usepgflibrary{plotmarks}
\usepackage[normalem]{ulem}
\def\method{CryoLithe}
\def\icecream{{{\sc Icecream}}}
\def\FBPColor{red}
\def\methodColor{blue}
\def\methodWColor{teal}
\def\CcareColor{gray}

\def\isoCareColor{violet}
\def\ddwColor{pink}
\def\FBPmark{o}
\def\methodmark{square}
\def\methodWmark{square}

\def\Ccaremark{square*}
\def\isomark{star}
\def\isoCaremark{diamond}

\def\git{\href{https://github.com/swing-research/CryoLithe}{https://github.com/swing-research/CryoLithe}}
\def\dataref{\href{available-soon.data}{available-soon.data}}

%% file: commands.tex
\def\R{\mathbb{R}}          									 	          

\newcommand{\yb}{\mathrm{\mathbf{y}}}

\newcommand{\gammab}{\mathrm{\bm{\gamma}}}

\newcommand{\thetab}{\mathrm{\bm{\theta}}}

\newcommand{\Pc}{\mathcal{P}}

\usepackage{mdframed}
\usepackage{lipsum}

\newcommand{\overlayerimage}[5]{%
	\begin{tikzpicture}
		\node[inner sep=0] (img) {\includegraphics[width=#2]{#1}};
		\fill[white] (-2.3,-2.2) rectangle (-2.3+#3,-2.2+0.1);
		\node[anchor=north, text=white, font=\footnotesize] at (-2,-2.3) {#4};
		\node[anchor=north, text=white, ] at (-2.4,2.7) {#5};
		\begin{scope}[x={(img.south east)}, y={(img.north west)}]
			
			\fill[white] (0.,0) rectangle (0.1,0.1);
		\end{scope}
	\end{tikzpicture}%
}

%% file: NatureMethods.bbl
\begin{thebibliography}{10}
\expandafter\ifx\csname url\endcsname\relax
  \def\url#1{\burl{#1}}\fi
\expandafter\ifx\csname urlprefix\endcsname\relax\def\urlprefix{URL }\fi
\providecommand{\bibinfo}[2]{#2}
\providecommand{\eprint}[2][]{\url{#2}}
\providecommand{\doi}[1]{\url{https://doi.org/#1}}
\bibcommenthead

\bibitem{navarro2022quantitative}
\bibinfo{author}{Navarro, P.~P.}
\newblock \bibinfo{title}{Quantitative cryo-electron tomography}.
\newblock \emph{\bibinfo{journal}{Frontiers in Molecular Biosciences}} \textbf{\bibinfo{volume}{9}}, \bibinfo{pages}{934465} (\bibinfo{year}{2022}).

\bibitem{mccafferty2024integrating}
\bibinfo{author}{McCafferty, C.~L.} \emph{et~al.}
\newblock \bibinfo{title}{Integrating cellular electron microscopy with multimodal data to explore biology across space and time}.
\newblock \emph{\bibinfo{journal}{Cell}} \textbf{\bibinfo{volume}{187}}, \bibinfo{pages}{563--584} (\bibinfo{year}{2024}).

\bibitem{schaffer2019cryo}
\bibinfo{author}{Schaffer, M.} \emph{et~al.}
\newblock \bibinfo{title}{A cryo-fib lift-out technique enables molecular-resolution cryo-et within native caenorhabditis elegans tissue}.
\newblock \emph{\bibinfo{journal}{Nature methods}} \textbf{\bibinfo{volume}{16}}, \bibinfo{pages}{757--762} (\bibinfo{year}{2019}).

\bibitem{klumpe2021modular}
\bibinfo{author}{Klumpe, S.} \emph{et~al.}
\newblock \bibinfo{title}{A modular platform for automated cryo-fib workflows}.
\newblock \emph{\bibinfo{journal}{Elife}} \textbf{\bibinfo{volume}{10}}, \bibinfo{pages}{e70506} (\bibinfo{year}{2021}).

\bibitem{kelley2022waffle}
\bibinfo{author}{Kelley, K.} \emph{et~al.}
\newblock \bibinfo{title}{Waffle method: A general and flexible approach for improving throughput in fib-milling}.
\newblock \emph{\bibinfo{journal}{Nature Communications}} \textbf{\bibinfo{volume}{13}}, \bibinfo{pages}{1857} (\bibinfo{year}{2022}).

\bibitem{eisenstein2023parallel}
\bibinfo{author}{Eisenstein, F.} \emph{et~al.}
\newblock \bibinfo{title}{Parallel cryo electron tomography on in situ lamellae}.
\newblock \emph{\bibinfo{journal}{Nature Methods}} \textbf{\bibinfo{volume}{20}}, \bibinfo{pages}{131--138} (\bibinfo{year}{2023}).

\bibitem{buchholz2019cryo}
\bibinfo{author}{Buchholz, T.-O.}, \bibinfo{author}{Jordan, M.}, \bibinfo{author}{Pigino, G.} \& \bibinfo{author}{Jug, F.}
\newblock \bibinfo{editor}{IEEE} (ed.) \emph{\bibinfo{title}{Cryo-{CARE}: Content-aware image restoration for cryo-transmission electron microscopy data}}.
\newblock (ed.\bibinfo{editor}{IEEE}) \emph{\bibinfo{booktitle}{2019 IEEE 16th International Symposium on Biomedical Imaging (ISBI 2019)}}, \bibinfo{pages}{502--506} (\bibinfo{organization}{IEEE}, \bibinfo{year}{2019}).

\bibitem{liu2022isotropic}
\bibinfo{author}{Liu, Y.-T.} \emph{et~al.}
\newblock \bibinfo{title}{Isotropic reconstruction for electron tomography with deep learning}.
\newblock \emph{\bibinfo{journal}{Nature communications}} \textbf{\bibinfo{volume}{13}}, \bibinfo{pages}{6482} (\bibinfo{year}{2022}).

\bibitem{wiedemann2024deep}
\bibinfo{author}{Wiedemann, S.} \& \bibinfo{author}{Heckel, R.}
\newblock \bibinfo{title}{A deep learning method for simultaneous denoising and missing wedge reconstruction in cryogenic electron tomography}.
\newblock \emph{\bibinfo{journal}{Nature Communications}} \textbf{\bibinfo{volume}{15}}, \bibinfo{pages}{8255} (\bibinfo{year}{2024}).

\bibitem{kishore2025icecream}
\bibinfo{author}{Kishore, V.}, \bibinfo{author}{Debarnot, V.}, \bibinfo{author}{Righetto, R.~D.}, \bibinfo{author}{Engel, B.~D.} \& \bibinfo{author}{Dokmani{\'c}, I.}
\newblock \bibinfo{title}{Icecream: High-fidelity equivariant cryo-electron tomography}.
\newblock \emph{\bibinfo{journal}{bioRxiv}} \bibinfo{pages}{2025--10} (\bibinfo{year}{2025}).

\bibitem{tegunov2021multi}
\bibinfo{author}{Tegunov, D.}, \bibinfo{author}{Xue, L.}, \bibinfo{author}{Dienemann, C.}, \bibinfo{author}{Cramer, P.} \& \bibinfo{author}{Mahamid, J.}
\newblock \bibinfo{title}{Multi-particle cryo-em refinement with m visualizes ribosome-antibiotic complex at 3.5 {\aa} in cells}.
\newblock \emph{\bibinfo{journal}{Nature methods}} \textbf{\bibinfo{volume}{18}}, \bibinfo{pages}{186--193} (\bibinfo{year}{2021}).

\bibitem{burt2024image}
\bibinfo{author}{Burt, A.} \emph{et~al.}
\newblock \bibinfo{title}{An image processing pipeline for electron cryo-tomography in relion-5}.
\newblock \emph{\bibinfo{journal}{FEBS Open Bio}} \textbf{\bibinfo{volume}{14}}, \bibinfo{pages}{1788--1804} (\bibinfo{year}{2024}).

\bibitem{kak2001principles}
\bibinfo{author}{Kak, A.~C.} \& \bibinfo{author}{Slaney, M.}
\newblock \emph{\bibinfo{title}{Principles of computerized tomographic imaging}}  (\bibinfo{publisher}{SIAM}, \bibinfo{year}{2001}).

\bibitem{harauz1986exact}
\bibinfo{author}{Harauz, G.} \& \bibinfo{author}{van Heel, M.}
\newblock \bibinfo{title}{Exact filters for general geometry three dimensional reconstruction.}
\newblock \emph{\bibinfo{journal}{Optik.}} \textbf{\bibinfo{volume}{73}}, \bibinfo{pages}{146--156} (\bibinfo{year}{1986}).

\bibitem{lehtinen2018noise2noise}
\bibinfo{author}{Lehtinen, J.} \emph{et~al.}
\newblock \bibinfo{title}{Noise2noise: Learning image restoration without clean data}.
\newblock \emph{\bibinfo{journal}{arXiv preprint arXiv:1803.04189}}  (\bibinfo{year}{2018}).

\bibitem{bepler2020topaz}
\bibinfo{author}{Bepler, T.}, \bibinfo{author}{Kelley, K.}, \bibinfo{author}{Noble, A.~J.} \& \bibinfo{author}{Berger, B.}
\newblock \bibinfo{title}{Topaz-denoise: general deep denoising models for cryoem and cryoet}.
\newblock \emph{\bibinfo{journal}{Nature communications}} \textbf{\bibinfo{volume}{11}}, \bibinfo{pages}{5208} (\bibinfo{year}{2020}).

\bibitem{shkolnisky2012viewing}
\bibinfo{author}{Shkolnisky, Y.} \& \bibinfo{author}{Singer, A.}
\newblock \bibinfo{title}{Viewing direction estimation in cryo-em using synchronization}.
\newblock \emph{\bibinfo{journal}{SIAM journal on imaging sciences}} \textbf{\bibinfo{volume}{5}}, \bibinfo{pages}{1088--1110} (\bibinfo{year}{2012}).

\bibitem{quinto2009electron}
\bibinfo{author}{Quinto, E.~T.}, \bibinfo{author}{Skoglund, U.} \& \bibinfo{author}{{\"O}ktem, O.}
\newblock \bibinfo{title}{Electron lambda-tomography}.
\newblock \emph{\bibinfo{journal}{Proceedings of the National Academy of Sciences}} \textbf{\bibinfo{volume}{106}}, \bibinfo{pages}{21842--21847} (\bibinfo{year}{2009}).

\bibitem{purnell2023rapid}
\bibinfo{author}{Purnell, C.} \emph{et~al.}
\newblock \bibinfo{title}{Rapid synthesis of cryo-{ET} data for training deep learning models}.
\newblock \emph{\bibinfo{journal}{bioRxiv}}  (\bibinfo{year}{2023}).

\bibitem{hendriksen2021tomosipo}
\bibinfo{author}{Hendriksen, A.~A.} \emph{et~al.}
\newblock \bibinfo{title}{Tomosipo: fast, flexible, and convenient 3d tomography for complex scanning geometries in python}.
\newblock \emph{\bibinfo{journal}{Optics Express}} \textbf{\bibinfo{volume}{29}}, \bibinfo{pages}{40494--40513} (\bibinfo{year}{2021}).

\bibitem{harastani2024template}
\bibinfo{author}{Harastani, M.}, \bibinfo{author}{Patra, G.}, \bibinfo{author}{Kervrann, C.} \& \bibinfo{author}{Eltsov, M.}
\newblock \bibinfo{title}{Template learning: Deep learning with domain randomization for particle picking in cryo-electron tomography}.
\newblock \emph{\bibinfo{journal}{bioRxiv}} \bibinfo{pages}{2024--03} (\bibinfo{year}{2024}).

\bibitem{gubins2020shrec}
\bibinfo{author}{Gubins, I.} \emph{et~al.}
\newblock \bibinfo{title}{Shrec 2020: Classification in cryo-electron tomograms}.
\newblock \emph{\bibinfo{journal}{Computers \& Graphics}} \textbf{\bibinfo{volume}{91}}, \bibinfo{pages}{279--289} (\bibinfo{year}{2020}).

\bibitem{burley2017protein}
\bibinfo{author}{Burley, S.~K.} \emph{et~al.}
\newblock \bibinfo{title}{Protein data bank (pdb): the single global macromolecular structure archive}.
\newblock \emph{\bibinfo{journal}{Protein crystallography: methods and protocols}} \bibinfo{pages}{627--641} (\bibinfo{year}{2017}).

\bibitem{martinez2023simulating}
\bibinfo{author}{Martinez-Sanchez, A.}, \bibinfo{author}{Lamm, L.}, \bibinfo{author}{Jasnin, M.} \& \bibinfo{author}{Phelippeau, H.}
\newblock \bibinfo{title}{Simulating the cellular context in synthetic datasets for cryo-electron tomography}.
\newblock \emph{\bibinfo{journal}{bioRxiv}} \bibinfo{pages}{2023--05} (\bibinfo{year}{2023}).

\bibitem{lamm2024membrain}
\bibinfo{author}{Lamm, L.} \emph{et~al.}
\newblock \bibinfo{title}{Membrain v2: an end-to-end tool for the analysis of membranes in cryo-electron tomography}.
\newblock \emph{\bibinfo{journal}{bioRxiv}} \bibinfo{pages}{2024--01} (\bibinfo{year}{2024}).

\bibitem{ishemgulova2024endosome}
\bibinfo{author}{Ishemgulova, A.} \emph{et~al.}
\newblock \bibinfo{title}{Endosome rupture enables enteroviruses from the family picornaviridae to infect cells}.
\newblock \emph{\bibinfo{journal}{Communications Biology}} \textbf{\bibinfo{volume}{7}}, \bibinfo{pages}{1465} (\bibinfo{year}{2024}).

\bibitem{chaillet2023extensive}
\bibinfo{author}{Chaillet, M.~L.} \emph{et~al.}
\newblock \bibinfo{title}{Extensive angular sampling enables the sensitive localization of macromolecules in electron tomograms}.
\newblock \emph{\bibinfo{journal}{International Journal of Molecular Sciences}} \textbf{\bibinfo{volume}{24}}, \bibinfo{pages}{13375} (\bibinfo{year}{2023}).

\bibitem{TLGJCM_2023}
\bibinfo{author}{Chaillet, M.~L.} \emph{et~al.}
\newblock \bibinfo{title}{{cryo-ET tutorial dataset for template matching}} (\bibinfo{year}{2023}).

\bibitem{khatter2015structure}
\bibinfo{author}{Khatter, H.}, \bibinfo{author}{Myasnikov, A.~G.}, \bibinfo{author}{Natchiar, S.~K.} \& \bibinfo{author}{Klaholz, B.~P.}
\newblock \bibinfo{title}{Structure of the human 80s ribosome}.
\newblock \emph{\bibinfo{journal}{Nature}} \textbf{\bibinfo{volume}{520}}, \bibinfo{pages}{640--645} (\bibinfo{year}{2015}).

\bibitem{chaillet2025pytom}
\bibinfo{author}{Chaillet, M.~L.}, \bibinfo{author}{Roet, S.}, \bibinfo{author}{Veltkamp, R.~C.} \& \bibinfo{author}{F{\"o}rster, F.}
\newblock \bibinfo{title}{pytom-match-pick: a tophat-transform constraint for automated classification in template matching}.
\newblock \emph{\bibinfo{journal}{Journal of Structural Biology: X}} \bibinfo{pages}{100125} (\bibinfo{year}{2025}).

\bibitem{peck2025aretomolive}
\bibinfo{author}{Peck, A.} \emph{et~al.}
\newblock \bibinfo{title}{Aretomolive: Automated reconstruction of comprehensively-corrected and denoised cryo-electron tomograms in real-time and at high throughput}.
\newblock \emph{\bibinfo{journal}{bioRxiv}} \bibinfo{pages}{2025--03} (\bibinfo{year}{2025}).

\bibitem{kelley_toward_2026}
\bibinfo{author}{Kelley, R.} \emph{et~al.}
\newblock \bibinfo{title}{Toward community-driven visual proteomics with large-scale cryo-electron tomography of {Chlamydomonas} reinhardtii}.
\newblock \emph{\bibinfo{journal}{Molecular Cell}} \textbf{\bibinfo{volume}{86}}, \bibinfo{pages}{213--230.e7} (\bibinfo{year}{2026}).

\bibitem{russo_cryomicroscopy_2022}
\bibinfo{author}{Russo, C.~J.}, \bibinfo{author}{Dickerson, J.~L.} \& \bibinfo{author}{Naydenova, K.}
\newblock \bibinfo{title}{Cryomicroscopy in situ: what is the smallest molecule that can be directly identified without labels in a cell?}
\newblock \emph{\bibinfo{journal}{Faraday Discussions}}  (\bibinfo{year}{2022}).

\bibitem{khorashadizadeh2024glimpse}
\bibinfo{author}{Khorashadizadeh, A.}, \bibinfo{author}{Debarnot, V.}, \bibinfo{author}{Liu, T.} \& \bibinfo{author}{Dokmani{\'c}, I.}
\newblock \bibinfo{title}{Glimpse: Generalized local imaging with mlps}.
\newblock \emph{\bibinfo{journal}{arXiv preprint arXiv:2401.00816}}  (\bibinfo{year}{2024}).

\bibitem{kingma2014adam}
\bibinfo{author}{Kingma, D.~P.}
\newblock \bibinfo{title}{Adam: A method for stochastic optimization}.
\newblock \emph{\bibinfo{journal}{arXiv preprint arXiv:1412.6980}}  (\bibinfo{year}{2014}).

\bibitem{zaheer2017deep}
\bibinfo{author}{Zaheer, M.} \emph{et~al.}
\newblock \bibinfo{title}{Deep sets}.
\newblock \emph{\bibinfo{journal}{Advances in neural information processing systems}} \textbf{\bibinfo{volume}{30}} (\bibinfo{year}{2017}).

\bibitem{mohan2019robust}
\bibinfo{author}{Mohan, S.}, \bibinfo{author}{Kadkhodaie, Z.}, \bibinfo{author}{Simoncelli, E.~P.} \& \bibinfo{author}{Fernandez-Granda, C.}
\newblock \bibinfo{editor}{ICLR} (ed.) \emph{\bibinfo{title}{Robust and interpretable blind image denoising via bias-free convolutional neural networks}}.
\newblock (ed.\bibinfo{editor}{ICLR}) \emph{\bibinfo{booktitle}{International Conference on Learning Representations}} (\bibinfo{year}{2019}).

\bibitem{Brilot2012}
\bibinfo{author}{Brilot, A.~F.} \emph{et~al.}
\newblock \bibinfo{title}{Beam-induced motion of vitrified specimen on holey carbon film}.
\newblock \emph{\bibinfo{journal}{Journal of Structural Biology}} \textbf{\bibinfo{volume}{177}}, \bibinfo{pages}{630--637} (\bibinfo{year}{2012}).

\bibitem{bertiaux_luminal_2025}
\bibinfo{author}{Bertiaux, E.} \emph{et~al.}
\newblock \bibinfo{title}{The luminal ring protein {C2CD3} acts as a radial in-to-out organizer of the distal centriole and appendages}.
\newblock \emph{\bibinfo{journal}{PLOS Biology}} \textbf{\bibinfo{volume}{23}}, \bibinfo{pages}{e3003519} (\bibinfo{year}{2025}).
\newblock \bibinfo{note}{Publisher: Public Library of Science}.

\bibitem{verbeke2024self}
\bibinfo{author}{Verbeke, E.~J.}, \bibinfo{author}{Gilles, M.~A.}, \bibinfo{author}{Bendory, T.} \& \bibinfo{author}{Singer, A.}
\newblock \bibinfo{title}{Self fourier shell correlation: properties and application to cryo-et}.
\newblock \emph{\bibinfo{journal}{Communications Biology}} \textbf{\bibinfo{volume}{7}}, \bibinfo{pages}{101} (\bibinfo{year}{2024}).

\end{thebibliography}
